\documentclass[a4paper,11pt]{book}

\usepackage[brazil]{babel}
\usepackage[latin1]{inputenc}

\usepackage{makeidx}

\advance\topmargin by -2.5cm
\newcommand{\be} { \begin{enumerate} }
\newcommand{\ee} { \end{enumerate} }
\newcommand{\bi} { \begin{itemize} }
\newcommand{\ei} { \end{itemize} }
\newcommand{\bd} { \begin{description} }
\newcommand{\ed} { \end{description} }

\newcommand{\fh}{
\setlength{\unitlength}{1cm}
\begin{picture}(0.2, 0.1)
\put( 0, 0.2 ){\line(1,0){0.2}}
\put(0,0){\line(0,1){0.2}}
\put(0,0){\line(1,0){0.2}}
\put(0.2, 0){\line(0,1){0.2}}

\put( 0.01, 0.21){\line(1,0){0.2}}
\put( 0.02, 0.22){\line(1,0){0.2}}
\put( 0.03, 0.23){\line(1,0){0.2}}
\put( 0.04, 0.24){\line(1,0){0.2}}

\put( 0.21, 0.01){\line(0,1){0.2}}
\put( 0.22, 0.02){\line(0,1){0.2}}
\put( 0.23, 0.03){\line(0,1){0.2}}
\put( 0.24, 0.04){\line(0,1){0.2}}

\end{picture}
}

\newtheorem{assumption}{Assumption}[section]
\newtheorem{prop}{Proposition}[section]

\title{The Green Language}
\author{
José de Oliveira Guimarães\\
Departamento de Computação\\
UFSCar\\
São  Carlos - SP\\
Brasil\\
e-mail: jose@dc.ufscar.br}
\date{\today}

\begin{document}
\bibliographystyle{alpha}
\maketitle
\tableofcontents

\label{deflang}
\label{langdef}

\newpage

\chapter{Introduction}

This Report defines the Green Language, an object-oriented language
being designed at the Computer Science Department of the Federal
University of  São Carlos ({\it Universidade Federal de São Carlos -
\mbox{UFSCar}}). Language Green  separates the subtype
from the subclass hierarchy and supports garbage collection, classes
as first-class objects, parameterized classes,  introspective  reflection and a kind
of run-time metaobjects called shells.

Green will support compile and link-time metaobjects and will have a
special module system. These features have not been completely defined
and are not discussed in this report.

This is not an introduction to object-oriented concepts. We assume the
reader knows at least one object-oriented language well.

\chapter{Basic Elements}

\section{Comments}
In Green, anything between \verb@/*@ and \verb@*/@ is a
comment. Nested comments are

allowed. So the line,\\
\verb@     i = 10; /* comment /* i = 5; old code */ still a comment */@\\
is legal and contains one assignment.

There is another way to specify a comment:
anything after \verb@//@ till the end of the line is a comment.

\section{Basic Types}  \label{basic}

The language basic types are shown in the table below together with
the operators they support.  The comparison operators are supported by
all types.

\begin{center}
\vspace{4ex}
\begin{tabular}{|| l|l ||} \hline
basic type & operators \\ \hline  \hline
char       & \verb@++  --@ \\ \hline
boolean    & \verb@and  or  not xor@ \\ \hline
byte       & \verb@+  -  *  /  %  &  |  ^  ~  ++  --  <<  >>@ \\ \hline
integer    & \verb@+  -  *  /  %  &  |  ^  ~  ++  --  <<  >>@ \\ \hline
long       & \verb@+  -  *  /  %  &  |  ^  ~  ++  --  <<  >>@ \\ \hline
real       & \verb@+  -  *  /  ++  --@  \\ \hline
double     & \verb@+  -  *  /  ++  --@ \\ \hline
\end{tabular}
\end{center}

\verb@+@, \verb@-@, \verb@*@, and \verb@/@ are the arithmetic
operators. The remainder of division of {\tt p} by {\tt t} is given by
``\verb@p%t@''. Operators ``\verb@&@'', ``\verb@|@'', and ``\verb@^@''
 are the bitwise ``and'', ``or'', and ``xor'' (exclusive or).  ``\verb@~@'' is the bit
to bit complement.

Operators {\tt ++} and {\tt --} are only applied to variables and are
prefixed.
They increase (\verb@++@) or decrease ({\tt --}) their operands by
one and do not return a value, unlike C++.

The operators \verb@<<@ and \verb@>>@ are right and left shift of
bits. In an expression \\
\verb@     k << n@\\
{\tt k} may be {\tt byte}, {\tt integer}, or {\tt long} and {\tt n}
may be {\tt byte} or {\tt integer}.

The Green compiler translates source code into C.
The table below shows the mapping between Green and C basic types.

\begin{center}
\begin{tabular}{|| l|l ||} \hline
Green & C \\ \hline  \hline
{\tt char}  & {\tt signed char} \\ \hline
{\tt boolean} & {\tt char} \\ \hline
{\tt byte}    & {\tt unsigned char} \\ \hline
{\tt integer} & {\tt int} \\ \hline
{\tt long}    & {\tt long} \\ \hline
{\tt real}    & {\tt float} \\ \hline
{\tt double} & {\tt double} \\ \hline
\end{tabular}
\end{center}

\vspace{2ex}
So, the semantics of the Green basic
types depends on the semantics of the basic types in the C compiler
and  machine used. We hope to fix the size of {\tt integer} in 32 bits
and introduce 16-bit {\tt short} integers. Chapter~\ref{basictypes}
discusses other features of basic types.

Table~\ref{opp} shows the  precedence of the operators. Operators in the same
box have the same precedence. Unary operators and the assignment
operator are right associative. The operators\\
\verb@     <<   >>   <   <=   >   >=   ==   <>@\\
are neither left or right associative. That means expressions like\\
\verb@     (a << 5 << 1) == b == c@\\
are illegal. Every other operator is left associative. The higher in
the table, the higher the precedence. Any similarities between this
table and one of Stroustrup book \cite{str91} is not a coincidence,
although the operator precedence order in Green is rather different
from C++.

\begin{figure}
\begin{center}
\begin{tabular}{|| l l ||} \hline

\verb@()@ &  method call\\
\verb@[]@ &  subscripting\\
\verb@.@ &  method/variable selection \\ \hline

\verb@~@ &  bitwise complement\\
\verb@+@ & unary plus \\
\verb@-@ & unary minus\\
\verb@not@ & logical not \\
\verb@++@ & increment\\
\verb@--@ & decrement\\
\hline

\verb@<<@ & shift left \\
\verb@>>@ & shift right \\ \hline

\verb@&@ & bitwise and\\ \hline
\verb@^@ & bitwise xor \\ \hline
\verb@|@ & bitwise or \\ \hline

\verb@/@ & divide\\
\verb@&@ & multiply \\
\verb@%@ & remainder \\ \hline

\verb@+@ & binary plus \\
\verb@-@ & binary minus \\ \hline

\verb@==@ & equal\\
\verb@<>@ & not equal \\
\verb@<@ & less than\\
\verb@<=@ & less than or equal \\
\verb@>@ & greater than\\
\verb@>=@ & greater than or equal \\ \hline

\verb@and@ & logical and\\ \hline

\verb@xor@ & exclusive or\\ \hline

\verb@or@ & logical or \\ \hline

\verb@= @ & assignment \\ \hline

\end{tabular}
\end{center}
\caption{Operator precedence table}
\label{opp}
\end{figure}

Operators {\tt and} and {\tt or} finish their execution as soon as
possible. Then, in \\
\verb@     ok = false and good();@\\
\verb@     ok = true or good();@\\
the {\tt good} method would not be called.

\section{Literal Values}

A character constant should be enclosed between \verb@'@ as in language
C. All escape characters of C are valid in Green.

The {\tt boolean} type has two pre-defined values: {\tt true} and {\tt
false}. These are constants that can be cast to {\tt 1} and {\tt 0}, respectively.

A byte value is a literal integer with a postfixed ``{\tt b}''. For
example,\\
\verb@    2b, 32b, 255b@\\
are byte values.

Any literal number without a floating point or exponent is an integer literal
number that can be postfixed with an {\tt i}: {\tt 3200i}, {\tt -1i}.

{\tt long} values must be postfixed with an upper case {\tt L}:\\
\verb@    3L,   3000000L@

Any literal number with a floating point or an exponent is a {\tt real}
number. Example:\\
\verb@    1.1   320E+5    1e-10    1r@\\
A real number may have a trailing ``{\tt r}'' as in {\tt 1r} above.

A double real value must have a ``{\tt d}'' at its end:\\
\verb@    1d    2.1e-10d    445.076d@\\

The rules for valid numbers ({\tt byte}, {\tt integer}, {\tt long}, {\tt real} and
{\tt double}) are the rules of the C compiler used.

No automatic conversion is made among values of the different basic
types. Any conversion must be explicitly made. For example, to assign
a real variable {\tt r} to an integer variable {\tt i} one should
write:\\
\verb@     i = integer.cast(r);@\\
This subject is further discussed in Chapter~\ref{basictypes}.

\section{Identifiers}

An identifier is a sequence of one letter followed by any number of
letters, digits, and underscore (``{\tt \_}''). There is no limit for the
size of an identifier, although there may be problems when translating
long names into C source code. Upper and lower case are considered
different. We expect that the Green compiler will issue a warning if
two identifiers  differing only in the case of the letters are used
in the same scope.

\section{Assignments}

Green uses {\tt =} for assignments which are considered expressions. Then the code\\
\begin{verbatim}
a = b = 1;
c = (i = getInt()) + 4;
while (ch = readCh()) <> '\0' do
  ;
\end{verbatim}
is legal.

\section{Control Statements}

The {\tt if} statement of Green has two legal forms:\\
\verb@     if expr then stat-list endif@\\
or\\
\verb@     if expr then stat-list else stat-list endif@\\
For example,
\begin{verbatim}
if i > 0 and a < b
then
  k = 0;
else
  super.put(k);
endif
\end{verbatim}
is a valid {\tt if} statement.

There is also a {\tt case} command, as shown below:
\begin{verbatim}
case i of      // i = case expression
  0 :          // label expression
    readData();
  1, 2, 3 :    // label expressions
    writeDate();
  4 :
    begin
    k = read();
    write(k + 1);
    end
  end // case
\end{verbatim}

After the keyword {\tt case} there may appear an expression that we
will call ``case expression''. The constant expression of each option
(like {\tt 0}, {\tt 1}, {\tt 2}, {\tt 3}, {\tt 4}) will be called
``label expression''. The case expression should result in:
\bi
\item an object of any type.\footnote{Types and classes are defined
elsewhere.} Then the label expressions should be class names;
\item a value of type {\tt char}, {\tt boolean}, {\tt byte}, {\tt integer}, or {\tt
long}. The type of each label expression should be equal to the type
of the case expression.
\ei

The values ``\verb@0@'', ``\verb@1, 2, 3@'', and ``{\tt 4}'' in the
{\tt case} example just given are called ``{\it labels}''. There may
be a label ``{\tt otherwise}'' chosen when no other label applies.
\begin{verbatim}
i = 3;
case i of
  1 :
    ++j;
  2 :
    --j;
  otherwise :
    error();
end
\end{verbatim}

\section{Loop Statements}

Green supports four different types of loop statements. Since they are
very simple and similar to other language constructions, we do not
believe they complicate the language in any way.

The {\tt for} statement has the syntax:\\
\verb@     for Id = expr1 to expr2 do UnStatBlock@\\
where {\tt Id}  is a {\tt char}, {\tt byte}, {\tt integer}, or {\tt
long}  variable and {\tt expr1} and {\tt expr2}
are  expressions of the same type as {\tt Id}. {\tt UnStatBlock} is either a single
statement or a block of statements delimited by {\tt
begin-end}. Variable {\tt Id} may be declared in the {\tt for} statement:\\
\verb@     for Id : integer = expr1 to expr2 do UnStatBlock@\\

{\tt for} makes {\tt Id} assume the values from {\tt expr1} to {\tt
expr2} while iterating through {\tt UnStatBlock}. Therefore, {\tt
expr2} should be greater than {\tt expr1}. To {\tt Id} is added {\tt
1} in each iteration.
If {\tt expr2} \verb@<@ {\tt expr1}, the {\tt UnStatBlock} is not
executed. The value of {\tt Id} after the {\tt for} is undefined. Of
course, {\tt expr1} and {\tt expr2} are evaluated before the first
iteration. {\tt expr1} is evaluated first. {\tt Id} should be a local
non-array variable.

The {\tt while} statement has the form:\\
\verb@     while expr do@\\
\verb@       UnStatBlock@\\
{\tt UnStatBlock} is executed until the boolean expression {\tt expr}
becomes {\tt false}. If {\tt expr} is {\tt false} the first time it is evaluated,
{\tt UnStatBlock} is never executed.

There is a {\tt repeat}-{\tt until} statement:\\
\verb@     repeat@\\
\verb@       read(month);@\\
\verb@     until month >= 1 and month <= 12;@\\
The statement\\
\verb@     repeat@\\
\verb@       stat-list@\\
\verb@     until expr;@\\
is defined as:\\
\verb@     stat-list@\\
\verb@     while not (expr) do@\\
\verb@       stat-list@\\
That is, {\tt stat-list} is executed until {\tt expr} is {\tt true} and at

least one time.

The statement
\begin{verbatim}
loop
   stat-list
end
\end{verbatim}
is equivalent to
\begin{verbatim}
while true do
  begin
  stat-list
  end
\end{verbatim}
The loop is interrupted when a command {\tt break} is executed. Note
that {\tt break} can only be used within the {\tt loop-end}
statement. It cannot be inside a {\tt for}, {\tt while}, or {\tt
repeat-until} statement that is inside the {\tt loop-end} instruction.

\chapter{Classes and Methods}

\section{Syntax}

\begin{figure}
\begin{verbatim}
class ClassName
    // constructors
  public:
    // methods
  private:
    // methods and instance variables
end
\end{verbatim}
\caption{Syntax of class declaration}
\label{sy1}
\end{figure}

\begin{figure}
\verb@proc MethodName( parameter-list ) : ReturnTypes@\\
\verb@  // Local Variable Declarations @\\
\verb@begin@\\
\verb@  // method body @\\
\verb@end@\\
\caption{Syntax for method declaration}
\label{metn}
\end{figure}

\begin{figure}[t]
\begin{verbatim}
class Store
    proc init() begin end
  public:
    proc get() : integer
      begin
      return x;
      end
    proc set( px : integer )
      begin
      x = px;
      end
  private:
    var x : integer;
end
\end{verbatim}
\caption{Class {\tt Store} in Green}
\label{po1}
\end{figure}

A class is a type declaration used to create objects of that
class. The syntax employed for class declaration is shown in
Figure~\ref{sy1}. A class can declare methods and instance variables,
which are called ``function members'' and ``data members'' in the C++
jargon.

A class can have a public and a private part, as indicated by ``{\tt
public:}'' and ``{\tt private}'', respectively. The public part can
only declare methods. The private part can declare methods and
instance variables and is visible only inside the class
code. There may be just only public and one private part and the
public  should appear before the private part. Before the public part
there may be one or more constructors, all of them with the same name
{\tt init} but with different number and/or parameter types.

A method is declared as shown in Figure~\ref{metn}
where the return value types are optional.
{\tt parameter-list} is a sequence of variables with their types,
as in Pascal and all parameter passings are by value.
The declaration of local variables and instance variables is preceded by the keyword
{\tt var}. The declaration of methods is preceded by \mbox{\tt proc}.
An example of class declaration is in Figure~\ref{po1}.

A local variable can be declared in any place a statement can appear
using the syntax\\
\verb@    var k : integer = 0; @\\
The scope of the variable is the point of declaration till the end of
the method.

In this case the
keyword {\tt var} is used to declare a single variable. Then, the code\\
\verb@    var i : integer = 0, ch : char = '?';@\\
in a method body is illegal.


The return value of a method is set by
the keyword \mbox{\tt return} as in language C. After the command {\tt return} is executed, the method finish its
execution and the control is returned to the caller of the method.
Even if the method is
not supposed to return anything, one can use {\tt return} although
without an argument.


Variables and parameters refer to objects and are similar to
pointers in Pascal/C. Therefore in the code
\begin{verbatim}
...
  var a, b : A;
begin
...
b = A.new();
a = b;
...
end.
\end{verbatim}
the declaration of {\tt a} and {\tt b} does
not lead to the creation of
two objects of class {\tt A}. These are created by the
method
\verb@new@, using dynamic memory allocation:\\
\verb@   b = A.new();@\\
There is no way to deallocate memory, which is made
automatically by the garbage collector.

The statement \verb@a = b@ makes {\tt a} and {\tt b}
 refer to the same object,
thus creating an aliasing. Any alteration in the {\tt a}
object by means
of a message send will be reflected in the object
referred to by {\tt b}, since they refer to the same object.
A variable
of one of the basic types does  contain a value --- it is not a
pointer.

\section{Message Send}

The statement\\
\verb@    a.set(1)@\\
is the sending of the message ``\verb@set(1)@'' to the object {\tt
a}.\footnote{We will use object ``{\tt a}'' instead of ``object {\tt a}
refer to''.} ``{\tt set}'' is the message name and {\tt 1} is the real
parameter.

Suppose {\tt a} was declared as\\
\verb@       var a : Store;@\\
and refers to a {\tt Store} object at run time. Then, the statement
``{\tt a.set(1)}'' will order the run-time system to search for a
method {\tt set} taking one integer parameter in the class of the
object {\tt a}, which is {\tt Store}. This method is found and
executed. Inside method {\tt set} of {\tt Store} (here on, {\tt
Store::set} --- see Figure~\ref{po1}), the statement\\
\verb@     x = px@\\
means ``\verb@a.x = px@'' because message {\tt set} was sent to {\tt
a}.  All references to the instance variables
refer to the instance variables of object {\tt a}.

In a message send the value returned by the executed method need not
to be used, as in C++. Then, the statement\\
\verb@     a.get();@\\
is legal. Of course, the compiler should issue a warning message.

\section{self}

The keyword {\tt self} is a predefined variable that points to the
object that received the message that caused the execution of the
method. So, {\tt a.set(1)} caused the execution of {\tt Store::set}
and inside this method {\tt self} refer to the same object as {\tt
a}. Then, the statement ``{\tt x = px}'' can be rewritten as ``{\tt
self.x = px}''. Variable {\tt self} cannot be used in the left-hand side of an
assignment.

\section{Method Overloading}

Method overloading is supported. A class may
define two methods with the same name if they differ in the parameter
types or number of parameters. Then
the methods
\begin{verbatim}
proc print();
proc print( width : integer );
proc print( w : integer; y : integer );
proc print( x : Figure );
proc print( x : Circle );
proc print( w : Person );
\end{verbatim}
may all belong to the same class.  For limitations on the use of
overloaded methods, read Section~\ref{typeo}.

\section{Assertions}

Assertions are expressions that must be true before and after the
execution of a method. For example, the method
\begin{verbatim}
proc push( x : integer )
  assert
    before not full();
    after  not empty();
  end   // end of the assert clause
  begin
    // push x in the stack
    ...
  end
\end{verbatim}
defines pre and postconditions between the {\tt assert} and the first
{\tt end} keyword. The pre and postcondition are the expressions after the
keywords {\tt before} and {\tt after}. If this method belongs to class
{\tt Stack}, the message send to {\tt x} in\\
\verb@     x = Stack.new();@\\
\verb@     x.push(5);@\\
 makes the system test ``{\tt not full()}''.  If this evaluates to
{\tt false}, one of two things will occur:
\bi
\item method\\
\verb@     proc correctAssertionBefore( mi : MethodInfo )@\\
of class {\tt Stack} will be executed, if class {\tt Stack} has such a
method. ``{\tt mi}'' is an object that describes the method that
caused the exception --- see Chapter~\ref{inr}. This method should be
public. Of course, it is inherited by subclasses;
\item exception {\tt AssertionBeforeException} will be throw if class
{\tt Stack} does not have a method {\tt correctAssertionBefore}.
\ei
After method {\tt push} is executed, the expression ``{\tt not empty()}''
is tested. If this expression evaluates to {\tt false}, one of two
things will occur: if {\tt Stack} has a method {\tt
correctAssertionAfter}, this is executed. Otherwise, an exception {\tt
AssertionAfterException} will be thrown. Note the expressions of parts
{\tt before} and {\tt after} may call public (not private) class methods.

Variables can be declared between  the before and after clauses. There
should be one variable for each {\tt var} keyword and the variable
should be initiated in the declaration. As an example, suppose the
stack should store only positive numbers and {\tt getSize} is the
{\tt Stack} method that returns the number of elements in the
stack. The new method {\tt push} could be:
\begin{verbatim}
proc push( x : integer )
  assert
    before
      x >= 0 and not full();
    var oldSize : integer = getSize();
    after
      not empty() and getSize() == oldSize - 1;
  end

  begin
    // body of push
    ...
    end
\end{verbatim}
{\tt oldSize} is like a local variable whose scope is from its
declaration to the end of the assert clause. All assert variables are
initiated before the method is called. The value returned by a method,
if there is only one, is accessed by variable {\tt result}:
\begin{verbatim}
proc cast( ch : char ) : integer
  assert
    after result >= 0 and result <= 127;
  end
\end{verbatim}

\section{Abstract Classes}
\label{abs}  \label{absc}

A class in which some method bodies may not defined is called {\it
abstract}. The class declaration is preceded by the keyword {\tt
abstract} and so are the methods not fully defined. Although the body
of a method (its local  variables and statements) may not be
specified, its header should. As an example, the class {\tt Container}
of Figure~\ref{container} is abstract and the method bodies of {\tt put}, {\tt get}, {\tt
getSize}, and {\tt getMaxSize} are missing.

\begin{figure}
\begin{verbatim}
abstract class Container
  public:
    abstract proc add( x : integer )

    abstract proc get() : integer

    proc empty() : boolean
      begin
      return getSize() == 0;
      end

    proc full() : boolean
      begin
      return getSize() == getMaxSize();
      end

    abstract proc getSize() : integer

    abstract proc getMaxSize() : integer;
end
\end{verbatim}
\caption{Class {\tt Container} for storing objects}
\label{container}
\end{figure}

To fully define abstract classes we need to specify some features and
restrictions, which is made below.
\bi
\item constructors, defined in Chapter~\ref{classobj}, cannot be
abstract;

\item some methods and instance variables may be fully defined and the
methods may use other class methods;

\item an abstract class may not have any abstract method. It still is
an abstract class as long as the keyword {\tt abstract} precedes the
word {\tt class} in the class declaration;

\item constructors may be defined although an object of an abstract
class cannot be created --- abstract classes are made only to play the
role of superclasses and as a design specification. A variable whose
type is an abstract class can be declared and it can refer to objects
of subclasses of the abstract class;

\item all abstract methods must be in the public or subclass
section. See Section~\ref{subclass} for the definition of subclass section.
\ei

\section{Why Green Does Not Support Operator Overloading ?}

Operator overloading is the use of operators of the basic classes
(\verb@+@, \verb@*@, \verb@<=@, \verb@and@, ...) as method names in
normal classes. In general this makes the language more complex
without adding no power to it. Other reasons for not supporting this
feature are  discussed next.

When one defines a method\\
\verb@     proc +(other : T) : T@\\
in a class {\tt T}, we assume {\tt +} has the same semantics as the
basic-class operator ``{\tt +}''.  That is, ``{\tt +}'' is free from
side effects. The evaluation of ``\verb@x + y@''  does not change the
value of {\tt x} or {\tt y} if they are integers.

It is tempting to introduce a slightly change to the semantics of a
method\\
\verb@     proc +(other : Matrix) : Matrix@\\
of class {\tt Matrix} in such a way the return value is {\tt self}
then saving the creation of a large {\tt Matrix} object that would be
returned by ``{\tt +}''. Parameter {\tt other} would be added to {\tt
self} by {\tt Matrix::x} then producing a side effect. This is the
reason why normal classes cannot declare operators as methods.

Besides the side-effect problem (semantics differences), there is
another related to this: there is no simple way a method can emulate
some operators like {\tt and} and {\tt or} of class boolean. These
methods have short-circuit evaluation. This means an expression\\
\verb@     i < n and v[i] > 0@\\
evaluates to {\tt false} if \verb@i >= n@, regardless of the result of
\verb@v[i] > 0@.

Suppose operator {\tt and} was added to a class {\tt Correct} and {\tt
ok} is a variable of this class. The expression\\
\verb@     ok and v[i] > 0@\\
would be equivalent to\\
\verb@     ok.and( v[i] > 0 )@\\
and should return {\tt false} if {\tt ok} if {\tt false}, without
evaluating ``\verb@v[i] > 0@''. But by the semantics of method call
the real parameter is always evaluated.

For short, Green does not support operator overloading because of the
semantic differences between basic classes and reference classes and
because of the impossibility of simulating some operators.

\section{Expanded Variables}

An instance or local variable may be declared as\\
\verb!var clock : @Clock;!\\
to mean it obeys value instead of reference semantics, like the basic
classes. The declaration alone allocates memory for the object, which
need not to be dynamically allocated.

Variables declared with {\tt @} are constants --- it is a compiler
error to use them in the left-hand side of an assignment. These {\tt
@}-variables correspond to variables of expanded classes in Eiffel or
to non-pointer variables in C++.  They are called {\it expanded}
variables in Green.

An expanded variable or {\tt @}-variable  should be initiated by
sending to it an {\tt init}\footnote{See Section~\ref{init}.} message:\\
\verb@     clock.init(13, 52);@\\
This code replaces a call to method {\tt new} of class {\tt Clock}
that would be made if {\tt clock} were a regular variable:\\
\verb@     clock = Clock.new(13, 52);@\\
Messages {\tt init} can only be sent to expanded variables. That will
never cause an error because an expanded variable will always refer to
an object of its declared class --- an expanded variable cannot
receive objects in assignments. Then, in particular, they cannot
receive objects of subclasses which do not have a particular {\tt
init} method.

If {\tt clock} were a regular variable and ``{\tt clock.init(13,
52)}'' were legal, a runtime type error could occur. Variable {\tt
clock} could refer to an object whose class does not define an {\tt
init} method. These methods never belong, by definition, to a class
type, which is the set of public method signatures.

Expanded variables always refer to the same object at runtime. This
allows the compiler to optimize message sends to them. No runtime
search for a method is necessary --- a method call is made as
efficient as a regular procedure call in procedural languages.

There should not be any cycles in the classes of expanded
variables. That is, it is illegal a class {\tt A} declare an expanded
variable of class {\tt B} and class {\tt B} declare an expanded
variable of class {\tt A}.

\chapter{Class Objects}  \label{classobj}

\section{Classes as Objects}

In Green classes are also objects. They can be passed as parameters,
stored in variables, and used in any place another object can. When we
declare\\
\verb@     class A@\\
\verb@       ...@\\
\verb@     end@\\
we are specifying the methods and instance variables that objects of
class {\tt A} will have. To declare the object that represents class
{\tt A} we write
\begin{verbatim}
object A
    // constructor
  public:
    // methods
  private:
    // methods and instance variables
end
\end{verbatim}
{\tt A} is the name of a classless object created before the program starts its
execution. The identifier {\tt A} is a read-only variable that refers
to the class object. Then {\tt A} is in fact a constant and can be
used as labels of case statements.

        Class objects play the role of metaclasses found in other
object-oriented languages. To keep the language as simple as
possible, we have chosen not to allow a class to have a class. This
causes the infinite regression problem: a class should have a class
that should have a class and so on.

\label{init}
Objects are allocated by methods called {\tt new} of class
objects. These methods are automatically created by the compiler. For
each method {\tt
init} in class {\tt A}\footnote{There may be several because of
overloading.} the compiler creates a {\tt new} method in class object
{\tt A} with the same number of parameters and same types.

Each method {\tt new} allocates memory for a class-{\tt A} object,
initializes some  variables,\footnote{Such as the hidden instance
variable that refers to an array with pointers to the methods of the
object class.} and
sends to this object the corresponding message {\tt init}. Then the
code\\
\verb@     f = Circle.new(30, 50);@\\
will call method \verb@Circle::init(integer, integer)@ after
allocating  memory to the object and doing some internal work.

All methods called {\tt init} are the constructors of the class. Like
other methods they may be overloaded providing alternative ways for
initializing objects of the class. They cannot have return value types
and cannot be abstract (see Section~\ref{absc}). {\tt init} methods
are always put before the {\tt public} section of the class. They have
special scope rules that do not match those of public or private sections.

If a class does not define any {\tt init} method, the compiler will
issue an error message since  no object of this class
can be created. If one wants this, she should declare the class as abstract.

A method called {\tt new} may be declared in the public section of a
{\it class object}. However, there should be no {\tt init} method in
the {\it class} with number of parameters and parameter types equal to
the {\tt new} method. If this were allowed, there would be a collision
between the programmer's method {\tt new} and the method {\tt new}
created by the compiler based on the {\tt init} method of the class.

It is interesting noting a method {\tt new} may be defined in a
class. It will not differ from any other method in any way and will
not be related to the constructors or to {\tt new} methods of the
class object.

The private part of a class object can be used inside the corresponding class:
\begin{verbatim}
object Person
  private:
    var minSalary : real = 800.0;
end

class Person
    proc init( pname : String; psalary : real )
      begin
      if psalary < Person.minSalary
      then
        salary = Person.minSalary;
      else
        salary = psalary;
      endif
      name = pname;
      end
  public:

    ...
  private:
    var name : String;
        salary : real;
    ...
end
\end{verbatim}
This will never cause any error because {\tt Person} is a read-only
variable --- it will always point to an object that has a {\tt
minSalary} variable.    This variable is initialized in its
declaration. This is allowed in class objects but not in
classes. A class object defines an object and as such there is memory
associated to it. Therefore variable {\tt minSalary} exists since
the start of the program. Instance variables in classes cannot be
initialized in the declaration because classes are just types ---
there is no memory associated to them. Besides that, a type is a
static declaration that should not be mixed with run-time commands
as assignments.

Green supports a shortcut to the common assignment \\
\verb@     b = B.new();@\\
in which the type of {\tt b} is {\tt B}. This assignment can be
written as \\
\verb@     b#init();@\\
This makes life easier in case of arrays (seen ahead) and long class
names.

  Although class object {\tt Person} is related to class {\tt Person}, they
represent different things. Therefore, to use variable {\tt minSalary} of
class object {\tt Person} inside class {\tt Person} it is necessary to use the dot:\\
\verb@     Person.minSalary@

In fact, there are only two special relationships between a class
object {\tt A} and class {\tt A}:
\be
\item the compiler adds a method {\tt new} to the class object for
each method {\tt init} of class {\tt A} and;
\item class {\tt A} can manipulate the private part of the class
object --- the opposite is not true.
\ee

Since a class object is created only once before the program starts,
there is only one instance of it at run time. A class object {\tt A}
cannot be abstract even if class {\tt A} is. A class object is an
object and therefore represents a real entity which is, of course, not
abstract.

A class object may declare a method {\tt init} without parameters
before its public section. This  method is called after the creation of the
class object before the program starts. There is no calling order
among the {\tt init} methods of the classes of a program. The {\tt
init} method of a class object  can be called by any class object
method,  although this will be rarely necessary. Since the {\tt init}
method of a class object is not public, it cannot be called outside
the class object itself.

\section{Constant Declaration}

Constants can be declared
in the {\tt public} or {\tt private} section of a class object as in
the example:
\begin{verbatim}
object Compass
  public:
    const
      North = 1,
      South = 2,
      East  = 3,
      West  = 4;
end
\end{verbatim}
These constants are accessed using the dot, as in
``\verb@Compass.North@''.

The type of a constant may be specified:\\
\verb@     const@\\
\verb@       MaxCh : integer = 64,@\\
\verb@       Last  : char = '\0';@\\
Otherwise, the constant type will be the same as the value type. Only
 constants of strings (Chapter~\ref{strings}) and basic types can be
 declared.  The value after ``{\tt =}'' may be an expression but the
 compiler should be able to evaluate it.

 Constants cannot be declared
 in classes. If this were allowed we would have two different ways of
 declaring constants (in classes and class objects), which is
 bad. Besides that, a type declaration like a class should specify
 only the properties of each of its objects, not constants shared by
 all objects of a class. This is the job of class objects that are
 equivalent to metaclasses of other languages.

\section{Enumerated Constants}

Enumerate constants are declared as\\
\verb@     enum(red, yellow, green, blue);@\\
in a class object.
This declaration is exactly the same as\\
\verb@     const@\\
\verb@       red = 0, yellow = 1, green = 2, blue = 3;@\\
or
\begin{verbatim}
const
  red    : integer = 0,
  yellow : integer = 1,
  green  : integer = 2,
  blue   : integer = 3;
\end{verbatim}
An integer value can be assigned to an {\tt enum} constant:\\
\verb@     enum(red, yellow = 5, green, blue);@\\
This is equivalent to \\
\verb@     enum(red, yellow = 5, green = 6, blue = 7);@\\

\section{Where the Program Execution Starts}

The execution of a program starts in a method {\tt run} of a class
object specified at link time. Method {\tt run} may take a {\tt
String} array as parameter:\\
\verb@     proc run( args : array(String)[] )@\\
This array is initialized with the program arguments before the
program starts. Method {\tt run} may not take parameters. Anyway, only
one method {\tt run} must be defined in the class object and in {\tt
one} of the signatures just described.

\chapter{Inheritance}

\section{Introduction}

\begin{figure}
\setlength{\unitlength}{1cm}
\begin{picture}(12, 3)(-3.3, 0)
\thicklines
\put( 1, 1.2 ){{\tt B}}
\put( 1.1, 1.5 ){\vector(0, 1){1}}
\put( 1, 2.65 ){{\tt A}}
\put( 7, 1.2 ){{\tt v}}
\put( 7.3, 1.3 ){\vector(1, 0){1.2}}
\put( 8.9, 1.3 ){\circle{0.5}}
\put( 8.75, 1.2 ){$\cal Q$}
\end{picture}
\caption{a) Class {\tt B} inherits from class {\tt A} \hspace*{3ex} b) Variable {\tt v} refer to object $\cal Q$}
\label{def1}
\end{figure}

Inheritance of a class {\tt A} by a class {\tt B} is set by the
keyword {\tt subclassOf}:\\
\verb@     class B subclassOf A ... end@\\   \label{metclash}
The inheritance of a class by another will be represented
graphically as shown in Figure~\ref{def1}. It will be used an
arrow from the subclass to the superclass. So, a superclass will
be always above its subclasses in a figure. An object will be
represented as a small circle and a reference of a variable to \label{graphrep}
an object
is represented as an arrow. See Figure~\ref{def1} where variable
{\tt v} refer to  object $\cal Q$.

Multiple inheritance is not allowed although multiple subtyping,
defined ahead, is.
In this report, we will call ``superclasses of {\tt A}''  all direct and
indirect superclasses of {\tt A}.

A class may be specified as ``final'' using a compiler
option.\footnote{Yes, this is similar to Java final classes.} Final
classes cannot be subclassed and they do not have subtypes (defined
elsewhere). That means a variable of a final class {\tt Store} can only
refer to {\tt Store} objects, making optimizations much easier --- there
is no polymorphism in this specific case.

Some of the Green built-in or standard classes are final: {\tt String} and
all wrapper classes ({\tt Char}, {\tt Boolean}, {\tt Byte}, {\tt
Integer}, {\tt Real}, {\tt Double}).

We have chosen not to introduce a {\tt final} keyword in Green. This
final mechanism is much like a compiler optimization and therefore
belongs to the realm of language implementation. Object-oriented
programming has no concept of ``final'' classes and  real-world
entities need not to be modeled as such.\footnote{Of course, one can always
invent some class that must not have any subclass but that will be
very rare.} Therefore,  we decided to have final classes but in the
domain of the compiler and not in the language. There may be even other
optimizations like this and I hope to discuss how they could be made
in a yet-to-be-made chapter about compile-time metaobjects.

\section{The {\tt Any} and {\tt AnyClass} Classes}

Class {\tt AnyClass} inherits from {\tt Any} and
    is automatically inherited by any class that
does not explicitly inherit from another class. So, {\tt Any} and {\tt
AnyClass} are
superclasses of all other classes in any program, including the array
classes. No object of these classes can be created --- they are abstract.
Class {\tt Any} is the equivalent of {\tt Object} in
Smalltalk and Java.

Class {\tt AnyClassObject} is an abstract class that inherits from
{\tt Any}. A variable of {\tt AnyClassObject} can refer to any class
object. For example, one can write\\
\verb@     var a : AnyClassObject;@\\
\verb@     a = Circle;@\\
in which {\tt Circle} is a class. In this example, {\tt
AnyClassObject} may be replaced by {\tt Any} although this is not
generally {\tt true} as will be seen elsewhere.

Class {\tt Any} cannot be directly inherited since it lacks some
methods all classes should inherit and that are defined in class {\tt AnyClass}.
Class {\tt Any} defines the methods specified below. These methods
will work correctly with any object, even though they do not know at compile time the
structure of the  object (its instance variables).

\be
\item \verb@toString() : String@

returns the empty string {\tt ""}. This should be redefined in
subclasses to return a string representation of the object.

\item \verb@isObjectOf( aClass : AnyClassObject ) : boolean@

returns {\tt true} if {\tt aClass} is the receiver class or its
superclass. If {\tt Circle} is subclass of {\tt Figure}, the {\tt if}
expression will always be true:
\begin{verbatim}
var f : Figure;
f = Circle.new(10, 30, 7);
if f.isObjectOf(Figure) and
   f.isObjectOf(Circle) and
   not f.isObjectOf(Square)
then
  ...
endif
\end{verbatim}

\item \verb@shallowClone() : Any@\\
      \verb@            ( exception : CatchOutOfMemoryException )@

creates and returns a new object of the same class as the
object that received this message. The instance variables of the
receiver are copied to the new object. If there is not sufficient
memory for the operation, exception {\tt OutOfMemoryException} is thrown.

\item \verb@deepClone() : Any@\\
      \verb@         ( exception : CatchOutOfMemoryException )@

creates and returns a new object of the same class as the
receiver of this message. It creates a copy of each object referred
directly or indirectly by the receiver and assembly the new object
 references to match that of the receiver object. If there is not sufficient
memory for the operation, exception {\tt OutOfMemoryException} is thrown.

\item {\tt shallowCopy( other : Any ) : boolean;}

copies the instance variables of the object pointed by {\tt other}
into {\tt self}. If the classes of {\tt self} and {\tt other} are
different from each other, this method returns {\tt false}.\footnote{
Note that {\tt  shallowCopy} returns {\tt false} even when the class
of object {\tt other} is a subclass of the class of {\tt self}.} No
object is created by this method.

\item {\tt shallowEqual( other : Any ) : boolean}

compares the instance variables of {\tt self} and {\tt other} using
the operator \verb@==@ and returns {\tt false} if some test returns
{\tt false}.

The operator {\tt ==}, when applied to two variables whose types are
classes, returns {\tt true} if the two variables point to the same
object.

\item {\tt deepEqual( other : Any ) : boolean}

makes a deep equality test between {\tt self}  and {\tt other}. This
method returns {\tt false} if the object layout of {\tt self} is
different from {\tt other} or if there is an instance variable
whose type is a basic type in an
object referred directly or indirectly by {\tt self} whose value is
different from the value of the corresponding instance variable in the
objects pointed directly or indirectly by {\tt other}.

\item \verb@getInfo() : AnyObjectInfo@

returns an object that describes {\tt self}. This method is explained
in Chapter~\ref{inr} and Appendix~\ref{irl}.

\item {\tt equals( other : Any ) : boolean}

returns {\tt true} if {\tt self} is equal to {\tt other}. The body of
this method is\\
\verb@     return self == other;@\\
which results in an address comparison. Each class should redefine
this method to give it a better meaning.

\ee

Note {\tt deepCopy} is missing from this
listing. We believe it would be rarely used.

Class {\tt AnyClass} defines the following methods:
\bi
\item \verb@getClassInfo() : ClassInfo@

  returns an object that describes the class of the object.

\item \verb@getClassObject() : AnyClassObject@

      returns the class of the receiver which is a classless
object. For example, the {\tt if} expression that follows will always
be true.
\begin{verbatim}
var c : Circle;
c = Circle.new(10, 20, 10);
if c.getClassObject() == Circle then ... endif
\end{verbatim}

\ei
Chapter~\ref{inr} and Appendix~\ref{irl} discusses these methods in depth.

\section{Method Look-up  and {\tt super}}

The statement\\
\verb@     a.m@{\tt (p$_{1}$, p$_{2}$, ... p$_{n}$)}\\
is the sending of the message {\tt m(p$_{1}$, p$_{2}$, ... p$_{n}$)}
to the object {\tt a} refer to. The message name is {\tt m} and {\tt
p$_{1}$}, {\tt p$_{2}$}, ... {\tt p$_{n}$} are its arguments
(parameters).

At run time, this message send orders the run-time system to look for a
method called {\tt m} in the class of the object {\tt a} refer to. If
no method with this name is found, the search continues in the
superclass of the object class  and  so on. As we will see in Chapter~\ref{gts}, the
compiler guarantees that a message will be found at run time if no
compiler error has occurred. This run-time search for a method is
called dynamic method look-up.

The statement\\
\verb@     super.m@{\tt (p$_{1}$, p$_{2}$, ... p$_{n}$)}\\
in a class {\tt B} makes the compiler look for a method {\tt m}
beginning in the superclass of {\tt B}, that we will call {\tt A}.
If {\tt A} does not define a
method {\tt m}, the search continues in the superclass of {\tt A}
and so on.
This search is made at compile time and results in a
direct call to a method {\tt m} of a direct or indirect superclass of
{\tt B}. The receiver of this message is {\tt self}. Of course, there
will be an error if the method found is abstract.

An {\tt init} method of a superclass can be called using {\tt
super}, since a message to {\tt super} is a message to {\tt self} with
a fixed place to begin the search for the method.

\section{The {\tt subclass} Section}  \label{subclass}

\begin{figure}
\begin{verbatim}
class Figure
  public:
    proc walkTo( newX, newY : integer )
      begin
      setX(newX);
      ...
      end
    ...

  subclass:
    proc setX( newX : integer )
      begin
      x = newX;
      end
    proc getX() : integer
      begin
      return x;
      end
    ...

  private:
    var x, y : integer;
end

class Circle subclassOf Figure
  public:
    proc move( newX, newY : integer )
      begin
      setX(newX);  // ok, calls Figure::setX
      ...
      end
    ...
end
\end{verbatim}
\caption{Use of {\tt subclass} section}
\label{sub1}
\end{figure}

Subclasses cannot access private instance variables of the
superclass. If it is necessary to reveal private information of a
class to its subclasses, the programmer should declare methods to
access this information in the {\tt subclass} section of the
class. See Figure~\ref{sub1}. Note class {\tt Circle} accesses method
{\tt setX} through a message to {\tt self}. If {\tt Circle} had a
method
\begin{verbatim}
proc medX( other : Circle ) : integer
  begin
  return (x + other. getX())/2;
  end
\end{verbatim}
there would be a compiler error in ``{\tt other.getX()}''. The reason,
as will be seen in Chapter~\ref{gts}, is that {\tt other} may be
referring to an object that is not subclass of {\tt Circle} and that
may not have a method {\tt getX}.

Dynamic look-up is  also made with methods of the subclass section. To show
that, suppose class {\tt Circle} defines a method {\tt setX} in its
subclass section. Then the code
\begin{verbatim}
var c : Circle = Circle.new(10, 50);
c.walkTo(20, 40);
\end{verbatim}
calls method {\tt Figure::walkTo} (since {\tt walkTo} is not redefined
in {\tt Circle}) which sends message ``{\tt setX(newX)}'' to {\tt
self}. The  executed method will be {\tt Circle::setX}.

There should be at most one subclass section in a class
declaration. It should be after the public and before the private
section.
When creating a subclass {\tt B} of a class {\tt A}, some restrictions
apply in the redefinition or definition of methods:

\be
\item a public method of {\tt A} may be redefined only as a
public method of {\tt B};
\item a subclass method of {\tt A} may be redefined only as a subclass
method of {\tt B};

\item a private method of {\tt A} may be redefined in any section
(public, subclass, or private) of {\tt B} since no one outside {\tt A}
should or could know of the existence of its private methods. Of
course, the {\tt A} and the {\tt B} method would be completely
unrelated to each other. As an example, suppose class {\tt A} defines
a public method {\tt m} and a private method {\tt p}. Inside {\tt m}
there is a message send ``{\tt self.p()}''. Class {\tt B} inherits
from {\tt A} and defines a public method {\tt p}. The code\\
\verb@     var b : B;@\\
\verb@     b = B.new();@\\
\verb@     b.m();@\\
will call method {\tt A::m} which sends message ``{\tt p()}'' to {\tt
self} that calls method {\tt A::p}. If {\tt p} were public, the method
called would be {\tt B::p}.
\ee

Restriction {\tt 2} could be relaxed to allow a subclass method of
{\tt A} to be redefined as a public method of {\tt B} but that would
not add any power to the language and would make it harder to
understand (so we believe).

The subclass section may declare abstract methods which  should be
redefined in subclasses. It does make sense to have abstract methods
in the subclass section because dynamic look-up is also made with methods
of this section.

\section{\tt nil}

{\tt nil}  is a special global variable that points to an object of the
predefined class {\tt Nil}. It can be assigned to any variable whose
type is a class. {\tt Nil} is considered to be a subtype of any
other type, which means {\tt nil} can be assigned to any variable
whose type is a class. Albeit that, class {\tt Nil} has no method and
{\tt Nil.new()} always returns {\tt nil} --- see class {\tt Nil}
below.
\begin{verbatim}
object Nil
  public:
    proc new() : Nil
      begin
      return nil;
      end
end

class Nil
end
\end{verbatim}

It is a run-time error to send
a message to any object of class {\tt Nil}. If this happens an
exception {\tt MessageSendToNilException} will be thrown.

\section{Abstract Classes and Inheritance} \label{absinh}

Two points are of interest here:
\bi
\item a subclass of an abstract class can call the superclass
constructors using {\tt super};

\item  if a class inherits from an abstract class and does not define all
method bodies of inherited abstract methods, then this class should
also be declared abstract.
\ei

\section{Abstract Classes, Assertions, and Inheritance}

Assertions may be specified even to abstract methods. Class {\tt
Container} defined in Figure~\ref{container} is redefined below with
some assertions. If a class
inherits from {\tt Container} and does not define new assertions for
an inherited method, this method will use the inherited assertions.

\begin{verbatim}
abstract class Container
  public:
    abstract proc add( x : integer )
      assert
        before not full();
      end

    abstract proc get() : integer
      assert
        before not full();
      end

    proc empty() : boolean
      begin
      return getSize() == 0;
      end

    proc full() : boolean
      begin
      return getSize() == getMaxSize();
      end

    abstract proc getSize() : integer
      assert
        after  result >= 0 and result <= getMaxSize();
      end

    abstract proc getMaxSize() : integer;
end
\end{verbatim}
Suppose  class {\tt List} inherits {\tt Container}:
\begin{verbatim}
class List subclassOf Container
  public:
    proc add( x : integer )
      begin
      n = n + 1;
      v[n] = x;
      end
    ...
end
\end{verbatim}
Since method {\tt put} does not define an assertion clause, it
inherits the assertion clause of {\tt Container} method {\tt put}.

As noted by Meyer \cite{meyer88}, the use of assertions with abstract
classes allows one to describe the program design and its
specification at the same time.

\chapter{Arrays}

\section{Introduction}

An array {\tt ac} of {\tt char} is declared as\\
\verb@    var ac : array(char)[];@\\
The number of array elements should not be specified. Arrays are
objects in Green and as such they are dynamically allocated with
{\tt new}:\\
\verb@     ac = array(char)[].new(10);@\\
Note that the number of elements should be specified as a parameter to
{\tt new}. An array of
any basic type has a {\tt fill} method  to fill
all array positions with a given value:\\
\verb@     v.fill(0)@

To access the i$^{th}$ array element, one should use {\tt
v[i]}. Indices range from {\tt 0} to {\tt n - 1} where {\tt n} is the
number of array elements.

A two-dimensional integer array  is also
declared without specifying the number of lines/columns:\\
\verb@     var mi : array(integer)[][];@\\
The dimensions are given in the allocation:\\
\verb@     mi = array(integer)[][].new(10, 30);@\\
It is also possible to give only the first array dimension:\\
\verb@     mi = array(integer)[][].new(10);@\\
This allocates ten empty lines that can be separately created:\\
\verb@     for i = 0 to mi.getSize() - 1 do@\\
\verb@         mi[i] = array(integer)[].new(30);@\\
{\tt getSize} returns the number of array elements.

Although the above code creates ten arrays with the same size (30), it
could have created arrays of different sizes. In fact, the number of
elements of each dimension does not belong to the array type. As a
consequence, a method \\
\verb@     proc mult( mr : array(real)[][] )@\\
accepts as parameter any two-dimensional {\tt real} array.

The operator {\tt []} is used to access array elements. Then,\\
\verb@     mr[0]@ is an array whose type is \verb@array(real)[]@\\
\verb@     mr[i][j]@ is a real number.

Since the number of elements of each dimension does not belong to the
array type,
\bi
\item the size of each dimension is never specified in a variable or
parameter declaration;
\item the dimensions are just instance variables of the array class
and therefore should be initialized by the {\tt new} method.
\ei

Arrays are a kind of parametrized classes specially tuned for
efficiency. By the discussion above, an array depends only on its
element type and the number of dimensions and these would be
parameters to a non-existing generic class {\tt Array}.

\section{Methods of Array Classes}  \label{mac}

An array of a basic class, like \verb@array(char)[]@, inherits from
abstract class {\tt AnyArray}. An array of a non-basic class, like
\verb@array(Person)[]@, inherits from abstract class {\tt
AnyClassArray}, which inherits from {\tt AnyArray}. Class {\tt
AnyArray} defines the following methods:

\vspace*{3ex} \noindent \verb@getSize() : integer@

returns the array size.

\vspace*{3ex} \noindent
\verb@set( v : Any; i : integer )@\\
\verb@   ( exception : CatchAnyArrayException )@

set to {\tt v} the {\tt i$^{th}$} position of the array. The {\tt
exception} variable is used to throw exceptions, described in
Chapter~\ref{exceptions}.
The methods
that follow work similarly with arrays of two or more dimensions.
These methods may throw the following exceptions:
\bi
\item {\tt TypeErrorException}. This exception is thrown if
there is a type error; that is, if
the array element type is {\tt T} and  the run-time type of
object {\tt v} is not a subtype of {\tt T};

\item {\tt TooManyDimensionsException}, which is thrown  if the array has
less dimensions than those specified. For example, if one uses the
next method {\tt set} with a one-dimensional array. There should have been
specified only only index, {\tt i} or {\tt j}.

\ei

\vspace*{3ex} \noindent
\verb@set( v : Any; i, j : integer )@\\
\verb@   ( exception : CatchAnyArrayException )@

\vspace*{3ex} \noindent \verb@set( v : Any; i, j, k : integer; others : ... array(integer)[] )@\\
\verb@   ( exception : CatchAnyArrayException )@

\vspace*{3ex} \noindent \verb@get( i : integer ) : Any@\\
\verb@   ( exception : CatchAnyArrayException )@

returns the element of the array index {\tt i}. The methods that
follow work similarly with arrays of two or more dimensions. The {\tt
get} methods may throw exception {\tt TooManyDimensionsException}.

\vspace*{3ex} \noindent \verb@get( i, j : integer ) : Any@\\
\verb@   ( exception : CatchAnyArrayException )@

\vspace*{3ex} \noindent \verb@get( i, j, k : integer, others : ... array(integer)[] )@\\
\verb@   ( exception : CatchAnyArrayException )@

See Section~\ref{vnp} for the meaning of ``{\tt ...}''.

\vspace*{3ex} \noindent \verb@toString() : String@

returns a string like ``{\tt array(T)[][]}'' if the array has two
dimensions and {\tt T} as the element type.

\vspace*{3ex}
Each array of a basic class defines the following methods and
constructors. {\tt T} is
the array element type. Then, in \verb@array(char)[]@, {\tt T}
represents {\tt char}.

\bi

\item \verb@init( first, second, third, ... : integer )@

Constructor of an array in which {\tt first}, {\tt second}, ... are
the array dimensions. An array of {\tt n} dimensions has {\tt init}
methods from {\tt 1} to {\tt n} parameters. As in the example\\
\verb@     mi = array(integer)[][].new(10);@\\
in which method\\
\verb@     proc init( first : integer )@\\
is called by {\tt new}, the minor dimensions need not to be specified
in the array creation. Later on, the program must allocate memory for
these minor dimensions in order to use the array;

\item {\tt getIter() : DS.Iter(T)}
returns an iterator for the array. Iterators are presented in
Chapter~\ref{stand}. {\tt T} is the type of the array elements. ``{\tt
DS.Iter(T)}'' is the type of the iterator;

\item {\tt forEach( f : Function(T) )}

calls ``\verb@f.exec@'' on each array element.
 Class {\tt Function} is defined in
Chapter~\ref{stand}. {\tt T} is the type of the array elements;

\item {\tt replaceBy( cmd : Command(T) )}

replaces each array element {\tt x} by ``\verb@cmd.doIt(x)@''.
 Class {\tt Command} is defined in
Chapter~\ref{stand}. {\tt T} is the type of the array elements;

\item {\tt collect( f : Filter(T) ) : array(T)[]}

collects all array elements {\tt x} such that ``{\tt f.test(x)}''
evaluates to {\tt true}. These elements are
inserted in a new array returned by this method. {\tt T} is the type
of the array elements. {\tt Filter} is defined in Chapter~\ref{stand};

 \item {\tt remove( f : Filter(T) )}

  removes all array elements {\tt x} such that ``{\tt f.test(x)}''
evaluates to {\tt true};

\item {\tt reset( up : boolean )}

initiates the array iterator. If {\tt up} is true, method {\tt next}
will yield array elements from index {\tt 0} to {\tt getSize() - 1}. If
{\tt up} is {\tt false}, the elements will be yield from {\tt getSize()
- 1} to {\tt 0};

\item {\tt reset()}

the same as {\tt reset(true)};

\item {\tt more() : boolean}

returns {\tt true} if there is more elements to be yield by the
iterator;

\item {\tt next() : T}

returns the next array element according to the order set by calling
{\tt reset}. The counter of the iterator is incremented. An example of
use of an iterator is:
\begin{verbatim}
var vi : array(integer)[];
     k : integer;
...
vi.reset();
while vi.more() do
  Out.writeln( vi.next() );
\end{verbatim}

\item {\tt fill(value : T)}

fill all array positions with value.

\ei

Class {\tt AnyClassArray} defines the following methods:

\bi

\item {\tt getIter() : DS.Iter(Any)}

\item \verb@forEach( f : Function(Any) )@\\
      \verb@(exception : TypeErrorException)@

\item \verb@replaceBy( cmd : Command(Any) )@
      \verb@(exception : TypeErrorException)@

\item \verb@collect( f : Filter(Any) )@\\
      \verb@(exception : TypeErrorException) : array(Any)[]@

\item {\tt remove( f : Filter(T) )}

  removes all array elements {\tt x} such that ``{\tt f.test(x)}''
evaluates to {\tt true};

\item {\tt reset( up : boolean )}

\item {\tt reset()}

\item {\tt more() : boolean}

\item {\tt next() : Any}

\ei

Some methods will throw exception {\tt TypeErrorException} if a type
error occurs. Each of these methods play a role equal to the method of
same name added to every basic-class array.

The compiler adds a {\tt init} method, as defined to basic-class
arrays, to all arrays.

\section{Initialization and Creation of Arrays}

An array can be initialized in its declaration through a special
syntax:
\begin{verbatim}
proc selectFriend( day : integer ) : String
  begin
  var friends : array(String)[] = #( "Tom", "Jo", "Anna", "Peter" );

  return friends[ day%friends.getSize() ];
  end
\end{verbatim}
Note that {\tt selectFriend} {\it could not} have been written as
\begin{verbatim}
proc selectFriend( day : integer ) : String
  var friends : array(String)[] = #( "Tom", "Jo", "Anna", "Peter" );
  begin
  return friends[ day%friends.getSize() ];
  end
\end{verbatim}
since assignments are not allowed in the local variable declaration
section before {\tt begin}.

Arrays can also be initiated with constants in the private section of
a class object:
\begin{verbatim}
object Month
  public:
    proc get( i : integer ) : String
      begin
      return strMonth[i];
      end
  private:
    var strMonth : array(String)[] = #( "Jan", "Feb", "Mar", "Apr",
         "May", "Jun", "Jul", "Ago", "Sep", "Oct", "Nov", "Dez" );
end
\end{verbatim}

An array object, like any other, can be created using \verb@#@~:\\
\verb@     var v : array(integer)[];@\\
\verb@     v#init(10);@\\
This is the same as\\
\verb@     v = array(integer)[].new(10);@\\

\section{Arrays and Inheritance}  \label{anyarray}

As already seen, class {\tt AnyArray} defines methods for
getting and setting array elements, as shown in the following example.

\begin{verbatim}
var aa : AnyArray;
var vi : array(integer)[];
var mr : array(real)[][];

vi#init(40);
mr#init(30, 50);
vi[0] = 1;
mr[0][0] = 7.0;

aa = vi;
var i : integer = integer.cast(aa.get(0));

aa = mr;
var r : real = real.cast(aa.get(0, 0));

aa.set(3.0, 0, 0);          // aa[0][0] = 3.0
Out.writeln( mr[0][0] );    // 3.0
\end{verbatim}
Methods {\tt get} and {\tt set}  throw an exception if their
parameters are not type correct.

Although array classes are much like other classes, they cannot be
subclassed. If they could, arrays could not be efficiently compiled.

Every array class inherits from {\tt AnyArray} and is subtype only of
{\tt AnyArray}. Therefore an array ``{\tt array(Figure)[]}'' is not a
supertype or superclass of ``{\tt array(Circle)[]}'', even {\tt
Circle} being a subclass of {\tt Figure}. If it were, as
in Java, there could be a run-time type error --- see the code below.
\begin{verbatim}
var vf : array(Figure)[];
var vc : array(Circle)[];

vc#init(10);
  // suppose this is allowed
vf = vc;
vf[0] = Square.new(10);
  // now vc[0] refer to a square
Out.writeln( vc[0].getRadius() );   //  message not found
\end{verbatim}

\section{Methods with Variable Number of Parameters}  \label{vnp}

A method that  takes an undetermined number of parameters may be
specified as:\\
\verb@     proc print( whiteSpace : integer; v : ... array(Any)[] )@\\
In this example {\tt print} has one fixed parameter {\tt whiteSpace}
and any number of parameters following it, represented by array {\tt
v}. This array should always be at the end of the parameter list.
Each of the variable number
of real parameters should be a subtype of {\tt Any}, which includes any
object. Basic type objects as integers and characters are packed in
wrapper objects of classes {\tt Integer}, {\tt Char}, etc. These
classes are defined in Chapter~\ref{basictypes}. In a call to {\tt
print}, the  real parameters after the first one  are used to
initialize an array passed as a parameter to this method.

A complete example of definition and use of method {\tt print} is given
next.
\begin{verbatim}
object MyScreen
  public:
    proc print( whiteSpace : integer;
                v : ... array(Any)[] )
      assert
        before whiteSpace >= 0 and whiteSpace < 80 and v.getSize() > 0;
      end

        var i : integer;
      begin
      for i = 0 to v.getSize() - 1 do
        begin
        Out.write( getWhiteSpace(whiteSpace) );
        Out.writeln( v[i].toString() );
        end
      end
   ...
end
\end{verbatim}
This example uses object {\tt Out} and its method {\tt writeln}  to do
the output to the screen. Class {\tt MyScreen} can be used as\\
\verb@     MyScreen.print( 5, "i = ", i, "(", 2, ")" );@\\
in which {\tt whiteSpace} would receive {\tt 5} and the other
parameters would be packed in an array. Parameter {\tt i} (which
belongs to type {\tt integer}) and number {\tt 2} would be used to
create {\tt Integer} objects that would then be inserted in the
array. Classes like {\tt Integer} are called {\it wrapper classes} and
are discussed in Section~\ref{wrapper}.

It may be interesting to restrict the parameter of the undefined part
to subtypes of a given class as is made in the next example.
\begin{verbatim}
abstract class Drawable
  public:
    abstract proc draw()
end

object MyScreen
  public:
    proc drawAll( v : ...array(Drawable)[] )
        var i : integer;
      begin
      for i = 0 to v.getSize() - 1 do
        v[i].draw();
      end
    ...
end
\end{verbatim}
Note that a call\\
\verb@     Screen.drawAll( circle, 5 );@\\
would result in a compile time error since class {\tt integer} ({\tt
5}) is not subtype of {\tt Drawable}.

There may not be in the same class two methods like\\
\verb@     proc m( i : integer; ch : char; v : ... array(Any)[] )@\\
\verb@     proc m( j : integer; ch : char )@\\
They differ only in the last parameter {\tt v} that indicates a
variable number of parameter. If this were allowed there would be an
ambiguity in the call\\
\verb@     x.m(5, #'A');@\\
Either method {\tt m} could be used.

\section{Expanded Arrays}

Array variables can be declared with {\tt @} if the array dimensions
are specified:\\
\verb2     var clockArray : @array(Clock)[100];2\\
\verb2     var matrix     : @array(integer)[MaxLin][MaxCol];2\\
{\tt MaxLin} and {\tt MaxCol}  must be constants. We expect
expanded array variables will be very rarely used. They are supported
to make the language orthogonal.

In the declaration of {\tt clockArray}, {\tt Clock} is not preceed by
{\tt @}. That means the {\tt Clock} objects should be allocated with
{\tt new}:\\
\verb@     clockArray[0] = Clock.new(13, 52);@\\
One can choose to allocate all {\tt Clock} objects by using {\tt
@Clock} as the element type:\\
\verb!     var clockArray : @array(@Clock)[100];!\\
\verb!     clockArray[0].init(13, 52);!\\
\verb$     clockArray[0] = Clock.new(0, 0); // error !!$\\
The last line is a compile-time error since {\tt @}-variables cannot
receive objects in assignments.

One may also declare\\
\verb!     var clockArray : array(@Clock)[];!\\
Here {\tt clockArray} is a dynamically-allocated array of constant
objects. This should be preferable to the declaration\\
\verb!     var anotherClockArray : @array(@Clock)[100];!\\
{\tt clockArray} should be created with {\tt new}:\\
\verb!     clockArray = array(@Clock)[].new(max);!\\
\verb!     clockArray[0].init(13, 52);!\\

\chapter{The Green Type System}  \label{gts}

\section{Types and Subtypes} \label{subtype}

The {\it type} of a class is the set of signatures of
its public methods.
The signature of a method is its name, return value
type (if it returns a value), and formal parameter types
(parameter names are discarded). For example, the type of the
class \verb@Store@ of Figure~\ref{po1} is given by\\
\verb@     typeOf(Store)@ = {\tt \{ set(integer), get() : integer \}}\\
Throughout this report \verb@typeOf(A)@ will be  the type of
 class {\tt A}.

A type {\it S} = {\tt \{n$_{1}$, n$_{2}$, \ldots n$_{p}$\}}
is equal to type
\verb@@{\it T} = {\tt \{m$_{1}$, m$_{2}$, \ldots m$_{q}$\}} ({\it S} $=$ {\it T})
if $p = q$ and {\tt n$_{i}$} = {\tt m$_{i}$}, $1 \leq i \leq p$.
The relation  $=$ for methods is defined
as follows.

Let \\
\verb@     @{\tt n({\it T}$'_{1}$, {\it T}$'_{2}$, \ldots {\it T}$'_{k}$) : {\it U}$'$}\\
\verb@     @{\tt m({\it T}$_{1}$, {\it T}$_{2}$, \ldots {\it T}$_{t}$) : {\it U}}\\
be the signatures of two methods. {\it U}  and {\it U}$'$
are return value types. We say that {\tt n} $=$
{\tt m} if
\bi
\item {\tt m}  and {\tt n} have the same name.
\item $t = k$.
\item {\tt {\it T}$_{i}$ $=$ {\it T}$'_{i}$}, $1 \leq i \leq t$.
\item {\tt {\it U}$'$ $=$ {\it U}}.
\ei
If necessary, the method signatures of {\it S} and {\it T}
should be arranged in an order such that the relation above becomes
true.  That is, the order the signatures appear in the set
does not matter, although the order of parameter/return value types of a signature does matter.

The type equality is a recursive definition because it uses the method
definition and vice-versa. An algorithm to test if two types are equal
could never finish its execution because of an endless
recursion. However, that never happens according to the Proposition
below.

\begin{prop}
An algorithm that tests if two types are equal according to the
language definition will always finish its execution.
\end{prop}

\noindent {\bf Proof:}

\begin{figure}
\verb@proc isEqual( S, T : Type )@\\
\verb@  var Ig : Set of Tuples (X, Y) where X and Y are types;@\\
\verb@@\\
\verb@begin@\\
\verb@Ig = empty set;@\\
\verb@return isEqualTo( S, T, Ig );@\\
\verb@end@\\
\verb@@\\
\verb@proc isEqualTo( S, T : Type; Ig : Set )@\\
\verb@begin@\\
\verb@@\\
\verb@  // (S,T) is inserted into Ig when S = T@\\
\verb@if (S, T) belongs to set Ig@\\
\verb@then@\\
\verb@  return true;@\\
\verb@else@\\
\verb@  if S or T is a basic type@\\
\verb@  then@\\
\verb@    return S = T;@\\
\verb@  endif@\\
\verb@endif@\\
\verb@insert @{\tt ({\it S}, {\it T}) into Ig;}\\
\verb@for each method m@{\tt ({\it U$_{1}$}, {\it U$_{2}$}, ... {\it
U$_{k-1}$}) : {\it U$_{k}$} of S, do}\\
\verb@  if @{\tt {\it T} does not have method with the same name, number
of parameters}\\
\verb@    and return values as method m of S@\\
\verb@  then@\\
\verb@    return false;@\\
\verb@  else@\\
\verb@    assume @{\tt assume {\it V$_{i}$} is the type of the method
m of {\it T} corresponding to {\it U$_{i}$} of {\it S} }\\
\verb@    for i = 1 to k do@\\
\verb@      if not isEqualTo@{\tt ( {\it V$_{i}$}, {\it U$_{i}$}, Ig)
}\\
\verb@      then@\\
\verb@        return false;@\\
\verb@      endif@\\
\verb@  endif@\\
\verb@return true;@\\
\verb@end@\\
\caption{Algorithm to discover if two types are equal}
\label{ie}
\end{figure}

The algorithm {\tt isEqual} of Figure~\ref{ie} compares  two types and
returns {\tt true} if they are equal.
Algorithm {\tt isEqual} calls {\tt isEqualTo} that
performs a Depth-First Search in both types at
the same time, comparing the corresponding vertices. Therefore, it
will finish its execution. \fh

A type
\verb@@{\tt {\it S} $=$  \{n$_{1}$, n$_{2}$, \ldots n$_{p}$\}}
is a subtype of type
\verb@@{\tt {\it T} $=$  \{m$_{1}$, m$_{2}$, \ldots m$_{q}$\}} (we will
use {\it S $\prec$ T} for that)
if $p \ge q$ and {\tt n$_{i}$} $=$ {\tt m$_{i}$} for $1 \le i
\le q$.   By this definition, {\it S $\prec$ T} implies that
{\it T $\subset$ S}. That is, a subtype has at least the same method
signatures as its supertype.
Since {\it X $\subset$ X} for any type {\it X}, any type is
also subtype of itself.
We usually say ``class {\tt B} is a subtype of class {\tt A}''
instead of ``the type of class {\tt B} is a subtype of the type
of class {\tt A}".

When class {\tt B} inherits from class {\tt A}, {\tt B}
is a subclass of {\tt A}. So, {\tt B} inherits all public
methods of {\tt A}, implying that {\tt B} is a subtype of {\tt A}.
Class {\tt B} can redefine an inherited method from {\tt A} but its
signature should be the same as in the superclass.
By this type definition, any subclass is also a subtype, but it is
possible to have a subtype that is not a subclass.

This type system is a restriction of that of \mbox{POOL-I} \cite{Lin90}
language.
The programs that are type correct according to the type system
defined above are also type correct according to the \mbox{POOL-I} type
system since this has less restrictions than that. \mbox{POOL-I} follows the
Cardelli \cite{Car84} rules for subtyping. To explain
Cardelli's  rules, suppose class {\tt B} is subtype of class {\tt A}
that defines a method\\
\verb@     @{\tt proc m( x : C ) : D}\\
Class {\tt B} can define a method\\
\verb@     @{\tt proc m( x : C$'$ ) : D$'$}\\
such that {\tt C $\prec$ C$'$} and {\tt D$'$ $\prec$ D}. In our type
system, {\tt C} must be equal to {\tt C$'$} and {\tt D} must be equal
to {\tt D$'$}.

Some object-oriented languages such as C++ \cite{str91} and Eiffel
\cite{meyer88} \cite{meyer87} associate subtype with subclass. A
subtype of a class can be created only by subclassing the class. This
results in a less flexible type system than that of Green. As an example,
suppose there is a method\\
\verb@     @{\tt proc m( x : A )}\\
of some
class (it does not matter). In Eiffel and C++, method {\tt m} accepts
only objects of class {\tt A} and its subclasses as parameters. In
the type system of Green, method {\tt m} accepts objects of any class that is
subtype of {\tt A}. Since there are potentially more subtypes than
subclasses of {\tt A}, the Green type system  supports a
higher degree of polymorphism than the Eiffel/C++ type system. The
separation between the subtype and subclass hierarchy gives to a
statically typed language much of the flexibility of untyped
languages. This kind of type system is employed in the languages
\mbox{POOL-I} \cite{Lin90}, Sather \cite{sather} \cite{zorn94}, Java
\cite{java},\footnote{In fact, a restricted version of it.}
 and School \cite{rodriguez93}.  The language Emerald
\cite{raj91} also uses this subtype relationship although it does not
support inheritance.

Snyder \cite{snyder86} \cite{snyder} asserts that there is a violation
of encapsulation if subtyping is tied to subclassing. To show that,
suppose class {\tt B} inherits from {\tt A} and a class-{\tt B} object
is used as a parameter to method {\tt m} whose signature was shown in
the previous paragraph. If class {\tt B} is modified to not inherit
from {\tt A}, although keeping its interface, the code that passes
class-{\tt B} objects to {\tt m} becomes type incorrect. This means
that the inheritance of {\tt A} by {\tt B} is not private to {\tt B}:
it is public and cannot be changed without program modifications.

The type of a class
 can be discovered
without compiling its file if there is no syntax error. This is
possible since the type of a class is just its interface. It does not
depend on the semantic correctness of the code inside the
class methods. Then we have the
following assumption:
\begin{assumption} \label{known}
 It is possible to find the type of a class by doing a syntactic analysis in it if
 there is no syntax error in the file the class is.
\end{assumption}

\section{Rules for Type Checking}

The statement\\
\verb@   aa = bb@\\
where the type of {\tt aa} is a class,
makes the variables {\tt aa} and {\tt bb} refer to the
same object. The declared type of {\tt aa} and {\tt bb} must
be in the relationship\\
\verb@   type(bb)@ $\prec$ \verb@type(aa)@\\
where {\tt type(x)} is the type of class {\tt Z} considering {\tt x} is
declared as\\
\verb@     var x : Z;@\\
That is,  assignments
of the kind\\
 \verb@   Type = Subtype@\\
are valid.

Assignments of the kind\\
\verb@     Type = Supertype@\\
are allowed if made through method {\tt cast}:\\
\verb@     b = B.cast(a);@\\
This method is automatically added by the compiler to each class
object (see Section~\ref{cast}) and it throws exception {\tt
TypeErrorException} if object {\tt a} cannot be cast to class {\tt
B}. Method {\tt cast}  supplied by the compiler will succeed if the
class of object {\tt a} at run-time is a subtype of {\tt B}.

The statement\\
\verb@       @{\tt a.m(p$_{1}$, p$_{2}$, ... p$_{n}$)}\\
sends the message {\tt m(p$_{1}$, p$_{2}$, ... p$_{n}$)} to the object
{\tt a} refer to and
is legal if:
\bi
\item method {\tt m({\it T}$_{i}$, {\it T}$_{2}$, ... {\it T}$_{n}$) :
{\it U}$_{n+1}$} belongs to {\tt type(a)}, where {\tt type(a)}
is the declared type of {\tt a} and;
\item for $1 \leq i \leq n$,
  \bi
  \item {\tt type(p$_{i}$)} $\prec$ {\it
T$_{i}$}, where {\tt type(p$_{i}$)} is the declared type of {\tt
p$_{i}$} and {\tt T$_{i}$} is a class name or
\item {\tt p$_{i}$} is a basic value ({\tt char}, {\tt integer}, ...)
and can be automatically converted to {\tt
T$_{i}$}.\footnote{Currently no automatic cast between basic types is allowed.}
  \ei

\ei

The  run-time system  searchs for
a method named {\tt m} in the class of the object. If
no method by this name is found, it searchs in the object superclass,
then in the superclass of  superclass and so on. If method
{\tt m} is  not found, an error occurs. To send a
message to an object that cannot respond to it is
a {\it type error}.

The keyword {\tt self} inside a method code evaluates to the
object that received the message that caused the execution of the
method.
A message send to {\tt self} is considered by the type system as a
normal message send in which {\tt type(self)} is the type of the class in
which this statement is with the addition of all methods of the:
\bi
\item  {\tt private} section;
\item {\tt subclass} sections of this class and all superclasses.
\ei

The message send\\
\verb@       @{\tt super.m(p$_{1}$, p$_{2}$, ... p$_{n}$)}\\
in a method of a class {\tt B}
 is considered type correct if
\bi
\item the superclass of {\tt B} is {\tt A} and {\tt m} does not belong
to the private section of {\tt A} and;
\item the message send\\
\verb@       @{\tt self.m(p$_{1}$, p$_{2}$, ... p$_{n}$)}\\
inside an {\tt A} method would be correct.
\ei

The {\tt a} variable in the statement \mbox{\tt a.m(p$_{1}$, p$_{2}$,
... p$_{n}$)} may  refer to an object whose
type is a subtype of its declared type because of assignments of
the form \verb@Type = Subtype@. But that does not cause any
error, since a subtype has all methods of a class with exactly the
same method signatures.

The {\tt m} method of class
\begin{verbatim}
class A
  public:
    proc m( x : Store )
      begin
      x.set(5);
      v = x.get();
      end
    ...
  private:
    var v, k : integer;
end
\end{verbatim}
takes a formal parameter {\tt x} of class \verb!Store!. Using the above
rules, the compiler enforces that:
\bi
\item class {\tt Store} has methods corresponding to the message
sends to {\tt x}\footnote{We say ``message send to {\tt x}" to
mean ``message send to the object {\tt x} will refer to at run
time" since a message is sent to an object, not to a variable.} inside method {\tt m}, that is, {\tt set}
and {\tt get}.
\item the types of the real parameters can be converted into the types
of the formal parameters. Remember there is no automatic coercion
among the basic types.

\ei

\section{Discussion on Types}

In parameter passing to methods/procedures, there is an
implicit assignment\\
 \verb@         formal parameter = real parameter@\\
since any parameter is passed by value. The same is valid for
method return values.
Therefore, any type analysis can be restricted to
assignments since this embodies parameter passing and return values
of methods.

When a variable is declared, the programmer must give a class
name that represents its type, as in\\
\verb@     var a : A;@\\   \label{cty}
Then variable {\tt a} can refer to objects of any subtype of
class {\tt A}. In fact, {\tt A} does not represent the class
{\tt A} but the type \verb@type(A)@. That is the reason we
used ``{\tt A}" as the type of the variable instead of
\verb@typeOf(A)@.

Since this variable can refer to objects
of classes that are subtypes of {\tt typeOf(A)}, the
compiler does not know the size of the object that the variable
will refer to at run time.  This
is the reason the  declaration of variable {\tt a} does not automatically
allocate  memory for a class-{\tt A} object as in
other languages.

This is also the reason why, inside a class-{\tt A} method, we
cannot access an instance variable {\tt x} of a parameter {\tt p} that
has type {\tt A}, as in\\
\verb@      proc m( p : A ) // method of A @\\
\verb@        begin@\\
\verb@        p.x = 10;@\\
\verb@        end@\\
{\tt x} can refer at run time
to an object that does not belong to class {\tt A} or its subclasses
and therefore does not have instance variable {\tt x}.

\section{Assertions and Subclassing}
\label{asssub}

               Assertions are inherited by subclasses. If a subclass

overrides a superclass method and does not define an {\tt assert}
clause for this method, the assert clause of the superclass method is
added to the subclass method by the compiler.

               The {\tt before} and {\tt after} expressions of an
assert clause of a
subclass method should be semantically related to the corresponding
expressions of the superclass method. This can be better understood by
studying the relationship among types in inherited subclass methods,
described next.

It could be possible to redefine the parameter types of methods in
subclasses. For example, a method\\
\verb@     proc whoEats( food : Vegetable ) : Animal@\\
of a class {\tt AnimalWorld} could be redefined in a subclass {\tt MammalWorld} as\\
\verb@     proc whoEats( food : Food ) : Mammal@\\
in which {\tt Vegetable} is subtype of {\tt Food} and {\tt Mammal} is
subtype of {\tt Animal}. Green does not support this feature because
\bi
\item it is not largely used. In most cases we want to use a subtype
as the parameter type of a subclass method, like use {\tt Tomato}
instead of {\tt Food} as the type of {\tt food} in a subclass
method. This is not allowed --- we
could use a supertype as the parameter type. When redefining the return
value type in a subclass method, we can use a subtype of the return
value type of the superclass method (as {\tt Mammal} and {\tt Animal});
\item it would conflict with overloading. In Green two methods with
different parameter types are considered different.
\ei

To understand why the sub and superclass method types should be
related as described previously, we will study the example below.

\begin{verbatim}
var a : AnimalWorld;
var b : MammalWorld;
a#init();
b#init();
var v : Vegetable = Vegetable.new("Grass");
var animal : Animal;

animal = a.whoEats(v);
a = b;
animal = a.whoEats(v);
\end{verbatim}
The last statement is a message send to a {\tt MammalWorld} object. Then it
should take a {\tt Food} object as parameter and there is an implicit
assignment\\
\verb@     food = v;@\\
This is correct since {\tt food} type ({\tt Food}) is a supertype of
{\tt v} type ({\tt Vegetable}). Method {\tt whoEats} of {\tt MammalWorld}
returns a {\tt Mammal} object that can be assigned to an {\tt Animal}
variable.

Returning to assertions, a subclass method should make weaker
assumptions on its parameters and stronger assumptions on its returned
value \cite{meyer88}. As an example, a class {\tt Function} defines a
mathematical function and has a method
\begin{verbatim}
proc findRoot( firstGuess : real ) : real
  assert
    before Math.abs( f(firstGuess) ) < 1;
    after  Math.abs( f(result) ) < 1E-5;
  end
\end{verbatim}
to find a function root. {\tt Math} is a class object that defines
mathematical functions like {\tt sin}, {\tt cos}, and {\tt abs} which
returns the absolute value of a number. Method {\tt f} of {\tt
Function} evaluates the function with the given parameter.

  {\tt firstGuess} parameter of {\tt findRoot} is a first guess of the
root  and the function value at this point should be at a distance
less than {\tt 1} from {\tt 0}. Variable {\tt result} holds the value
returned by the function and therefore {\tt f(result)} should be very
close to zero.

A subclass {\tt Polynomial} of {\tt Function} could define a method
\begin{verbatim}
proc findRoot( firstGuess : real ) : real
  assert
    before Math.abs( f(firstGuess) ) < 5;
    after  Math.abs( f(result) ) < 1E-20;
  end
\end{verbatim}
that uses a much better algorithm to find the root. It requires a
weaker assumption on the parameter and produces a better result
(stronger assumption). One can match superclasses with weaker
assumptions because a superclass is more abstract than its
subclass. A superclass may has (and generally has) less methods and represents
more concepts than its subclasses.

  One   can match subclasses with stronger assumptions by the same
reasons. The result is that {\tt before} and {\tt after} expressions
should obey similar requirements of parameter and return value type
redefinition in subclass methods. As an example, the code
\begin{verbatim}
class Example
  public:
    proc calc( f : Function )
        var r : real;
      begin
      r = f.findRoot(1.0);
      Out.writeln(r);
      end
    ...
end  // Example

...
var e : Example;
var p : Function;

e#init();
p = Polynomial.new("x*x - 2*x + 1");
e.calc(p);
\end{verbatim}
makes variable {\tt p} refer to a polynomial object and calculates a
root with a better precision than the creator of class {\tt Example} expected. Since
the programmer is using a {\tt Function} parameter, she should expect
the worst precision {\tt 1E-5} than what she really gets, {\tt
1E-20}. Similar reason applies to the parameter value: any value good
for {\tt Function::findRoot} is also good for {\tt Polynomial::findRoot}.

\section{Constructors, Inheritance, and Types}

A constructor ({\tt init} method) of a class {\tt A} can be called by

\bi
\item a compiler-created {\tt new} method of class object {\tt A};
\item sending a message to {\tt self} inside
class {\tt A};
\item sending a message to {\tt super} in a subclass of {\tt A};
\item sending a message to an expanded variable.
\ei

Polymorphism {\it does work} with constructors. In any message send
``{\tt self.init(...)}'' or ``{\tt super.init(...)}'' the search for a
{\tt init} method is made at runtime. Then,
a constructor of a superclass can call a subclass
constructor. This is probably wrong because subclass
constructors can (and should) expect the superclass instance variables
have been initiated. Besides that,
endless loops can  occur since subclass constructors
usually call superclass constructors  unconditionally.
 In this case the superclass constructor calls the
subclass constructor creating thus a cycle.

Unlike other languages as C++ or Java, Green does not require a
subclass to call the superclass constructor. Albeit that, not to call
the superclass constructor is  a bad practice and we expect the
compiler will  issue a warning if this will never or may not occur at run
time. The reason for not introducing this feature is that it would
require a special syntax. Then one could call a method {\tt init} by
two different ways and that would be confusing. Anyway, in most of the cases
a subclass {\tt init} method will call the superclass {\tt init} in
its first statement.

Since the programmer cannot send a {\tt init} message to any object
but {\tt self} (even when through
 {\tt super}), in fact {\tt init} does not belong to the
type of the class. This makes sense because the parameters of the
constructors can reveal a lot about the class implementation.

The implementation of a class (its instance variables and method
bodies) is highly subject to changes, and therefore the class {\tt
init} methods are too. The {\tt init} methods are closely related to
instance variables --- in general the {\tt init} parameters are
assigned directly to instance variables without any processing.
If the {\tt init} methods belonged to the class type, a
change to the class implementation would invalidate working code since
this depends mainly on the class types.  As the {\tt init} methods do not
belong to the types, any change in them requires only modification in
the code
\bi
\item that creates (with {\tt new}) objects of the class;
\item of subclasses of the class that calls the superclass
constructor.
\ei

As an example of the problem above, suppose there are two {\tt Queue}
classes, {\tt SQueue} and {\tt DQueue}, with equal types. If the
implementation of {\tt SQueue} were modified from a linked list to an
array, there could be added a method \\
\verb@     proc init( maxSize : integer )@\\
to set the maximum number of elements in the Queue. Now {\tt SQueue}
would have a type different from {\tt DQueue} if {\tt init} belonged
to the class type.

All {\tt init} methods must be declared before the class public section. This
prevents dirt tricks as to put some or all constructors in the
{\tt subclass} section. Then only subclasses could call the {\tt init}
methods. If all {\tt init} methods were put in the subclass section,
only subclasses could create objects of the class.\footnote{But where the
corresponding {\tt new} methods would be added~? They could not be put
in the public section of the class object ...} This is not allowed because:
\bi
\item it would be redundant with the module system that already offers
visibility control;
\item it makes the code hard to understand since it adds to the
constructs more responsibilities than they were supposed to have.
\ei

When a message is sent to {\tt self} inside a class, the method called may
belong to a subclass of this class. In a {\tt init} method, a message
send to {\tt self} could call a method of a subclass and this most probably
will result in an error because:
\bi
\item the subclass instance variables may be accessed by this subclass
method called by {\tt init} and;
\item the subclass instance variables may not have been initialized
since in the normal order of initialization superclasses are
initialized before subclasses. In most cases the first statement of a
{\tt init} method is a call to a superclass {\tt init} method.
\ei

To avoid this kind of error we hope the compiler will issue a warning
if there is a message send to {\tt self} inside a method {\tt
init}. This error may occur even if the message send refer to a
private method because this method may call a public or subclass
method with which dynamic look-up is made. Passing {\tt self} as
parameter in a message send inside {\tt init} is also dangerous
because the methods called can send a  message to the {\tt self}
passed as parameter. And then a subclass method can be called.

\section{Types and Overloaded Methods}  \label{typeo}

Suppose the type of variable {\tt screen} is {\tt MyScreen} that
defines several methods called {\tt print} --- an overloaded method.

In a message send\\
\verb@     screen.print(x)@\\
the method to be called, among all overloaded {\tt print} methods, is
defined at compile time. The declared type of {\tt x} must be equal
to a parameter type of one of the {\tt print} methods. As an example,
suppose class of {\tt screen} has the following {\tt print} methods\\
\verb@     proc print( f : Figure )@\\
\verb@     proc print( c : Circle )@\\
\verb@     proc print( i : integer )@\\
and {\tt Square} is a subclass of {\tt Figure}. The call\\
\verb@     screen.print( Square.new(10) );@\\
would be illegal because there is no method\\
\verb@     proc print( s : Square )@\\
There should be used a cast:\\
\verb@     screen.print( Figure.cast( Square.new(10) ) ) ;@\\
The definition of subtype of Green prevents the compiler from choosing
the nearest method to be used when the real parameter type does not
match the formal parameter type of any overloaded method. To
understand that, suppose methods\\
\verb@     proc print( f : Figure )@\\
\verb@     proc print( w : Window )@\\
belong to {\tt MyScreen}. Class {\tt Frame} is subtype
of both {\tt Figure} and {\tt Window}. Now, which method should be
called in the code below~?
\begin{verbatim}
var screen : MyScreen;
var f : Frame;

screen#init();
f#init();
...
screen.print(f);
\end{verbatim}
If there was an order among all types, the compiler could choose the
type nearest to {\tt Frame}. But there may be no coherent or rational
order --- the definition of subtype is too loose to allow that. For
example, {\tt Figure} and {\tt Window} may have the same set of method
interfaces --- the same type. In this case, which one should come
first in the total ordering of the types~?

There may be a semantic error at run
time  if the class of the object {\tt x} is a subtype (see
Section~\ref{subtype}) of its declared type:
\begin{verbatim}
    var x : Figure;
    x = Circle.new(20, 45);
    screen.print(x);
\end{verbatim}
In this case the compiler will generate code that will call method
{\tt print(Figure)} instead of {\tt print(Circle)}.

\section{Assertions and Subtyping}

The observation on assertions and subclassing (Chapter~\ref{asssub})
also applies to assertions and subtypes but with one big difference:
it would be very difficult to enforce any rule about the relationship
assertion/subtype method because the compiler does not know who are
the supertypes of a class when compiling it. A class {\tt A} is
considered a supertype of {\tt C} if {\tt A} has at least all {\tt C}
method signatures. For the linker and compiler, however, a class {\tt
A} is supertype of {\tt C} if:
\bi
\item {\tt C} is subclass of {\tt A} or;
\item there is an assignment\\
\verb@     x = y@\\
somewhere in the whole program such that the declared types of {\tt x}
and {\tt y} are {\tt A} and {\tt C}, respectively or;
\item {\tt A} is a supertype of a class {\tt B} which is supertype of
{\tt C}.
\ei

When one uses a type {\tt A} in a variable declaration like\\
\verb@     var a : A;@\\
she expects the methods executed when sending messages to {\tt a} to
obey the assertions of {\tt A} methods, even when variable {\tt a}
refers to a subtype object. To enforce this rule would be difficult
because of the dynamic nature of Green which separates classes and
types (and therefore objects and types since objects are created from
classes and types are used to declare variables).

This problem could be solved if the compiler put a shell (see
Chapter~\ref{shells})  in object
{\tt x} each time an assignment \\
\verb@     x = exp;@\\
were found and the declared type of {\tt exp} were a subtype, but not a
subclass, of the declared type of {\tt x}. The shell would check the
assertions of the methods. This assignment would be replaced by\\
\verb@     x = exp;@\\
\verb@     Meta.attachShell( x, Assert_A.new() );@\\
in which {\tt A} is the declared class of {\tt x} and \verb@Assert_A@
a shell class with the \verb@class_A@ assertions.

Variable {\tt x} might be used, after this assignment, in a command \\
\verb@     y = x;@\\
in which the declared type of {\tt y} is a supertype, but not a
superclass, of the declared type of {\tt x}. The the shell on {\tt x}
should be removed (should it~?) and another shell with class-of-{\tt
y} assertions should be attached to {\tt x}.

The idea of putting shells to check assertions is very complex and
would require the creation of great numbers of shell
classes. Therefore this feature is not supported by Green.

\chapter{The Basic Classes} \label{basictypes}

All the basic types are classes with value semantics. This mean a
variable of a basic type does not point to an object of that type. The
variable contains the object of that type. Because of this, some restrictions applies
to basic types: they cannot be subclassed and there is no polymorphism
with them. A polymorphic variable of a type {\tt T} may refer to
objects of {\tt T} and its subtypes, which may have different
sizes. Then the variable should be a pointer which has a fixed size in
bytes but can point to objects of any size. From here on, we will use
``basic classes'' for ``basic types''.

\section{The Classes}

The basic classes and their methods are shown below. {\tt AnyValue} is
superclass of all basic classes and neither {\tt AnyValue} nor a basic
class can be subclassed. {\tt AnyValue}, like {\tt Any}, does not
inherit from any class. The basic classes use low level representation of characters,
booleans, etc. These are the types of the machine to which the code is
generated. For example, {\tt lowLevelChar} is the machine type used to
represent characters. Objects of basic classes cannot be dynamically
allocated because they have value semantics. Albeit this limitation, a
class object of a basic class is an object like any other. The
arithmetic and logical operators are represented as methods in the
following code although Green does not support operator overloading
for normal classes. Note operator ``='' for assignment is not a method
of any class.

\begin{verbatim}
abstract class AnyValue
  public:
    abstract proc toString() : String
    abstract proc getInfo() : AnyObjectInfo
    abstract proc getClassInfo() : ClassInfo
    abstract proc getClassObject() : AnyClassObject
end


class char subclassOf AnyValue
  public:
    proc toString() : String
    proc equals( other : char ) : boolean
      // returns true if self is equal to other

      /* these operators can only be used with variables. Unlike C++, they
        don't return anything. */
    proc ++()
    proc --()

      // comparison
    proc == ( other : char ) : boolean
    proc <> ( other : char ) : boolean
    proc <  ( other : char ) : boolean
    proc <= ( other : char ) : boolean
    proc >  ( other : char ) : boolean
    proc >= ( other : char ) : boolean

  private:
    var value : lowLevelChar;
end


class boolean subclassOf AnyValue
  public:
    proc toString() : String
    proc equals( other : boolean ) : boolean
      // returns true if self is equal to other

    proc and( other : boolean ) : boolean
    proc or(  other : boolean ) : boolean
    proc xor( other : boolean ) : boolean
    proc not() : boolean

      // comparison
    proc == ( other : boolean ) : boolean
    proc <> ( other : boolean ) : boolean
    proc <  ( other : boolean ) : boolean
    proc <= ( other : boolean ) : boolean
    proc >  ( other : boolean ) : boolean
    proc >= ( other : boolean ) : boolean

  private:
    var value : lowLevelBoolean;
end


class byte subclassOf AnyValue
  public:
    proc toString() : String
    proc equals( other : byte ) : boolean
      // returns true if self is equal to other

      /* these operators can only be used with variables. Unlike C++, they
        don't return anything. */
    proc ++()
    proc --()

      // unary
    proc +  () : byte
    proc -  () : byte
    proc ~  () : byte
      // binary
    proc +  ( other : byte ) : byte
    proc -  ( other : byte ) : byte
    proc *  ( other : byte ) : byte
    proc /  ( other : byte ) : byte
    proc %  ( other : byte ) : byte
    proc &  ( other : byte ) : byte
    proc |  ( other : byte ) : byte
    proc ^  ( other : byte ) : byte
    proc << ( other : integer ) : byte
    proc >> ( other : integer ) : byte

      // comparison
    proc == ( other : byte ) : boolean
    proc <> ( other : byte ) : boolean
    proc <  ( other : byte ) : boolean
    proc <= ( other : byte ) : boolean
    proc >  ( other : byte ) : boolean
    proc >= ( other : byte ) : boolean

  private:
    var value : lowLevelByte;
end


class integer subclassOf AnyValue
  public:
    proc toString() : String
    proc equals( other : integer ) : boolean
      // returns true if self is equal to other

      /* these operators can only be used with variables. Unlike C++, they
        don't return anything. */
    proc ++()
    proc --()

      // unary
    proc +  () : integer
    proc -  () : integer
    proc ~  () : integer
      // binary
    proc +  ( other : integer ) : integer
    proc -  ( other : integer ) : integer
    proc *  ( other : integer ) : integer
    proc /  ( other : integer ) : integer
    proc %  ( other : integer ) : integer
    proc &  ( other : integer ) : integer
    proc |  ( other : integer ) : integer
    proc ^  ( other : integer ) : integer
    proc << ( other : integer ) : integer
    proc >> ( other : integer ) : integer

      // comparison
    proc == ( other : integer ) : boolean
    proc <> ( other : integer ) : boolean
    proc <  ( other : integer ) : boolean
    proc <= ( other : integer ) : boolean
    proc >  ( other : integer ) : boolean
    proc >= ( other : integer ) : boolean

     /* in the method that follow we assume an integer has at least
       four bytes */
    proc getByte1() : byte
      /* returns the value of the less significant byte of the integer. In some
         machines this will be the first byte. In others, it will be
         the last. The following methods work similarly. */
    proc getByte2() : byte
    proc getByte3() : byte
    proc getByte4() : byte

  private:
    var value : lowLevelInteger;
end


class long subclassOf AnyValue
  public:
    proc toString() : String
    proc equals( other : long ) : boolean
      // returns true if self is equal to other

      /* these operators can only be used with variables. Unlike C++, they
         don't return anything. */
    proc ++()
    proc --()

      // unary
    proc +  () : long
    proc -  () : long
    proc ~  () : long
      // binary
    proc +  ( other : long ) : long
    proc -  ( other : long ) : long
    proc *  ( other : long ) : long
    proc /  ( other : long ) : long
    proc %  ( other : long ) : long
    proc &  ( other : long ) : long
    proc |  ( other : long ) : long
    proc ^  ( other : long ) : long
    proc << ( other : integer ) : long
    proc >> ( other : integer ) : long

      // comparison
    proc == ( other : long ) : boolean
    proc <> ( other : long ) : boolean
    proc <  ( other : long ) : boolean
    proc <= ( other : long ) : boolean
    proc >  ( other : long ) : boolean
    proc >= ( other : long ) : boolean

     /* in the method that follow we assume a long has at least
        eight bytes */
    proc getByte1() : byte
      /* returns the value of the less significant byte of the long. In some
         machines this will be the first byte. In others, it will be
         the last. The methods getByte2, getByte3, ... getByte8 work similarly. */


  private:
    var value : lowLevelLong;
end


class real subclassOf AnyValue
  public:
    proc toString() : String
    proc equals( other : real ) : boolean
      /* returns true if self is equal to other. A warning should be
         issued if this method is used */

      /* these operators can only be used with variables. Unlike C++, they
         don't return anything. */
    proc ++()
    proc --()

      // unary
    proc +  () : real
    proc -  () : real
      // binary
    proc +  ( other : real ) : real
    proc -  ( other : real ) : real
    proc *  ( other : real ) : real
    proc /  ( other : real ) : real

      /* comparison. A warning should be issued whenever == or <> is used. */
    proc == ( other : real ) : boolean
    proc <> ( other : real ) : boolean
    proc <  ( other : real ) : boolean
    proc <= ( other : real ) : boolean
    proc >  ( other : real ) : boolean
    proc >= ( other : real ) : boolean

  private:
    var value : lowLevelReal;
end


class double subclassOf AnyValue
  public:
    proc toString() : String
    proc equals( other : double ) : boolean
      /* returns true if self is equal to other. A warning should be
         issued if this method is used */

      /* these operators can only be used with variables. Unlike C++, they
        don't return anything. */
    proc ++()
    proc --()

      /* unary
    proc +  () : double
    proc -  () : double
      /* binary */
    proc +  ( other : double ) : double
    proc -  ( other : double ) : double
    proc *  ( other : double ) : double
    proc /  ( other : double ) : double

      // comparison. A warning should be issued whenever == or <> is used.
    proc == ( other : double ) : boolean
    proc <> ( other : double ) : boolean
    proc <  ( other : double ) : boolean
    proc <= ( other : double ) : boolean
    proc >  ( other : double ) : boolean
    proc >= ( other : double ) : boolean

  private:
    var value : lowLevelDouble;
end



object char
  public:
    proc getSizeInBits() : integer
      // returns the size of a char in bits
    proc getSize() : integer
      // return the size of a char variable in bytes

    proc cast( value : String ) : char
    proc cast( any : AnyClass )
             ( exception : CatchTypeErrorException ) : char
      begin
      if not castOk(any) or the value of any cannot be converted to char
      then
        exception.throw( TypeErrorException.new() );
      endif
        // converts any to char
      ...
      end
    proc cast( value : byte ) : char
      assert
        before castOk(value);
      end
    proc cast( value : integer ) : char
      assert
        before castOk(value);
      end

    proc castOk( any : AnyClass ) : boolean
      begin
      var aClass : Any = any.getClassObject();
      return aClass == Char or aClass == Byte or
             aClass == Integer;
      end
    proc castOk( value : byte ) : boolean
      begin
      return value <= char.getMaxIntegerChar();
      end
    proc castOk( value : integer ) : boolean
      begin
      return value >= 0 and value <= char.getMaxIntegerChar();
      end


    proc getMinValue() : char
      begin
      return '\0';
      end
    proc getMaxValue() : char
      begin
      return '\x7F';
      end
    proc getMaxIntegerChar() : integer
        // returns maximum char in an integer value
      begin
      return 127;
      end
    proc getMinIntegerChar() : integer
      begin
      return 0;
      end
end


object boolean
  public:
    proc getSizeInBits() : integer
      // returns the size of a boolean in bits
    proc getSize() : integer
      // return the size of a boolean variable in bytes

    proc cast( value : String ) : boolean

    proc cast( any : AnyClass )
             ( exception : CatchTypeErrorException ) : boolean
      begin
      if not castOk(any) or the value of any cannot be converted to boolean
      then
        exception.throw( TypeErrorException.new() );
      endif
        // convert any to boolean
      ...
      end

    proc castOk( any : AnyClass ) : boolean
      begin
      var aClass : Any = any.getClassObject();
      return aClass == Boolean or aClass == Byte or
             aClass == Integer;
      end

    proc cast( value : integer ) : boolean
      // cast 0 to false and anything else to true

    proc cast( value : byte ) : boolean
      // cast 0 to false and anything else to true

    proc getMinValue() : boolean
      begin
      return false;
      end
    proc getMaxValue() : boolean
      begin
      return true;
      end
end


object byte
  public:
    proc getSizeInBits() : integer
      // returns the size of a byte in bits
    proc getSize() : integer
      // return the size of a byte variable in bytes


    proc cast( any : AnyClass )
             ( exception : TypeErrorException ) : byte
      begin
      if not castOk(any) or the value of any cannot be converted to byte
      then
        exception.throw( TypeErrorException.new() );
      endif
        // converts
      ...
      end

    proc cast( value : String ) : byte

    proc cast( value : boolean ) : byte
      assert
        before castOk(value);
      end

    proc cast( value : char ) : byte
      assert
        before castOk(value);
      end


    proc cast( value : integer ) : byte
      assert
        before castOk(value);
      end

    proc cast( value : long ) : byte
      assert
        before castOk(value);
      end

    proc cast( value : real ) : byte
      assert
        before castOk(value);
      end

    proc cast( value : double ) : byte
      assert
        before castOk(value);
      end

    proc castOk( any : AnyClass ) : boolean
      begin
      var aClass : Any = any.getClassObject();
      return aClass == Boolean or aClass == Byte or
             aClass == Integer;
      end

    proc castOk( value : integer ) : boolean
      begin
      return value >= byte.getMinValue() and
             value <= byte.getMaxValue();
      end

    proc castOk( value : long ) : boolean
      begin
      return value >= byte.getMinValue() and
             value <= byte.getMaxValue();
      end

    proc castOk( value : real ) : boolean
      begin
      return value >= byte.getMinValue() and
             value <= byte.getMaxValue();
      end

    proc castOk( value : double ) : boolean
      begin
      return value >= byte.getMinValue() and
             value <= byte.getMaxValue();
      end

    proc getMinValue() : byte
      begin
      return 0;
      end
    proc getMaxValue() : byte
      begin
      return 255;
      end
end


object integer
  public:
    proc getSizeInBits() : integer
      // returns the size of an integer in bits
    proc getSize() : integer
      // return the size of an integer variable in bytes

    proc cast( any : AnyClass )
             ( exception : TypeErrorException ) : integer
      begin
      if not castOk(any) or the value of any cannot be converted to integer
      then
        exception.throw( TypeErrorException.new() );
      endif
        // converts
      ...
      end

    proc cast( value : String ) : integer
    proc cast( value : char ) : integer
    proc cast( value : boolean ) : integer
    proc cast( value : byte ) : integer

    proc cast( value : long ) : integer
      assert
        before castOk(value);
      end

    proc cast( value : real ) : integer
      assert
        before castOk(value);
      end

    proc cast( value : double ) : integer
      assert
        before castOk(value);
      end

    proc castOk( any : AnyClass ) : integer
      begin
      var aClass : Any = any.getClassObject();
      return aClass == Char or aClass == Boolean or aClass == Byte or
             aClass == Integer or aClass == Long or
             aClass == Real or aClass == Double;
      end

    proc castOk( value : double ) : boolean
      begin
      return value >= integer.getMinValue() and
             value <= integer.getMaxValue();
      end

    proc castOk( value : real ) : boolean
      begin
      return value >= integer.getMinValue() and
             value <= integer.getMaxValue();
      end

    proc castOk( value : long ) : boolean
      begin
      return value >= integer.getMinValue() and
             value <= integer.getMaxValue();
      end

    proc getMinValue() : integer
      begin
      return -2147483647 - 1;
      end
    proc getMaxValue() : integer
      begin
      return 2147483647
      end
end


object long
  public:
    proc getSizeInBits() : integer
      // returns the size of a long in bits
    proc getSize() : integer
      // return the size of a long variable in bytes

    proc cast( any : AnyClass )
             ( exception : TypeErrorException ) : long
      begin
      if not castOk(any) or the value of any cannot be converted to long
      then
        exception.throw( TypeErrorException.new() );
      endif
        // converts
      ...
      end

    proc cast( value : String ) : long
    proc cast( value : byte ) : long
    proc cast( value : integer ) : long
    proc cast( value : real ) : long
      assert
        before castOk(value);
      end
    proc cast( value : double ) : long
      assert
        before castOk(value);
      end

    proc castOk( any : AnyClass ) : boolean
      begin
      var aClass : Any = any.getClassObject();
      return aClass == Long or aClass == Integer or aClass == Byte
             aClass == Real or aClass == Double;
      end

    proc castOk( value : double ) : boolean
      begin
      return value >= long.getMinValue() and
             value <= long.getMaxValue();
      end

    proc castOk( value : real ) : boolean
      begin
      return value >= long.getMinValue() and
             value <= long.getMaxValue();
      end

    proc getMinValue() : long
      begin
      return -2147483647*2147483647 - 1;
      end
    proc getMaxValue() : long
      begin
      return 2147483647*2147483647;
      end
end


object real
  public:
    proc getSizeInBits() : integer
      // returns the size of a real in bits
    proc getSize() : integer
      // return the size of a real variable in bytes

    proc cast( any : AnyClass )
             ( exception : TypeErrorException ) : real
      begin
      if not castOk(any) or the value of any cannot be converted to real
      then
        exception.throw( TypeErrorException.new() );
      endif
        // converts
      ...
      end

    proc cast( value : String ) : real
    proc cast( value : byte ) : real
    proc cast( value : integer ) : real
    proc cast( value : long ) : real
    proc cast( value : double ) : real
      assert
        before castOk(value);
      end

    proc castOk( any : AnyClass ) : boolean
      begin
      var aClass : Any = any.getClassObject();
      return aClass == Byte or aClass == Integer or aClass == Long
             or aClass == Real or aClass == Double;
      end

    proc castOk( value : double ) : boolean
      begin
        // platform-dependent code
      end


      /* each of the following methods is based on a similar constant
         found in the header file "limits.h" of the standard C
         library. The method description is a  copy of the
         description of the corresponding constant as found in K&R. The
         method bodies are dependent on the platform and therefore not
         defined. */

    proc getRadix() : integer
      // radix of exponent representation
    proc getRounds() : integer
      // mode for addition
    proc getPrecision() : integer
      // decimal digits of precision
    proc getEpsilon() : real
      // smaller number e such that 1.0 + e <> 1.0
    proc getMantDig() : integer
      // number of base ``getRadix()'' in mantissa
    proc getMinValue() : real
      // minimum real number
    proc getMaxValue() : real
      // maximum real number
    proc getMaxExp() : real
      // greater k such that k^n is representable
    proc getMinExp() : real
      // smaller k such that k^n is a normalized number

end



object double
  public:
    proc getSizeInBits() : integer
      // returns the size of a double in bits
    proc getSize() : integer
      // return the size of a double variable in bytes

    proc cast( any : AnyClass )
             ( exception : TypeErrorException ) : double
      begin
      if not castOk(any) or the value of any cannot be converted to double
      then
        exception.throw( TypeErrorException.new() );
      endif
        // converts
      ...
      end

    proc cast( value : String ) : double
    proc cast( value : byte ) : double
    proc cast( value : integer ) : double
    proc cast( value : long ) : double
    proc cast( value : real ) : double

    proc castOk( any : AnyClass ) : double
      begin
      var aClass : Any = any.getClassObject();
      return aClass == Byte or aClass == Integer or aClass == Long or
             aClass == Real or aClass == Double;
      end

      /* each of the following methods is based on a similar constant
         found in the header file "limits.h" of the standard C
         library. The method description is a  copy of the
         description of the corresponding constant as found in K&R. The
         method bodies are dependent on the platform and therefore not
         defined. */

    proc getRadix() : integer
      // radix of exponent representation
    proc getRounds() : integer
      // mode for addition
    proc getPrecision() : integer
      // decimal digits of precision
    proc getEpsilon() : double
      // smaller number e such that 1.0 + e <> 1.0
    proc getMantDig() : integer
      // number of base ``getRadix()'' in mantissa
    proc getMinValue() : double
      // minimum double number
    proc getMaxValue() : double
      // maximum double number
    proc getMaxExp() : double
      // greater k such that k^n is representable
    proc getMinExp() : double
      // smaller k such that k^n is a normalized number
end
\end{verbatim}
Notice that
\bi
\item an object can be cast to a basic class through methods like \\
\verb@     proc cast( any : AnyClass )@\\
\verb@              ( exception : TypeErrorException ) : char@\\
of class object {\tt char}. This will only succeed if the class of
{\tt any} at run time is {\tt Char}, {\tt Byte}, or {\tt Integer} and
the value can be converted to a character;
\item  every basic value can be transformed into a string:
\begin{verbatim}
s = 5.toString();         // s will point to "5"
s = #'A'.toString();      // s will point to "A"
\end{verbatim}

\item {\tt real} and {\tt double} numbers are transformed into strings
using the format \verb@[-]m.ddddddE+-xx@ where {\tt m} is the number
before the decimal point;

\item the assertions used in the basic class definitions make use of
some implicit conversions such as
from {\tt integer} to {\tt long} in method \verb@castOk( value : long )@ of
class object {\tt integer};

\item hopefully the compiler will able to enable/disable the
assertions for basic class methods. If an assertion of a {\tt cast}
method is violated, the run-time system will throw an exception of one
of the following subclasses of {\tt AssertionException}:
\begin{verbatim}
AssertionCastCharException
AssertionCastBooleanException
AssertionCastByteException
AssertionCastIntegerException
AssertionCastLongException
AssertionCastRealException
AssertionCastDoubleException
\end{verbatim}
Class {\tt AssertionCastCharException} has methods
\begin{verbatim}
proc init( originalValueClass : AnyClassObject; value : Any )
proc getOriginalValueClass() : AnyClassObject;
proc getOriginalValue() : Any
\end{verbatim}
{\tt originalValueClass} is the class of the object that could not
have been converted to {\tt char}. {\tt value} is transformed into an object of a wrapper
class (see next Section). To understand this better, consider the code
\begin{verbatim}
var i : integer = 300;
var ch : char;
ch = char.cast(i);  // throws exception
\end{verbatim}
At run time, exception  {\tt AssertionCastCharException} will be
thrown by a call
\begin{verbatim}
exception.throw( AssertionCastCharException.new( integer,
                                                 Integer.new(i) ) )
\end{verbatim}
Class {\tt Integer} is a wrapper class described next.
\ei

\section{Wrapper Classes} \label{wrapper}

There are classes {\tt Char}, {\tt Boolean}, {\tt Byte}, {\tt
Integer}, {\tt Long}, {\tt Real}, and {\tt Double} that obey the
reference semantics and mirror the basic classes. Of course, all
inherit from {\tt AnyClass}. Automatic conversion
is provided between an object of one of these classes and its
corresponding basic class value. So, the code below is legal.
\begin{verbatim}
    var I : Integer;
        i : integer;
begin
i = 1;
I = 1;
i = I;
I = i;
I = Integer.new(5);
i = Integer.new(5);
i = i + I;
i = 6*I + 10 - I;
I = 3*i*I;
end
\end{verbatim}
The compiler inserts code to create objects of class {\tt Integer} and
to convert these objects in {\tt integer} values. For example, the
statements \\
\verb@     I = 1;@\\
\verb@     i = 5*I;@\\
would be implemented as\\
\verb@     I = Integer.new(1);@\\
\verb@     i = 5*I.get();@

Classes like {\tt Integer} are called wrapper classes because they
pack basic values in objects. Since class {\tt Integer} and the like
are subclasses of {\tt Any}, they allow the program to treat basic values as
 objects. For example, a method
\begin{verbatim}
proc print( any : Any )
  begin
  Out.write( any.toString() );
  end
\end{verbatim}
accepts values of any basic class and any object as parameter. So, \\
\verb@     a.print(5);@\\
is transformed into\\

\verb@     a.print( Integer.new(5) );@\\

Whenever a basic class value is used where an object is expected, a
wrapper object is automatically created.

The automatic conversion between basic classes and wrapper classes is
one of the novelties of Green. This allows one to see everything in
the program  as objects, a feature supported only by very dynamic
languages as Smalltalk and Self. On the other hand, the programmer can
use the basic classes for better performance and convert basic values to
objects when necessary. There is only one restriction on wrapper
classes: they cannot be subclassed. This is necessary to allow the
compiler to optimize their use.

Wrapper classes add a restriction on overloaded methods: a class cannot have two methods\\
\verb@     proc print( i : integer )@\\
\verb@     proc print( i : Integer )@\\
There would be an ambiguity in a call\\
\verb@     s.print(5)@\\
because {\tt 5} is converted to {\tt Integer} whenever necessary. In
the general case, a class has illegal method declarations if, after
changing the type of all parameters to the corresponding wrapper
classes, two methods have the same name, number of parameters, and
parameter types.

The {\tt Char} class is shown next. The other wrapper classes are
similar. After a wrapper object is created, its value is never modified.

\begin{verbatim}
class Char
    proc init( value : char )
      begin
      self.value = value;
      end
  public:
    proc get() : char
      begin
      return value;

      end
  private:
    var value : char;
end
\end{verbatim}

The {\tt ++} operator increments its operand and can be used with a
wrapper variable of any class but {\tt Boolean}. After the application
of {\tt ++} on a variable, it will refer to a different object since
wrapper objects never change its value. So, the code\\
\verb@     var I : Integer;@\\
\verb@     I = 1;@\\
\verb@     ++I;@\\
creates two wrapper objects.

\chapter{Strings} \label{strings}

Any sequence of characters between \verb@"@ is an object of type {\tt
String}, a predefined class. The character \verb@"@ itself can be put
in the sequence by prefixing it with \verb@\@. Examples of {\tt
String} handling are given below.
\begin{verbatim}
var s : String = "this is a string";
var p : String;

p = s + " yet " + "another" + " string " + "(YAS)" + "\n";
\end{verbatim}

The operator \verb@+@ is overloaded to concatenate two string objects
returning a {\it new} string and \verb@\n@ is the new line character.

A string can be initialized in its creation:
\begin{verbatim}
var s, t : String;
s = String.new("a string");
t = String.new(hour, "h ", min, "min ", "am");
\end{verbatim}
If  {\tt hour = 8}, {\tt min = 35}, {\tt t} would hold the string
\verb@"8h 35min am"@.
 A literal string such as ``a string" is a compiler
allocated object.

For performance sake, class {\tt String} cannot be subclassed, has no
subtypes, and  does not provide any method for modifying its
content. If you need to change a string you should use objects of
class {\tt DynString}. Then a variable of class {\tt String} will
always point to a string object that is never modified. The
definitions of classes {\tt String} and {\tt DynString} are given below.

\begin{verbatim}
object String
   public:
      proc cast( x : char ) : String
      proc cast( x : boolean ) : String
      proc cast( x : byte ) : String
      proc cast( x : integer ) : String
      proc cast( x : long ) : String
      proc cast( x : real ) : String
      proc cast( x : double ) : String
end

class String
    proc init( s : String )
      // initializes self with s
      assert
        before s <> nil;
      end

    proc init( v : ...array(Any)[] )
      // initializes self with strings corresponding to v values

  public:
    proc get( i : integer ) : char
        // get character at position i
      assert
        before i >= 0 and i < getSize();
      end

    proc getIter() : DS.Iter(char)
      // returns an iterator for the string. See Chapter "The Standard Library"

    proc cmp( other : String ) : integer
      /* compare strings self and other. Returns -1, 0, or 1 if self
         is less than, equal to, or greater than other, respectively.
         This is the same as the C strcmp function. */
      assert
        before other <> nil;
      end

    proc cmpIgnoreCase( other : String ) : integer
      /* the same as cmp but ignoring differences between upper and
         lower case letters */
      assert
        before other <> nil;
      end

    proc newConcat( other : String )
                  ( exception : CatchOutOfMemoryException ) : String
      /* creates and returns a new String that is the concatenation of
         self with other */
      assert
        before other <> nil;
      end

    proc tocharArray( copyto : array(char)[] )
                    ( exception : CatchOutOfMemoryException )
      // copies string into array copyto
      assert
        before copyto.getSize() >= getSize();
      end

    proc tocharArray( copyto : array(char)[]; i : integer )
                    ( exception : CatchOutOfMemoryException )
      /* copies string into array copyto beginning at position
         i of copyto */
      assert
        before copyto.getSize() - i >= getSize();
      end

    proc tobyteArray( copyto : array(byte)[] )
                    ( exception : CatchOutOfMemoryException )
      // copies string into array copyto
      assert
        before copyto.getSize() >= getSize();
      end

    proc tobyteArray( copyto : array(byte)[]; i : integer )
                    ( exception : CatchOutOfMemoryException )
      /* copies string into array copyto beginning at position
         i of copyto */
      assert
        before copyto.getSize() - i >= getSize();
      end

    proc getSize() : integer
      // return the number of elements of the string

    proc newToLowerCase()
                       ( exception : CatchOutOfMemoryException ) : String
      // returns a new string with all letters changed to lowercase

    proc newToUpperCase()
                       ( exception : CatchOutOfMemoryException ) : String
      // returns a new string with all letters changed to uppercase

    proc getSubset( from, to2 : integer )
                  ( exception : CatchOutOfMemoryException ) : String
      /* returns a new string with all characters of this string
         between positions from and to2. to is a reserved word, then to2. */
      assert
        before from >= to2 and 0 <= from and to2 < getSize;
        after  result.getSize() == to2 - from + 1;
      end

    proc search( s : String ) : integer
      /* searchs for string s in self.  Returns the index of s in self
         or -1 if it was not found. */
      assert
        before s <> nil;
        after  result >= -1 and result < getSize();
      end

    proc hashCode() : integer

    proc tobyte() : byte
      /* assumes the string is a number and converts it to a byte.
         The following methods work similarly. */
      assert
        before tobyteOk();
      end

    proc tointeger() : integer
      assert
        before tointegerOk();
      end

    proc tolong() : long
      assert
        before tolongOk();
      end

    proc toreal() : real
      assert
        before torealOk();
      end

    proc todouble() : double
      assert
        before todoubleOk();
      end

    proc toDynString()
                    ( exception : CatchOutOfMemoryException ) : DynString

end


class DynString
    proc init( s : String )
      // initializes self with s
      assert
        before s <> nil;
      end

    proc init( s : DynString )
      // initializes self with s
      assert
        before s <> nil;
      end

  public:

    proc get( i : integer ) : char
        // get character at position i
      assert
        before i >= 0 and i < getSize();
      end

    proc getIter() : DS.Iter(char)
      // returns an iterator for the string. See Chapter "The Standard Library"

    proc cmp( other : DynString ) : integer
      /* compare strings self and other. Returns -1, 0, or 1 if self
         is less than, equal to, or greater than other, respectively.
         This is the same as the C strcmp function. */
      assert
        before other <> nil;
      end

    proc cmpIgnoreCase( other : DynString ) : integer
      /* the same as cmp but ignoring differences between upper and
         lower case letters */
      assert
        before other <> nil;
      end

    proc concat( other : String )
               ( exception : CatchOutOfMemoryException )
        /*  creates and returns a new DynString that is the concatenation of
            self with other */
      assert
        before other <> nil;
      end

    proc tocharArray( copyto : array(char)[] )
                    ( exception : CatchOutOfMemoryException )
      // copies string into array copyto
      assert
        before copyto.getSize() >= getSize();
      end

    proc tocharArray( copyto : array(char)[]; i : integer )
                    ( exception : CatchOutOfMemoryException )
      /* copies string into array copyto beginning at position
         i of copyto */
      assert
        before copyto.getSize() - i >= getSize();
      end

    proc tobyteArray( copyto : array(byte)[] )
                    ( exception : CatchOutOfMemoryException )
      // copies string into array copyto
      assert
        before copyto.getSize() >= getSize();
      end

    proc tobyteArray( copyto : array(byte)[]; i : integer )
                    ( exception : CatchOutOfMemoryException )
        /* copies string into array copyto beginning at position
           i of copyto */
      assert
        before copyto.getSize() - i >= getSize();
      end

    proc getSize() : integer
      // return the number of elements of the string

    proc toLowerCase()
                    ( exception : CatchOutOfMemoryException )
      // returns a new string with all letters changed to lowercase

    proc toUpperCase()
                    ( exception : CatchOutOfMemoryException )
      // returns a new string with all letters changed to uppercase

    proc getSubset( from, to2 : integer )
                  ( exception : CatchOutOfMemoryException ) : DynString
        /* returns a new string with all characters of this string
           between positions from and to2. to is a reserved word, then to2. */
      assert
        before from >= to2 and 0 <= from and to2 < getSize;
        after  result.getSize() == to2 - from + 1;
      end

    proc search( s : DynString ) : integer
        /* searchs for string s in self.  Returns the index of s in self
           or -1 if it was not found. */
      assert
        before s <> nil;
        after  result >= -1 and result < getSize();
      end

    proc hashCode() : integer

    proc tobyte() : byte
      assert
        before tobyteOk();
      end

    proc tointeger() : integer
      assert
        before tointegerOk();
      end

    proc tolong() : long
      assert
        before tolongOk();
      end

    proc toreal() : real
      assert
        before torealOk();
      end

    proc todouble() : double
      assert
        before todoubleOk();
      end

    proc removeSpaceBegin()
      /* removes all spaces of the beginning of the string. A space is
         one of the followings: ' ', '\t', '\r', '\n' . */

    proc removeSpaceEnd()
      /* removes all spaces of the end of the string. A space is
         one of the followings: ' ', '\t', '\r', '\n' . */

    proc toString()
                 ( exception : CatchOutOfMemoryException )  : String
      // returns a String corresponding to this DynString

    proc prepend( toadd : DynString )
                ( exception : CatchOutOfMemoryException )
      // add toadd at the beginning of self

    proc removeAllCh( ch : char ) : boolean
      // removes all characters ch of the string

    proc remove( i : integer )
      // removes the character i of the string
      assert
        before i >= 0 and i < getSize();
      end

    proc insert( i : integer; ch : char )
               ( exception : CatchOutOfMemoryException )
      /* inserts character ch at position i of the string. The previous
         character at this position is right shifted */
      assert
        before i >= 0 and i < getSize();
      end

    proc add( i : integer; ch : char )
            ( exception : CatchOutOfMemoryException )
      // replaces character of position i by ch
      assert
        before i >= 0 and i <= getSize();
      end


    proc add( ch : char )
            ( exception : CatchOutOfMemoryException )
      // adds ch at the end of the string
end
\end{verbatim}

\chapter{Class Objects and Subtyping}  \label{classobjectssub}

\section{Types and Class Objects}
\label{tycobj}

\begin{figure}
\begin{verbatim}
object Window
  public:
    proc getDefaultColor() : integer
      begin
      return defaultColor;
      end
  private:
    var defaultColor : integer = 12;
end

class Window
    proc init()
      begin
      ...
      end

  public:
    ...
end
\end{verbatim}
\caption{Class object and class {\tt Window}}
\label{tyo1}
\end{figure}

A class object defines an object at compile time. A class specifies
the instance variables/methods  its objects will have at run time.
Class object {\tt Window} of Figure~\ref{tyo1} {\it defines} the
object {\tt Window} and as such it can assign a value to variable
{\tt defaultColor} in the line\\
\verb@     var defaultColor : integer = 12;@\\
This assignment  would be illegal if it were in class {\tt Window}
since a class declaration is a {\it type declaration}. It only
specifies  the shape of the objects of that class and could not have
an action such as an assignment attached to it.

Although class objects do not have a class, they do have a type, which
is just the set of public method signatures of the object.
Then, the type of class object {\tt Window} is\\
\verb@     { getDefaultColor() : integer,@\\
\verb@       new() : Window,@\\
\verb@       ...@\\
\verb@       }@\\
The dots (...) represent some methods the compiler adds to each class
object. These methods will be presented in
Section~\ref{otheradd}. Method {\tt new} is added by the compiler
since class {\tt Window} defines a method {\tt init}.

Now we are ready to define the compile-time function {\tt type}. This
function takes a variable as a parameter, which includes
classes.\footnote{If there is a class {\tt A}, the program has a
variable {\tt A} that points to the class object corresponding to this
class.} Function {\tt type} is defined as follows.
\bi
\item  {\tt type(a)} is the declared type of {\tt a} if {\tt a} is a
user-defined variable. For example,
\begin{verbatim}
var  x : integer;
     y : type(x);
     a : A;
     b : type(a);
\end{verbatim}
declares the type of {\tt y} as {\tt integer} and the type of {\tt b} as
{\tt A}. The statement \\
\verb@     var a : A = type(a).new()@\\
is equivalent to\\
\verb@     var a : A = A.new();@\\

\item {\tt type(A)} is the set of methods of class object {\tt A} if
{\tt A} is a class name. Then,\\
\verb@     var W : type(Window);@\\
declares a variable {\tt W} to which can be assigned an object whose
type is a subtype of {\tt type(Window)}.

To {\tt W} may be assigned {\tt Window}:\\
\verb@     W = Window;@\\
\verb@     i = W.getDefaultColor();  // calls method of Window@\\

In fact, the compiler will create an abstract class with method
signatures equal to those of {\tt Window}. This class will replace
``{\tt type(Window)}''.
\ei

If a class defines constants or enumerated constants in its public
section, these do not belong to the class type since a type is just a
set of {\it method} interfaces.

\begin{figure}
\begin{verbatim}
abstract class TypeA
  public:
    abstract proc m( x : TypeA ) : integer
    abstract proc n() : integer
end

object A
  public:
    proc m( a : TypeA ) : integer
      begin
      return a.n() + 1;
      end
    proc n() : integer
      begin
      return 0;
      end
end

object B
  public:
    proc m( a : TypeA ) : integer
      begin
      return A.m(a);
      end
    proc n() : integer
      begin
      return 1;
      end
end
\end{verbatim}
\caption{Delegation using class objects}
\label{tyo2}
\end{figure}

\begin{figure}
\begin{verbatim}
abstract class TypeA
  public:
    abstract proc m(x : TypeA) : integer
    abstract proc n() : integer
end

object A
  public:
    proc m() : integer
      begin
      return n() + 1;
      end
    proc n() : integer
      begin
      return 0;
      end
end

object B
  public:
    proc m() : integer
      begin
      return A.m();
      end
    proc n() : integer
      begin
      return 1;
      end
end
\end{verbatim}
\caption{Forwarding of messages using class objects}
\label{tyo3}
\end{figure}

\section{Delegation and Class Objects}

Since class objects are just objects, there is no inheritance
relationship among them. Sharing of methods among class objects is
made through a delegation mechanism. For example, class object {\tt B}
of Figure~\ref{tyo2} delegates the {\tt m} message to class {\tt
A}. Method  {\tt m} of {\tt B} should be called as\\
\verb@     B.m(B)@\\
that will execute\\
\verb@     A.m(B)@\\
that will call method {\tt n} of parameter {\tt a} in\\
\verb@    a.n() + 1@\\
Since {\tt a} refer to class object {\tt B}, this will call method
{\tt n} of {\tt B}.

This means method {\tt m} of {\tt A} knows the identity of the
original receiver of message {\tt m}, which was {\tt B}. This is the
true delegation.  If {\tt m} of {\tt A} did not know the original
receiver, there would be {\tt forwarding} of messages, as is shown in
Figure~\ref{tyo3}.

In this case,\\
\verb@     B.m();@\\
will call method {\tt m} of {\tt A} that will call method {\tt n} of
{\tt A}.

\section{Methods Added to Class Objects}
\label{cast}
\label{otheradd}

\begin{figure}
\begin{verbatim}

object Any
  public:
    proc equals( other : Any ) : boolean
    proc toString( p : Any ) : String
    proc isObjectOf( p : Any; aClass : AnyClassObject ) : boolean
    proc getInfo( p : Any ) : AnyObjectInfo
    proc shallowClone( p : Any ) : Any;
    proc deepClone   ( p : Any ) : Any;
    proc shallowCopy ( p : Any; other : Any ) : boolean;
    proc shallowEqual( p : Any; other : Any ) : boolean;
    proc deepEqual   ( p : Any; other : Any ) : boolean;
      // there are other methods
    ...
end
\end{verbatim}
\caption{Class object {\tt Any}}
\label{oth1}
\end{figure}

We have seen that class {\tt Any} defines the methods\\
\bi
\item {\tt getInfo() : AnyObjectInfo}

\item {\tt equals( other : Any ) : boolean}

\item {\tt toString() : String}

\item {\tt isObjectOf( aClass : AnyClassObject ) : boolean}

\item \verb@shallowClone() : Any@\\
      \verb@            ( exception : CatchOutOfMemoryException )@

\item \verb@deepClone() : Any@\\
      \verb@         ( exception : CatchOutOfMemoryException )@

\item {\tt shallowCopy( other : Any ) : boolean}

\item {\tt shallowEqual( other : Any ) : boolean}

\item {\tt deepEqual( other : Any ) : boolean}

\ei
that are inherited by all classes. Each of these methods calls a
similar method in class object {\tt Any}. As an example, method
{\tt shallowClone} of class {\tt Any} is defined as
\begin{verbatim}
class Any
  public:
    proc shallowClone() : Any;
      begin
      return Any.shallowClone(self);
      end
    ...
end
\end{verbatim}
These {\tt Any} methods are then inherited by all classes. Note {\tt
Any} of the expression\\
\verb@     Any.shallowClone(self)@\\
refers to the {\it class object} {\tt Any}.
Class object {\tt Any} defines a method with an extra parameter\\
\verb@     p : Any@\\
for each of the methods of class {\tt Any} listed
above --- see Figure~\ref{oth1}.

    To each {\it class object} are added methods with the same signatures
as those of class {\tt Any}. These methods are defined as follows.
\begin{verbatim}
object Person
  public:
    proc shallowClone() : Any
      begin
      return Any.shallowClone(Person);
      end
    ...
end
\end{verbatim}
By the above, class object {\tt Any} also has a method\\
\verb@     proc shallowClone() : Any@\\

In a message send\\
\verb@     a.shallowClone()@\\
where {\tt a} points to a class-{\tt A} object, method {\tt
shallowClone()} of class {\tt Any} is called. This method calls
method\\
\verb@     shallowClone(Any) : Any@\\
of class object {\tt Any} in the statement\\
\verb@     return Any.shallowClone(self);@\\
Pseudo-variable {\tt self}, which points to the same object as {\tt
a}, is passed as a
parameter. If this method of class object {\tt Any} needs to call
a method of the object being cloned, it will call it by sending a
message to its first parameter. Then the search for the method will
begin in  object {\tt a}. As we have seen, this is true delegation.

The statement \\
\verb@     p = Person.shallowClone();@\\
calls method {\tt shallowClone} of class object {\tt Person} that
executes \\
\verb@     Any.shallowClone(Person)@\\
creating a new class.

Class object {\tt Any} defines a method\\
\verb@     proc basicNew( aClass : AnyClassObject )@\\
\verb@                  ( exception : CatchCreationException ) : AnyClass@\\
which takes a class object as parameter an returns an uninitialized
object of that class. Exception {\tt CreationException} is thrown if
{\tt aClass} is not a class object\footnote{It may be a subtype of
{\tt AnyClassObject} without being a class object.} or if no object of
that class can be created. This last case occur if {\tt aClass} is an
abstract class or if it is not visible by some reason (this may occur
if private classes are to be supported by Green).

An example of use of {\tt basicNew} could be\\
\verb@     var a : Account;@\\
\verb@     a = Any.basicNew(Account);@\\
\verb@     a.set(myName, myBank, myBalance);@\\
{\tt basicNew} creates an object but it does not initialize it --- no
{\tt init} method of {\tt Account} is called. Then the object instance
variables should be initialized by other methods, as {\tt set} in the
above example.

{\tt basicnew} may be used to create a singleton class:
\begin{verbatim}
object Earth
  proc init()
    begin
    try(HCatchAll)
      earth = Any.basicNew(Earth);
    end
    end
  public:
    proc new() : Earth
      begin
        // always return the same object
      return earth;
      end
  private:
    var earth : Earth;
end

class Earth subclassOf Planet
    // no init methods
  ...
end
\end{verbatim}

To each class object {\tt A} the compiler adds  methods
\bi
\item
\begin{verbatim}
proc cast( x : Any )
         ( exception : CatchTypeErrorException ) : A
  begin
  if x.isObjectOf(A)
  then
    return x;   // converts x to A
  else
    exception.throw( TypeErrorException.new() );
  endif
  end
\end{verbatim}
that casts object {\tt x} into type {\tt A}. It throws an exception
{\tt TypeErrorException} if the {\tt x} class is not a subtype of {\tt A}.

\item {\tt getAssociateClassInfo() : ClassInfo}

returns information about the class associated to this class object.

\item {\tt getInitMethod() : ObjectInitMethodInfo}

returns information on the {\tt init} method of the class object.

\item \verb@ castObject( any : Any )@\\
\verb@      ( exception : CatchTypeErrorException ) : type(A)@\\
returns the object {\tt any} cast into type \verb@type(A)@. Exception {\tt CatchTypeErrorException} is thrown if this is not possible.

\ei
The last two methods are also defined in the abstract class {\tt
AnyClassObject} that inherits from {\tt Any}. A variable of type {\tt
AnyClassObject} can refer to any class object. {\tt AnyClassObject} is
{\it supertype} of the type of any class object.

The definition of class objects and of class {\tt Any} implies
that:
\bi
\item each class inherits from {\tt Any} and is therefore a subtype
of it;
\item to each class object are added methods with the same signatures
as those of class {\tt Any}. Therefore, the type of each class
object is a subtype of the type of {\tt Any}.
\ei
We conclude that the type of each object, whatever it is, is a subtype
of the type of class {\tt Any}. Note the definition of type was
extended to cope with objects and not only classes.

If one wants to modify a method automatically added to a class object
as {\tt shallowClone}, she should define this method in the class
object by herself:
\begin{verbatim}
object MyWindow
  public:
    proc shallowClone() : Any
      begin
        // my code
        ...
      end
end
\end{verbatim}
Then the compiler will not add a {\tt shallowClone} method to this
class object.

      If one wants to redefine all {\tt shallowClone} methods of all
class objects, she should attach a shell (Chapter~\ref{shells}) with a
method\\
\verb@     proc shallowClone( p : Any ) : Any@\\
to class object {\tt Any}. This is the Green way of getting the
functionality of other languages that support metaclasses. In one of
these languages, there could be a metaclass {\tt MetaClass} from which
all metaclasses inherit. In a language like this, one could redefine
all {\tt shallowClone} methods of all class objects by changing method
{\tt shallowClone} of {\tt MetaClass}.

\chapter{Exceptions} \label{exceptions}

\section{Definition}

Suppose one wants to open, read, and close a text file with a class
{\tt File}:
\begin{verbatim}
var f : File;
var text : array(char)[];

f = File.new();
f.open("mytext");
text = array(char)[].new( f.getSize() );
f.read(text);
f.close();
\end{verbatim}
There may be some error when allocating memory, opening, reading, or
closing  the file. In languages without an exception system, there are
two main patterns of taking care of these errors:

\be
\item requiring every method call that may fail to return a value
(e.g. {\tt false}) showing it failed;
\item requiring every method call that may fail to set a global
variable with the error number.
\ee

In any case the programmer is forced to fill the code with a lot of
{\tt if} statements to test for errors. This makes the code obscure
because code for  two different jobs are mixed statement by statement:
reading a file and treating exceptional conditions.

This code can be rewritten as
\begin{verbatim}
var f : File;
var text : array(char)[];
var errorCatch : CatchFileException;

errorCatch = CatchFileException.new();
try(errorCatch)
    f = File.new();
    f.open("mytext");
    text = array(char)[].new( f.getSize() );
    f.read(text);
    f.close();
end
\end{verbatim}
Now any error signaled by the code between {\tt try} and {\tt end} is
handled by the methods of class\\ {\tt CatchFileException}. The methods {\tt
new}, {\tt open}, {\tt read}, and {\tt close} throw an error by
sending a message {\tt throw} to  object {\tt exception}:\\
\verb@     exception.throw( FileOpenException.new(fileName) );@\\
{\tt FileOpenException} is a direct or indirect  subclass of class {\tt
Exception}. When {\tt exception} receives a message {\tt throw}, the
run-time system looks for a {\tt try} clause in the stack of called
methods that  can handle this exception. In the stack of called
methods there may be several methods with {\tt try} statements, each one
with an object like {\tt errorCatch} of the previous example. The
run-time system will send a message {\tt throw} to the object of one
of the {\tt try} statements and remove all methods of the stack of
called methods to this point. Then all methods that could not have
handled the exception are terminated. Now there is a question: how the
run-time system chooses the {\tt try} statement to be used~? To
explain that, suppose the stack of called methods is {\tt m$_{1}$},
{\tt m$_{2}$}, ... {\tt m$_{k}$} of which {\tt m$_{k}$} is at the
top. Each of these methods may have one or more {\tt try} statements
(maybe nested) in such a way that \\
\verb@     @{\tt c$_{1}$}, {\tt c$_{2}$}, ... {\tt c$_{n}$}\\
is the stack of objects of the {\tt try} statements. Each time the
execution reaches a {\tt try} statement its object is pushed into the
stack. When the {\tt end} of the {\tt try} statement is reached the
object is removed from the stack.

When {\tt exception} receives a message\\
\verb@     exception.throw( exceptionObject )@\\
the run-time system looks in the {\tt try} stack for an
object that has a method\\
\verb@     proc throw( p : E )@\\
in such a way the type of object {\tt exceptionObject} is a subtype of
{\tt E}.\footnote{Types of objects are defined similarly to types of
classes. See Chapter~\ref{classobjectssub}.} Suppose the object found
is {\tt c$_{i}$}, {\it 1
$\le$ i $\le$ n}. Then all methods corresponding to the {\tt try} statements
of the objects {\tt c$_{i+1}$}, ... {\tt c$_{n}$} are terminated and
the statement\\
\verb@     @{\tt c$_{i}$.throw(exceptionObject)}\\
is executed. After this the execution continues after the {\tt end} of
the {\tt try} statement of {\tt c$_{i}$}, unless the above statement
also raises an exception.

Before discussing further details of exceptions, let us complete the
initial example. Class\\ {\tt CatchFileException} defines {\tt throw} methods for
handling exceptions {\tt FileOpenException},\\ {\tt FileReadException}, and {\tt
FileCloseException}. It is given below together with class {\tt
FileOpenException} and a skeleton of method {\tt open} of {\tt File}.

\begin{verbatim}
class CatchFileException subclassOf CatchUnchecked
  proc init()
    begin
    end

  public:

    proc throw( exc : FileOpenException )
      begin
      Out.writeln( "Cannot open file ", exc.getFileName() );
      Runtime.exit(1);
      end

    proc throw( exc : FileReadException )
      begin
      Out.writeln( "Error in reading the file ", exc.getFileName() );
      Runtime.exit(1);
      end

    proc throw( exc : FileCloseException )
      begin
      Out.writeln( "Error in closing the file ", exc.getFileName() );
      Runtime.exit(1);
      end
end


class File
  public:
      // for a while ignore the exception parameter
    proc open( fileName : String )
             ( exception : CatchFileException )
      begin
      ...
      exception.throw( FileOpenException.new(fileName) );
      ...
      end
    ...
end

class FileOpenException subclassOf Exception
  public:
    proc init( pfileName : String )
      begin
      fileName = pfileName;
      end
    proc getFileName() : String
      begin
      return fileName;
      end
  private:
    var fileName : String;
end
\end{verbatim}
The initial example is
\begin{verbatim}
var f : File;
var text : array(char)[];
var errorCatch : CatchFileError;

errorCatch = CatchFileError.new();
try(errorCatch)
  f = File.new();
  f.open("mytext");
    // the same as text = array(char)[].new(f.getSize())
  text#init( f.getSize() );
  f.read(text);
  f.close();
end
\end{verbatim}
Suppose there is an error in method {\tt open} of {\tt File} called by
statement\\
\verb@     f.open("mytext");@\\
and an exception is thrown by statement\\
\verb@     exception.throw( FileOpenException.new(fileName) );@\\
The execution of method {\tt File::open} is terminated and the
run-time system looks for a {\tt throw} method in object {\tt
errorCatch} since the call to {\tt open} is inside a ``{\tt
try(errorCatch) ... end}'' block. Variable {\tt errorCatch} refers to
an object of class {\tt CatchFileError} that does have a method\\
\verb@     proc throw( exc : FileOpenException )@\\
Then the statement\\
\verb@     errorCatch.throw( exceptionObject ) @\\
is executed in which {\tt exceptionObject} refers to the object ``{\tt
FileOpenException.new(fileName)}'' created in the message send\\
\verb@     exception.throw( FileOpenException.new(fileName) );@\\
in {\tt File::open}.

Although exceptions may look complex, they in fact are not if we take
another point of view: sending a message to object {\tt exception} is
to send the same message to an object of a {\tt try} statement. In
fact, the first object in a stack of try objects that can handle the
message.

\section{Hierarchy of Exceptions}

Now it is time to fill some gaps in the definition of
exceptions. First, every exception class like {\tt FileOpenException}
must be  {\it subtype} (not necessarily subclass) of class {\tt
Exception}. This class defines only method {\tt toString} which is
supposed to return a string describing the exception.

There  are some predefined exception classes in Green.  They are
arranged next in such a way a subclass is two columns ahead of its
superclass of the previous line. Subclasses of the same class are in
the same column. Then, {\tt StackOverflowException} and
\verb@IllegalArrayIndexException@ are subclasses of
\verb@UncheckedException@.
\begin{verbatim}
Exception
  TypeErrorException
  WrongParametersException
  NotFoundException
  PackedException
  TooManyDimensionsException

  MetaException
    ClassNotInAllowedSetException
    NoShellException
    NoExtensionException

  UncheckedException
    StackOverflowException
    IllegalArrayIndexException
    OutOfMemoryException
    InternalErrorException
    MessageSendToNilException

    NoReflectiveInfoException
      NoReflectiveBodyInfoException
      NoReflectiveCallInfoException

    ArithmeticException
      DivisionByZeroException
      RealOverflowException
      RealUnderflowException

    AssertionException
      AssertionAfterException
      AssertionBeforeException
      AssertionCastCharException
      AssertionCastBooleanException
      AssertionCastByteException
      AssertionCastIntegerException
      AssertionCastLongException
      AssertionCastRealException
      AssertionCastDoubleException
\end{verbatim}
User exception classes are usually direct subclasses of {\tt
Exception}.  Subclasses of {\tt UncheckedException} are used
by the compiler to signal some errors that programmers usually do not
care about.

\section{Typing Exceptions}

    Each method must declare the exceptions it can throw as the type
of the parameter {\tt exception} as in method {\tt Fill::open}:\\
\verb@     proc open( fileName : String )@\\
\verb@              ( exception : CatchFileException )@\\
Here {\tt CatchFileException} can be replaced by the more specific
class {\tt CatchFileOpenException}. To each direct or indirect
subclass of {\tt Exception}, say {\tt AnError}, the Green compiler will
automatically create  classes
\begin{verbatim}
class CatchAnError subclassOf CatchUncheckedException
  public:
    proc init()
      begin
      end

    proc throw( exc : AnError )
      begin
      end
end

class HCatchAnError subclassOf CatchUncheckedException
  public:
    proc init()
      begin
      end
    proc throw( exc : AnError )
      begin
      Out.writeln("Exception AnError not caught");
      Runtime.exit(1);
      end
end
\end{verbatim}
that can be readily used as the {\tt exception} type. The last class is a
{\it hard} catch exception class. If an object of this class
catches an exception it terminates the program.

      Class {\tt CatchUncheckedException} is a predefined class with a
{\tt throw} method  for each subclass of {\tt
UncheckedException}. Each of these methods just throws the exception
again:
\begin{verbatim}
class CatchUncheckedException subclassOf Catch
   public:
     proc throw( exc : StackOverflowException )
               ( exception : CatchUncheckedException )
       begin
       exception.throw(exc);
       end
     ...
end
\end{verbatim}
Class {\tt HCatchUncheckedException}  inherits from class {\tt
Catch} (Figure~\ref{catch}) and defines a method {\tt throw} for each subclass of {\tt
UncheckedException}.  Each of these methods prints an explaining message
in the default output device and exits the program. Class {\tt CatchUncheckedException}
 must be supertype of every catch
class or catch class object. A catch class is a class whose objects
are used as arguments to {\tt try} commands. A catch class object is a
class object used as argument to a {\tt try} command. Objects of class
{\tt Catch} keep a reference to the exception object that was thrown
in a statement\\
\verb@     exception.throw( exceptionObject );@\\
It also keeps a reference to the class of the object as considered by
the run-time system, which may be different from\\
\verb@     exceptionObject.getClassObject()@\\
See details ahead.

\begin{figure}
\begin{verbatim}
class Catch
  proc init()  begin  initialize();  end
  public:

    proc initialize()
      begin   exceptionObject = nil;   classException = nil;
      wasFixed = false;
      end
    proc set( p_exceptionObject : Exception;
              p_classException : AnyClassObject )
      begin
      exceptionObject = p_exceptionObject;
      classException = p_classException;
      end
    proc getException() : Exception
        // returns the exception object thrown by exception.throw(obj)
      begin
      return exceptionObject;
      end
    proc getClassException() : AnyClassObject
        // returns the class of the object exception
      begin
      return classException;
      end
    proc wasThrown() :  boolean
      begin
      return exceptionObject <> nil;
      end
    proc fixed() : boolean
      begin
      if wasThrown() and wasFixed  then wasFixed = false;  return true;
      else return false;  endif
      end
    proc setFixed( p_wasFixed : boolean )
      begin
      wasFixed = p_wasFixed;
      end
  private :
    var exceptionObject : Exception;  classException :  AnyClassObject;
        wasFixed : boolean;
end
\end{verbatim}
\caption{Class {\tt Catch}}
\label{catch}
\end{figure}

The run-time system may throw exceptions whose classes are subclasses
of {\tt UncheckedException}. For example, when compiling the command\\
\verb@     x = v[i];@\\
the compiler will generate code to test if {\tt i} is within the legal
array bounds. This test will belong to the program run-time system and
may throw exception {\tt IllegalArrayIndexException}. All the
exceptions the run-time system may throw belong to subclasses of {\tt
UncheckedException} and need not to be declared as part of the type of
the {\tt exception} method parameter. However, if the programmer wants
to throw this exception herself, she must declare it:
\begin{verbatim}
proc aMethod()
            ( exception : CatchIllegalArrayIndexException )
  begin
  ...

  exception.throw( IllegalArrayIndexException.new() );
  ..
  end
\end{verbatim}
The program can catch exceptions of subclasses of {\tt
UncheckedException} even when they are not declared, which is the
normal case. Example:
\begin{verbatim}
class FigureSet
  public:
    proc add( f : Figure )
      /* add a figure to the set. Since there may not be sufficient
         memory, exception OutOfMemoryException (not declared) may be
         thrown. */
    ...
end

...
var x : Square;
var fig : FigureSet;
...

try( CatchOutOfMemory.new() )
  fig.add(x);
end
\end{verbatim}
In the previous example, the program need not to use the {\tt
try}-{\tt end} statement since {\tt add} may only throw unchecked
exceptions.

Class object {\tt CatchAll} is subtype of {\tt Catch} and has a
method\\
\verb@     proc throw( exc : Exception )@\\
Then {\tt CatchAll} can be used to catch all exceptions:
\begin{verbatim}
try(CatchAll)
  f = File.new("article.doc");
  ...
end
if CatchAll.wasThrown()
then
  Out.writeln("There was an error ...");
endif
\end{verbatim}

Method {\tt wasThrown} returns {\tt true} if an exception was thrown
inside the {\tt try}-{\tt end} command.

Class object {\tt HCatchAll} is similar to {\tt CatchAll} but it
finishes the program when an exception is thrown. A message is printed
using object {\tt Out}.

The execution of a program begins in a compiler-created object called
{\tt MainProgram} that catchs all unchecked exceptions that may be
raised by the user program. This object is defined as
\begin{verbatim}
object MainProgram
  public:
    proc run( theArgs : String )
      begin
      var args : array(String)[];
        // parse theArgs resulting in args
      ...
      var catchUnchecked : HCatchUncheckedException;
      catchUnchecked#init();
      try(catchUnchecked)
        UserClass.run(args);
      end
      end
end
\end{verbatim}
{\tt UserClass}  is the class object in which the program must begin
its execution.

     The type of object {\tt exception} declared in a method {\tt m}
must have {\tt throw} methods with exceptions the {\tt m} method can
throw. Then, if method
\begin{verbatim}
  proc m()
        (exception : CatchX)
    begin
    ...
    end
\end{verbatim}

can throw exceptions {\tt FileOpenException} and {\tt ParseException},
class {\tt CatchX} must have at least methods\\
\verb@     proc throw( exc : FileOpenException )@\\
\verb@     proc throw( exc : ParseException )@\\

The class
\begin{verbatim}
class CatchFruitException subclassOf CatchUncheckedException
  public:
    proc init()
      begin
      end
    proc throw( exc : FruitException )
      begin
      ...
      end
    proc throw( exc : BananaException )
      begin
      ...
      end
end
\end{verbatim}
defines two methods {\tt throw} in which one of the parameter types
({\tt BananaException}) is subtype of the other ({\tt
FruitException}). This introduces a kind of redundancy in the class
since a {\tt BananaException} object can be passed as a parameter not
only to method ``\verb@proc throw(exc : BananaException)@'' but also
to ``\verb@proc throw(exc : FruitException)@''. Let us see an example
explaining how the compiler manages this.

In a code like
\begin{verbatim}
catchFruitException = CatchFruitException.new();
try(catchFruitException)
  makeJuice();
end
\end{verbatim}
in which {\tt makeJuice} executes the message send\\
\verb@     exception.throw( BananaException.new() )@\\
the run-time system will look for a {\tt throw} method with a parameter type
that is supertype of {\tt BananaException}. This search is made in the
stack of objects of {\tt try} statements. First the run-time system pops the
first object off the stack and starts a linear search in the object
class for a method \\
\verb@     proc throw( exc : E )@\\
such that {\tt E} is supertype of the poped object class.\footnote{If
this object is a class object, the test is to discover if {\tt E} is
supertype of the class object type.} This search is made in the order
given by the declarations of methods in the text of the catch
class. If no {\tt throw}
method of this first object is adequate, the next object is poped off
the stack of {\tt try} objects and the search continues.

This is the only Green feature in which the order of method
declaration does matter. Notice that catch objects can be class
objects --- the only requirement for one catch object is that it must
be subtype of predefined class {\tt CatchUncheckedException}.

Suppose class {\tt CatchA} declares a method\\
\verb@     throw( exc : BananaException )@\\
and class {\tt CatchB}, which inherits from {\tt CatchA}, defines a
method\\
\verb@     throw( exc : FruitException )@\\
If an object of {\tt CatchB} is used in a try block, the search for a
handler is made first in {\tt CatchB} and then in {\tt CatchA}. So, an
exception signalling\\
\verb@     try(aCatchB)@\\
\verb@       exception.throw( BananaException.new() );@\\
\verb@     end@\\
will cause a search in {\tt CatchB} and method {\tt
throw(FruitException)} will be used even though method {\tt
throw(BananaException)} of the superclass would be more specific.

In class {\tt CatchFruitException}, method {\tt throw} with a {\tt
FruitException} parameter appears before method with a parameter type
{\tt BananaException}. Therefore the call\\
\verb@     exception.throw( BananaException.new() )@\\
in {\tt makeJuice()} will execute method\\
\verb@     CatchFruitException::throw( exc : FruitException )@\\

Since the correct thing should be to call\\
\verb@     CatchFruitException::throw( exc : BananaException )@\\
this method should be put before \verb@throw(exc : FruitException)@ in
its class declaration.

Every catch class or catch class object should be subtype of class
{\tt CatchUncheckedException}. A catch object will keep a
reference to the class of the exception object that was thrown. This
reference can be got by method {\tt getClassException}. Using this feature our
fruit example could be rewritten as
\begin{verbatim}
catch = CatchFruitException.new();
try(catch)
  makeJuice();
end
case catch.getClassException() of
  FruitException :
    Out.writeln("Fruit rotted");
  BananaException :
    Out.writeln("Cannot make a juice of bananas");
end
\end{verbatim}

    We supposed in this example that the {\tt throw} methods of {\tt
CatchFruitException} had empty bodies. The real exception treatment
is made in the case statement after {\tt try}-{\tt end}. If {\tt
makeJuice} raises an exception {\tt OrangeException}, subclass of {\tt
FruitException}, method \\
\verb@     proc throw( exc : FruitException )@\\
of object {\tt catch} will be executed since the catch class, {\tt
CatchFruitException}, does not have a method\\
\verb@     proc throw( exc : OrangeException )@\\
and {\tt OrangeException} is a subtype of {\tt FruitException}. The
the call\\
\verb@     catch.getClassException()@\\
of the case statement will in fact return class {\tt FruitException}
instead of {\tt OrangeException}. However, \\
\verb@ catch.getException()@\\
would return the {\tt OrangeException} object thrown by {\tt makeJuice} and\\
\verb@     catch.getException().getClassObject()@\\
would return class {\tt OrangeException}.

The above syntax mimics the exception system of Java and C++. This
example can be written in Java as
\begin{verbatim}
try {
  makeJuice();
}
catch ( BananaException exc )
{
  System.Out.writeln("Fruit rotted");
}
catch ( FruitException exc )
{
  System.Out.writeln("Cannot make a juice of bananas");
}
\end{verbatim}
Note the order of the catch clauses is important here. The exception
object can be accessed in this Java code using parameter {\tt exc} of
the catch clauses. In the Green code the exception object has to be
retrieved  by the call ``{\tt catch.getException()}''.

In the previous Java example, the compiler will issue an error message
if the {\tt try} block may throw an exception not presenting in the
catch clauses and that was not declared in the method header. We hope
the Green compiler will also issue an error in the same situation when
simulating the {\tt catch} clauses of Java with the case statement as
shown in the last Green example. That is, the compiler should issue a
warning if {\tt CatchJuiceException} declared method\\
\verb@     proc throw( exc : ReadFileException )@\\
since there is no correspondent {\tt case} label to treat this case.

We hope every class object of a catch class will have a {\tt get}
method for returning an object of the catch class. Then only one
object of the catch class need to be allocated in the whole
program. The following code illustrates this mechanism.
\begin{verbatim}
object CatchFileOpenException
  public:
    proc init()
      begin
      aCatch = CatchFileOpenException.new();
      end

    proc get() : CatchFileOpenException
      begin
      aCatch.initialize();
      return aCatch;
      end
  private:
    var aCatch : CatchFileOpenException;
end

class CatchFileOpenException subclassOf CatchUncheckedException
  public:
    proc init()
      begin
      end

    proc throw( exc : FileOpenException )
      begin
      ...
      end
end
\end{verbatim}

\section{Fixing Errors with Catch Objects}

After an exception is thrown, it will be catch by a catch object that
appears in a {\tt try}-{\tt end} command. A method {\tt throw} of the
catch object is called passing the exception object as parameter. The
{\tt throw} method may try to fix the error using information from the
exception object. To illustrate that, we will use the following
classes
\begin{verbatim}
class WriteErrorException subclassOf Exception
  public:
    proc init( pf : File )
      begin
      f = pf;
      end
    proc getFile() : File
      begin
      return f;
      end
 private:
   var f : File;
end



class CatchWriteErrorException subclassOf CatchUncheckedException
  public:
    proc init()
      begin
      end
    proc throw( exc : WriteErrorException )
        var success : boolean;
            f       : File;
      begin
      f = exc.getFile();

        /* tries to fix the error by changing the file name, writing to
           another drive or directory, changing the attributes of the
           file. That means putting the state of the object exc in a
           state that will not produce this error anymore */
      ...
        // if successful, set fixed to true
      setFixed(success);
      end
end



class File
  public:
    ...
    proc save()
             ( exception : CatchWriteErrorException )
      begin
      ...
      if error
      then
        exception.throw( WriteErrorException.new(self) );
      endif
      end
  private:
    ...
end


class Test
  public:
    ...
    proc saveAll()
      begin
      ...
      catch = CatchWriteErrorException.new();
      repeat
        try(catch)
          f.save();
          ...
        end
      until catch.fixed();
      end  // saveAll
end
\end{verbatim}
Method  {\tt saveAll} saves files to disk through statement ``{\tt
f.save()}'' which may throw an exception {\tt
WriteErrorException}. The thrown exception keeps a reference to the
file that could not be saved to disk --- see {\tt File::save}. If the
exception is throw in the above example, method\\
\verb@     CatchWriteErrorException::throw@\\
will be executed. This method tries to fix the error by asking the
user directions on what to do. Method {\tt throw} may:
\bi
\item change the file name. The original name may be illegal;
\item write the file to another drive/directory;
\item change the file attributes. The file may be read only.
\ei

In general, a method {\tt throw} should try to put the object ({\tt f}
in this case) in an error-free state. The object whose defective state
caused the exception should be passed as a parameter to the exception
object. Then method {\tt throw} can access this object through the
exception object.

After {\tt CatchWriteErrorException::throw} is executed, the execution
continues after the {\tt try}-{\tt end} command. Then ``{\tt
catch.fixed()}'' is executed. Method {\tt fixed} returns true when an
exception was thrown and the error was fixed thus enabling the {\tt
try}-{\tt end} command to be executed again. The {\tt throw} method of
the catch object that fixes the error should call method {\tt
setFixed} of the catch object.

\section{Comparison with Other Exception Systems}

  By comparing the example in Java and Green we conclude the Green
exception system is at least equivalent to the Java system.
  In  fact, the Green system is much more. It casts catch clauses in
{\tt throw} methods of {\tt Catch} classes making them reusable. Green
shapes exceptions in a more object-oriented fashion by grouping what
would be Java catch clauses in Green {\tt Catch} classes. Not only
catch classes can be reused but also catch objects that can be passed
as parameters, stored in variables, and so on. One can imagine a
factory class \cite{gamma94} which returns catch objects appropriate
to the current environment. For example, in a command-line environment
the factory class would return catch objects that would print error
messages in the standard output. If a graphical environment is used,
the messages could be shown in windows and so on.

Catch classes can be subclassed and {\tt throw} methods overridden thus
providing code reuse of error treatment.

Java requires a method to declare the exceptions it may
throw\footnote{Some exceptions in Java need not to be declared ---
they are unchecked.} through a special syntax:
\begin{verbatim}
void makeJuice()
     throws FruitException, BananaException
{
   ...
   throw new FruitException();
}
\end{verbatim}
Green requires the declaration of exceptions in an object-oriented
way:
\begin{verbatim}
proc  makeJuice()
               ( exception : CatchFruitException )
begin
...
exception.throw( FruitException.new() );
...
end
\end{verbatim}
{\tt exception} is an implicit parameter of {\tt makeJuice}. This means
the thrown of an exception follows the  rules of message
sends. And, since one puts a class like {\tt CatchFruitException} as
the parameter type, it is not necessary to list all exception classes
in the method header. This may be cumbersome to do if a lot of
different kinds of exceptions can be thrown by a lot of methods,
although in general the number of exception classes a method uses is
small.

\section{Exception and Meta-Level Programming}

The {\tt exception} parameter in Green is a meta-level parameter
automatically managed by the compiler. We can imagine the program
being executed in a plane of a three-dimensional space called
``application plane''. There is a ``meta-level exception'' plane
parallel to the application plane. The execution in both planes are
synchronized. While at the application
plane there is a stack of called methods, in the meta-level exception
plane there is a stack of catch objects. The application plane manages
parameter passing to methods just as ordered by the program. These
parameters correspond to the first parameter list after the method
name, as {\tt elem} in
\begin{verbatim}
proc search( elem : integer )    // first parameter list
           ( exception : CatchNotFoundException ) : integer  // second list
  var i : integer;
begin
...
  // didn't find anything
exception.throw( NotFoundException.new( elem, i ) );
...
end
\end{verbatim}
The second parameter list,\\
\verb@     ( exception : CatchNotFoundException )@\\
corresponds to the meta level exception plane which is automatically
managed by the compiler.

The {\tt exception} parameter can only be used as a message
receiver. It cannot be used as a message-send real parameter or in the
left or right-hand side of an assignment. This restriction could be
lifted if we considered {\tt exception} has class {\tt CatchException}
that inherits from {\tt Catch} and defines the method\\
\verb@     proc throw( exc : Exception )@\\
However, this would make things confusing because the declared type of
{\tt exception} may be different (and generally is) from {\tt
CatchException}. The {\tt exception} type would be the declared type
in a message send and {\tt CatchException} anywhere else.

Whenever the program execution reaches a {\tt try(catch)} statement,
the catch object is pushed into the stack of catch objects. After the
{\tt try} statement finishes its execution, the object is removed from
the stack.

If a method does not declare an exception parameter, the compiler
introduces one whose type is {\tt CatchUncheckedException}. Then a
method \\
\verb@     proc set( i : integer )@\\
in fact is\\
\verb@     proc set( i : integer )@\\
\verb@             ( exception : CatchUncheckedException )@\\
Through this {\tt exception} parameter the compiler can throw
exceptions like {\tt OutOfMemoryException} or {\tt
MessageSendToNilException}, which  are unchecked. Since programmers
normally do not throw unchecked exceptions, they do not declare an
exception parameter in methods that can only throw this kind of
exceptions.

Object {\tt Runtime}, discussed in Chapter~\ref{sco} and
Appendix~\ref{irl}, allows one to inspect the catch object stack. At
run time the program can discover what exceptions can be caught and it
can even attach a shell to a catch object of the catch object
stack. See Chapter~\ref{shells} for definition of shells.

\section{Static Type Correctness of Exceptions}

Class {\tt CatchFruitException} can catch exceptions of classes {\tt
FruitException} and {\tt BananaException}. Assume a class {\tt
CatchBurnedFoodException} can catch exceptions of class {\tt
BurnedFoodException} and method {\tt heat} can {\tt throw} exceptions
of classes {\tt FruitException}, {\tt BananaException}, and\\ {\tt
BurnedFoodException}. Then in a method
\begin{verbatim}
proc prepareFood( f : Food )
                ( exception : CatchFruitException )
     var catch : CatchBurnedFoodException;
  begin
  catch = CatchBurnedFoodException.new();
  try(catch)
    heat(f);
    split(f);
  end
  end
\end{verbatim}
There will not be any type error relating to exceptions. Any exception
raise by {\tt heat} will be catch by the {\tt try} statement or by the
method that called {\tt prepareFood}.\footnote{Or in fact another
method that indirectly called {\tt prepareFood}.}

That means the type of {\tt exception} inside the {\tt try-end} statement is
enhanced by the type of {\tt CatchBurnedFoodException}. Inside this
{\tt try-end}, the type of {\tt exception} is:
\begin{verbatim}
proc throw( exc : FruitException )        // from the exception object
proc throw( exc : BananaException )       // from the exception object
proc throw( exc : BurnedFoodException )   // from the catch object
\end{verbatim}
What happened is that the {\tt exception} object passed as a parameter to
{\tt prepareFood} was the top element of the stack of catch
objects. The {\tt try} statement pushed object {\tt catch} into this
stack enhancing the type of {\tt exception}. This occur because inside
the {\tt try-end} statement, a message sent to {\tt exception} is searched
first in the object of the top of the catch stack. If the method is not
found there the search continues in the stack from top to bottom
making the run-time type of {\tt exception} to be the union of the
types of all objects in the catch stack. However, the compile-time
type is more restricted because an {\tt exception} parameter  assumes
the type it is declared with, at least outside any {\tt try-end}
statement.

\section{Exceptions and Enhancing of Objects at Runtime}

The casting of catch clauses in an object-oriented form reveals an
interesting aspect of exceptions: an {\tt exception} object has a dynamic
type which depends on the types of objects in an object stack. Since
the object stack grows and shrinks at run time, so does the {\tt
exception} type. When the program execution reaches the {\tt try}
command, a catch object is pushed into the stack thus adding method
signatures to the {\tt exception} type. When the execution of a {\tt
try-end} command ends, the catch object is poped off the stack thus
making the {\tt exception} type return to its previous value. This
dynamic type modification would cause run-time type errors if it were
not restricted to a block defined at compile time, which is the {\tt
try-end} block.

Exceptions cannot be cast in a pure object-oriented form for two
reasons:
\bi
\item the dynamic nature of the {\tt exception} type;
\item the urgence an exception should be treated: the first catch
object that can handle the thrown exception object is used.
\ei

This last item will become clear after we try to solve the former item
by using shells (Green  metaobjects --- see Chapter~\ref{shells}).
Suppose Green supports a modified kind of shells in which a shell
class
\begin{verbatim}
shell class Border(Window)
  public:
    proc init( c : integer )
      // methods not defined in Window
    proc setBorderColor( c : integer )
      ...
    proc getBorderColor() : integer
    ...
end
\end{verbatim}
can declare in its public section methods not defined in class {\tt
Window}. When a shell of {\tt Border} is attached to a {\tt Window}
object, the object can be cast to class {\tt BorderedWindow}, which is
a subclass of {\tt Window} and defines the methods of shell class {\tt
Border}. See the example.
\begin{verbatim}
var window : Window;

var borderedWindow : BorderedWindow;
...
 Meta.attachShell( window, Border.new( Color.blue ) );

borderedWindow = BorderedWindow.cast(window);
Out.writeln( borderedWindow.getBorderColor() );
...
\end{verbatim}
{\tt BorderedWindow} has no inheritance relationship with shell class {\tt Border}.

The message send ``{\tt borderedWindow.getBorderColor()}'' will call
method {\tt getBorderColor} of shell class {\tt Border}: {\tt
borderedWindow} points to a {\tt Window} object whose type was enhanced by
the shell attachment. This dynamic type modification can easily result
in type errors:\\
\verb@     Meta.removeShell(window);@\\
\verb@     borderedWindow.setBorderColor(Color.green);@\\
The last statement will result in a run-time type error since the
object pointed by {\tt window} and {\tt borderedWindow} does not have
a {\tt setBorderColor} anymore.

This kind of error does not occur if only one variable refers to the
object and its type is enhanced only in a specific region of the code
defined at compile time
such as the block of a {\tt try-end} command. Now we are going to
introduce a new syntax in which {\tt exception} objects will have
their types enhanced by catch classes. This syntax is not legal in
Green --- it is only used to exemplify the ideas. The code
\begin{verbatim}
proc prepare( f : Food )
            ( exception : CatchReadFileException )
  var catch : CatchBananaException;
begin
catch = CatchBananaException.new();
try(catch)
  prepareFood(food);
end
end
\end{verbatim}
is written as
\begin{verbatim}
proc prepare( f : Food )
            ( exception : CatchReadFileException )
  begin
  try(CatchBananaException.new() )
    prepareFood(food);
  end
  end
\end{verbatim}
and translated by the compiler into
\begin{verbatim}
proc prepare( f : Food )
            ( exception : CatchReadFileException )
  begin
  Meta.attachShell( exception, CatchBananaException_ReadFileException.new() );
    // new scope --- illegal in Green but legal in this example
    begin
    var exception : CatchBananaException_ReadFileException; // 1
    exception = CatchBananaException_ReadFileException.cast( prepare::exception );
    prepareFood(food);
    end
  end // prepare
\end{verbatim}
There are some observations to be made here. First,
\verb@CatchBananaException_ReadFileException@ is the compiler created
class
\begin{verbatim}
shell class CatchBananaException_ReadFileException(CatchReadFileException)
  public:
     // all the methods of CatchBananaException
end
\end{verbatim}
Shells of this class can be attached to {\tt CatchReadFileException}
objects as parameter {\tt exception}. The result is an object that has
methods of classes {\tt CatchBananaException} and {\tt
CatchReadFileException}. See the ``{\tt attachShell}'' message to {\tt
Meta} in {\tt prepare}. But that does not mean a message send \\
\verb@     exception.throw( BananaException.new() )@\\
is legal since the {\tt exception} type is {\tt
CatchReadFileException} which does not have the method\\
\verb@     proc throw( exc : BananaException )@\\
It is necessary to {\it cast} {\tt exception} into a type that does
have this method, which is made in the line following line \verb@// 1@.

We are assuming the {\tt begin-end} block inside the {\tt begin-end}
block of method {\tt prepare} introduces a new scope, a rule not valid
in Green. The {\tt exception} variable declared in this block (line
\verb@// 1@) is initiated with the {\tt exception} variable of the
outer scope.  That is, ``\verb@prepare::exception@'' means the
parameter {\tt exception} of {\tt prepare}. Again, this is not valid
Green syntax.

{\tt prepareFood} may have been declared as
\begin{verbatim}
proc prepareFood( f : Food )
                ( exception : CatchBananaReadException )
  begin
  ...
  exception.throw( BananaException.new() );
  ...
  end
\end{verbatim}
in which {\tt CatchBananaReadException} has all methods of {\tt
CatchBananaException} and\\ {\tt CatchReadFileException}.

The call \\
\verb@     prepareFood(food);@\\
in method {\tt prepare} is legal because it is translated to \\
\verb@     prepareFood(food)@\\
\verb@                (exception) // passes exception as a meta-level parameter @\\
and {\tt exception} has the same type as {\tt
CatchBananaReadException}. This last statement is true because of the cast\\
\verb@     exception = CatchBananaException_ReadFileException.cast(prepare::exception );@\\
Note this cast is not legal in Green because shell classes do not
exist at run time (they are not real classes) and therefore they do
not support method {\tt cast}.

The statement\\
\verb@     exception.throw( BananaException.new() );@\\
of {\tt prepareFood} would call method\\
\verb@     proc throw( exc : BananaException )@\\
of object {\tt exception} since this is a normal message send.

Now suppose {\tt prepare} was modified to
\begin{verbatim}
proc prepare( f : Food )
            ( exception : CatchReadFileException )
  begin
  try( CatchBananaException.new() )
    try( CatchFruitException.new() )
      prepareFood(f);
    end
  end
  end
\end{verbatim}
Using the implementation of exceptions just described, based in shell
attachements, the statement\\
\verb@     exception.throw( BananaException.new() )@\\
in {\tt prepareFood} continues to call the same method of object {\tt
exception} as before,\\
\verb@     proc throw( exc : BananaException )@\\
which means the inner {\tt try-end} command of {\tt prepare} was
skipped and the control transferred to the outer {\tt try-end}. That
is not the best alternative since the inner {\tt try-end} command can also
treat the exception and fewer {\tt try-end} commands should be poped
off the stack. In general one wants to treat an error as soon as
possible to avoid the premature ending of methods in the call stack.

As we have seen, this does not happen if shells are used to implement
exceptions because the following situation may occur:
\bi
\item the topmost {\tt try} can treat exception {\tt BananaException};
\item a {\tt BananaException} is thrown and treated by the bottom-most
{\tt try} command thus making a great number of methods finish their execution.
\ei

That justifies the claim we have make in the beginning of this section:
exceptions cannot be cast in a pure object-oriented form  because an
exception should be treated as soon as possible. when we cast
exceptions in a pure object-oriented form, all {\tt throw} methods are
put in a single class losing therefore the relative order of the {\tt
try} statements that furnishes these methods. Then a {\tt throw}
method may be invoked when another nearest could be called. Note we
assumed a message send \\
\verb@     exception.throw(exc)@\\
causes a search at run time for the most specific method that accepts
parameter {\tt exc}. This is not valid for overloaded methods in
Green.
Note that although we
have used shells to prove exceptions cannot be made completely object-oriented, there would be no difference if
we have used any other mechanism to enhance the type of an object. For
example, we could have created a class {\tt CatchBananaReadTwoException} in which some
{\tt throw} methods were forwarded to an instance variable:
\begin{verbatim}
class CatchBananaReadTwoException subclassOf CatchReadFileException
  public:
    proc init( f : CatchReadFileException )
      begin
      self.f = f;
      end
    proc throw( exc : ReadFileException )
      begin
      f.throw(exc);
      end
    proc throw( exc : BananaException )
      begin
        // treats the exception
      ...
      end
  private:
    var f : CatchReadFileException;
end
\end{verbatim}
Method {\tt prepare} would become:
\begin{verbatim}
proc prepare( f : Food )
            ( exception : CatchReadFileException )
    var exception : CatchReadFileTwoException;
  begin
  exception#init(prepare::exception);
  prepareFood(food);
  end
\end{verbatim}

\section{Assertions and Exceptions}

When an assertion of any method fails, an exception of class {\tt
AssertionBeforeException} or {\tt AssertionAfterException} is
thrown. Each of these classes has a constructor \\
\verb@     proc init( mi : MethodInfo )@\\
{\tt MethodInfo} is a class of the introspective class library that
gives information about the method such as its name, class, and so on.

\section{Subtyping and Exceptions}

It is time to define subtyping rules that include exceptions. The type
of a class is the set of its method signatures as before. But now a
method signature includes the type of parameter {\tt
exception}. Therefore, the signature of method\\
\verb@     proc get( i : integer )@\\
\verb@             ( exception : CatchIllegalIndexException ) : Any@\\
is\\
\verb@     get(integer)(CatchIllegalIndexException) : Any@\\

A type {\it S} = {\tt \{n$_{1}$, n$_{2}$, \ldots n$_{p}$\}}
is equal to type
\verb@@{\it T} = {\tt \{m$_{1}$, m$_{2}$, \ldots m$_{q}$\}} ({\it S} $=$ {\it T})
if $p = q$ and {\tt n$_{i}$} = {\tt m$_{i}$}, $1 \leq i \leq p$.
The relation  $=$ for methods is defined
as follows.

Let \\
\verb@     @{\tt n({\it T}$'_{1}$, {\it T}$'_{2}$, \ldots {\it
T}$'_{k}$)({\it E}$'$) : {\it U}$'_{1}$}\\
\verb@     @{\tt m({\it T}$_{1}$, {\it T}$_{2}$, \ldots {\it
T}$_{t}$)({\it E}) : {\it U}$_{1}$}\\
be the signatures of two methods. We say that {\tt n} $=$
{\tt m} if
\bi
\item {\tt m}  and {\tt n} have the same name.
\item $t = k$.
\item {\tt {\it T}$_{i}$ $=$ {\it T}$'_{i}$}, $1 \leq i \leq t$.
\item {\tt {\it U}$'$ $=$ {\it U}}.
\item {\it E$'$} $=$ {\it E}.
\ei
The definition of subtype remains the same. That is, {\it S} is
subtype of {\it T} is {\it T $\subset$ S}.

Since a subclass is a subtype, a superclass method overridden in a
subclass should keep its signature which includes keeping the type of
parameter {\tt exception}. This could be different: it could be legal
in Green to declare the type of {\tt exception} in an overridden
subclass method to be a supertype of the type of {\tt exception} in
the superclass method. That is, if\\
\verb@     proc get(integer)(CatchJuiceException) : Any@\\
is the superclass method it could be legal in Green to override this
method to\\
\verb@     proc get(integer)(CatchFruitException) : Any@\\
in a subclass in which {\tt CatchFruitException} is a supertype of
{\tt CatchJuiceException}. The sole reason this is illegal in Green is
that it would make the language more complex. There is no logical
error in this feature as explained in the following paragraphs.

Suppose class {\tt JuiceMachine} defines a method\\
\verb@     proc makeJuice()@\\
\verb@                   ( exception : CatchJuiceException )@\\
which is overridden in a subclass {\tt SpecialJuiceMachine} as \\
\verb@     proc makeJuice()@\\
\verb@                   ( exception : CatchFruitException )@\\
Suppose {\tt CatchJuiceException} has methods\\
\verb@     proc throw( exc : FruitException )@\\
\verb@     proc throw( exc : DryException )@\\
If {\tt CatchFruitException} has method\\
\verb@     proc throw( exc : FruitException )@\\
then the redefinition of {\tt makeJuice} interface in {\tt
SpecialJuiceMachine} would be
correct because {\tt CatchFruitException} is supertype of {\tt
CatchJuiceException}.  No run-time type error can occur because of
this redefinition of {\tt makeJuice} with a different interface.
To understand  that, let us
study the following code
\begin{verbatim}
catch = CatchJuiceException.new();
try(catch)
  var specialJuiceMachine : SpecialJuiceMachine;
    // ok, assignment of the kind "Type = subtype"
  juiceMachine = SpecialJuiceMachine.new();
  juiceMachine.makeJuice();
end
\end{verbatim}
The method {\tt makeJuice} called in this code will be \\
\verb@      SpecialJuiceMachine::makeJuice()@\\
The exceptions this method may throw will be catch by the enclosing
{\tt try-end} command because:
\bi
\item method {\tt SpecialJuiceMachine::makeJuice} can throw a subset
of exceptions that\\
 {\tt JuiceMachine::makeJuice} can since {\tt CatchFruitException},
type of {\tt exception} in\\
 {\tt SpecialJuiceMachine::makeJuice}, has
less {\tt throw} methods than {\tt CatchJuiceException}, the type of
{\tt exception} in {\tt JuiceMachine::makeJuice}. That is, {\tt
CatchFruitException} is supertype of {\tt CatchJuiceException}. As
stated before, the language
requires that from overridden subclass methods;
\item the {\tt try-end} command uses a {\tt catch} object appropriate
to handle exceptions of\\
 {\tt JuiceMachine::makeJuice}
  since there is a

statement\\
\verb@     juiceMachine.makeJuice()@\\
in which the declared type of {\tt juiceMachine} is {\tt
JuiceMachine}. Then the {\tt try} command in fact will catch
exceptions of less types than it is able to catch. Then  there cannot
be any run-time error.
\ei

\chapter{Shells and Dynamic Extensions: Metaprogramming in Green}
\label{shells}

This Chapter describes dynamic extensions and the Green metaobjects
called shells. A metaobject controls the behavior of a single
object. Any message sent to the object is packed into an object {\tt
Message} and delivered to the object metaobject. The metaobject can
then call the corresponding object method, forward the message to
another object, and so on. A dynamic extension can be attached to a
class at run time. It will replace the class methods by the
corresponding methods of the extension class.

Shells and dynamic extensions are discussed deeply by
Guimar~aes \cite{jose98}. In this report we will define these features
without further discussing  them. The definitions given here are
slightly different and more generic than that of \cite{jose98}.

\section{Shells}

\begin{figure}
\setlength{\unitlength}{1cm}
\begin{picture}(9, 2)(-5.5, 0)
\thicklines
\put( 1, 1 ){{\tt s}}
\put( 1.25, 1.1 ){\vector(1, 0){1}}
\put( 2.6, 1.1 ){\oval(0.5, 1.2)}
\put( 2.4, 0.9 ){$\cal F$}
\put( 2.95, 1.1 ){\vector(1, 0){1}}
\put( 4.4, 1.1 ){\circle{0.5}}
\put( 4.2, 0.95 ){$\cal Q$}
\end{picture}
\caption{A shell $\cal F$ attached to an object $\cal Q$ }
\label{f1}
\end{figure}

A shell is a pseudo-object with methods and instance variables that
can be attached to an object as graphically represented in
Figure~\ref{f1}.  Any message sent to the object will be
first searched in the shell and then in the object.

\begin{figure}
\begin{verbatim}
shell class Border(Window)
  proc init()
    begin
    end
  public:
    proc draw()
      begin
      self.drawBorder();
        // call object method
      super.draw();
      end
    private:
      proc drawBorder()
        ...
end
\end{verbatim}
\caption{A shell class to subtype-{\tt Window} objects}
\label{shell1}
\end{figure}

\begin{figure}
\begin{verbatim}
class Window
  proc init()
    begin
    ...
    end
  public:
    proc draw()
      ...

    ...
end
\end{verbatim}
\caption{Class {\tt Window} with at least method {\tt draw}}
\label{shell2}
\end{figure}

A shell class {\tt Border} is defined in Figure~\ref{shell1}. A shell
of class {\tt Border} can be attached to objects of {\tt Window} and
its subtypes as specified by the line\\
\verb@     shell class Border(Window)@\\
Class {\tt Window} is shown in Figure~\ref{shell2}. There may not be
any class object associate to a shell class. If this were allowed,
shells would be more complex since we could have a class object
without an associate class (shell classes are not classes !).

The set of public method interfaces of {\tt Border} must be a subset
of the set of method interfaces of {\tt Window}. If we consider a
shell class has a type, then the type of {\tt Border} should be
supertype of {\tt Window}.

\begin{figure}
\begin{verbatim}
object Meta
  public:
      // only the method interfaces are shown
    proc attachShell( any : IdentAST; exp : ExprAST )
    proc removeShell( any : IdentAST )
    proc attachExtension( aClass : ClassNo; dynExt : ExtensionClassNo )
    proc removeExtension( aClass : ClassNo )
end
\end{verbatim}
\caption{Interface of the {\tt Meta} module}
\label{meta}
\end{figure}

A shell of {\tt Border} can be attached to a {\tt Window} object {\tt window}
through the syntax\\
\verb@     Meta.attachShell( window, Border.new() )@\\
{\tt Meta} is a module and modules are compile-time objects. The {\tt
Meta} module interface is shown in
Figure~\ref{meta}. {\tt IdentAST}, {\tt ExprAST}, {\tt ClassNo},
and {\tt ExtensionClassNo} are the classes of the compiler Abstract
Syntax Tree (AST) that represent an identifier, an expression, a
class, and a dynamic extension class, respectively.

The message send of {\tt attachShell} to {\tt Meta} at compile time
will generate a run-time message send that will create a shell {\tt
Border} and attach it to object {\tt window}.

Class-{\tt Border} methods can call methods of the object the shell
is attached to by sending messages to {\tt super}. As an example, let
us study the code
\begin{verbatim}
window = Window.new("My Window", 30, 20, 120, 300);
window.draw(); // call Window::draw()
Meta.attachShell( window, Border.new() );
window.draw();
\end{verbatim}
The last statement will call {\tt Border::draw} that will draw a
border through its first statement\\ ``{\tt self.drawBorder()}'' and
execute ``{\tt super.draw()}'', which will call {\tt Window::draw}.

This is the same as if {\tt Border} inherited  class  {\tt
Window}. We can attach a {\tt Border} shell to an object of a {\tt
Window} subtype:
\begin{verbatim}
var window : Window;

  // ColorWindow is subtype (maybe subclass) of Window
window = ColorWindow.new("My Window", Color.green, 30, 20, 120, 300);
  // call ColorWindow::draw()
window.draw();
Meta.attachShell( window, Border.new() );
window.draw();
\end{verbatim}
The last message send will execute {\tt Border::draw} that will call
{\tt Border::drawBorder}\\ (``{\tt self.drawBorder()}'') and {\tt
ColorWindow::draw} (``{\tt super.draw()}'').

A shell of class {\tt Border} can be attached to objects of {\tt
Window} and its subtypes which may include classes that are not
subclasses of {\tt Window}. Therefore the calls to {\tt super} in {\tt
Border} methods can only refer to public {\tt Window} methods.

Shells can be removed by method {\tt removeShell}:\\
\verb@     Meta.removeShell(window)@\\
This command removes the last shell attached to object {\tt
window}. There may be more than one shell attached to an
object. Shells are attached and removed in a stack fashion.

If there is no shell attached to {\tt window}, this message send
throws exception {\tt NoShellException}.

The above code
may throw the following exceptions:\\
\verb@     OutOfMemoryException@\\
\verb@     ClassNotInAllowedSetException@\\
These exceptions can be caught by the class {\tt CatchMetaException}
that also catches\\
\verb@     NoShellException@\\
\verb@     NoExtensionException@\\
This last class is discussed ahead. In fact, these exception classes compose a hierarchy:
\begin{verbatim}
MetaException
  ClassNotInAllowedSetException
  NoShellException
  NoExtensionException
\end{verbatim}

The {\tt throw} methods of {\tt CatchMetaException} do nothing --- they do not
finish the program or print any error message. There is also a class
{\tt HCatchMetaException} whose methods print an error message and
finishes the program.

It is interesting to note that, in a statement \\
\verb@     Meta.attachShell( obj, ShellClass.new() )@\\
no compile time error can be issued  based on the declared type of
{\tt obj} and on {\tt ShellClass} type. The declared type of {\tt
obj}, say, {\tt ObjClass}, may have a subtype that has all methods of
{\tt ShellClass}. This is necessary, although not sufficient, to make
the statement legal at run time.

\section{A High-Level View of Shell Implementation}

The examples  of the previous section showed attachements of {\tt
Border} shells to objects of classes {\tt Window} and {\tt
ColorWindow}. These classes belong  to the ``allowed set'' of {\tt
Border}, the set of all classes to whose objects {\tt Border} shells
may be attached at run time. If a shell {\tt Border} may be attached to a {\tt
NiceWindow} object at run time, the programmer should specify this
fact through a compiler option. Of course, no class of the allowed set
can be abstract. For each class {\tt A} of the {\tt Border} allowed
set, the compiler creates a special class \verb@Border$s$A@ that
inherits from {\tt A} and has all class-{\tt Border} methods, including
method bodies.

The statement\\
\verb@     Meta.attachShell( window, Border.new() )@\\
is a compile time call to object {\tt Meta}.
This needs to  be a compile time message because it contains an
expression ``{\tt Border.new()}'' which is illegal in Green: shell
classes like {\tt Border} are not classes and cannot receive the {\tt
new} message. In fact the compiler will get around this problem by
executing method {\tt init()} of {\tt Border}\footnote{Method {\tt
new()} of class object {\tt Border} corresponds to class-{\tt Border}
method {\tt init}.} on object {\tt window} (``{\tt window.init()}'')
after the shell was attached to it. That is, the compile-time message
send \\
\verb@     Meta.attachShell( window, Border.new() )@\\
will be replaced by the statements  {\it
\bi
\item find the class of object {\tt window} which will be called {\tt
C};
\item check whether {\tt C} is in the allowed set of {\tt Border}. If
not, throw exception {\tt ClassNotInAllowedSetException};
\item there is a class \verb@Border$s$C@. Then change the class of {\tt
window} to \verb@Border$s$C@;
\item create an object {\tt obj} with all instance variables of {\tt
Border}. If there is not enough memory, throw exception {\tt
OutOfMemoryException}. Insert the pair ({\tt window}, {\tt obj}) in a
hash table. Later on, {\tt obj} (with the instance variables) will be
retrieved using {\tt window} as key to the table;
\item calls method {\tt Border::init} on {\tt window}. This is as if
there were a message send\\
\verb@     window.Border::init()@\\
in which the method to be executed is fixed at compile time --- {\tt
init} of {\tt Border}.
\ei
}

The compiler needs the source code of the shell class in order to
create a new class like \verb@Border$s$Window@. Shell classes are much
like C++ {\it templates} in which the parameter is the superclass;
that is, a class of the {\tt Border} allowed set.

\section{Shell Inheritance}

Shell classes can inherit from other shell classes. If a shell class
{\tt C} inherits shell class {\tt B} and class {\tt A} belongs to the
allowed set of {\tt C}\footnote{That means shells of {\tt C} can be
attached to {\tt A} objects.}, the compiler will create a class
\verb@C_B_A@ such that this class:
\bi
\item has the source code of {\tt C};
\item inherits from class \verb@B_A@.
\ei

\vspace*{3ex}
Class \verb@B_A@~:
\bi
\item has the source code of {\tt B};
\item inherits from {\tt A}.
\ei
Then any message sends to {\tt super} inside class {\tt C} will cause
a search for a method in {\tt B} and then in {\tt A}.

Maybe shell classes will be able to inherit from normal classes. This
opens the possibility of any class be treated as a shell class. This
can be seen in the example\\
\verb@     shell class Empty(Window) subclassOf ColorWindow@\\
\verb@     end@\\
in which {\tt ColorWindow} is a subclass of {\tt Window}. A shell class
{\tt Empty} has everything {\tt ColorWindow} has thus effectively
treating {\tt ColorWindow} as a shell class. Note the compiler would
need the source code of {\tt ColorWindow} and all of its superclasses
for the same reasons it needs the source code of the shell classes.

\section{Shell Initialization}

A shell class can define  a method {\tt init} to initiate the shell
object.\footnote{Although shells are not exactly objects, we will
treat them as such.} This may be used to communication between the
shell and some external control of the object. For example, the shell
\begin{verbatim}
shell class DrawCount(Window)
  proc init( counter : Counter )
    begin
    self.counter = counter;
    end
  public:
    proc draw()
      begin
      counter.add(1);
      super.draw();
      end
  private:
    var counter : Counter;
end
\end{verbatim}
can be used with the class
\begin{verbatim}
class Counter
  proc init()
    begin
    n = 0;
    end
  public:
    proc add( s : integer )
      begin
      n = n + s;
      end
    proc get() : integer
      begin
      return n;
      end
  private:
    var n : integer;
end
\end{verbatim}
to count the number of times the {\tt draw} method of a {\tt Window} object was
called:
\begin{verbatim}
var catch : CatchMetaException;
var c : Counter;
var w : Window;

w = Window.new(...);
c = Counter.new();
try(catch)
  Meta.attachShell( w, DrawCount.new(c) );
end
...
Out.writeln("draw was called ", c.get(), " times");
\end{verbatim}
Since shells are not really objects, they cannot be referenced
by any variable. Therefore there is no way of getting the value of an instance
variable of a shell object. So, {\tt DrawCount} could not have been
implemented as
\begin{verbatim}
shell class DrawCount(Window)
  proc init()
    begin
    n = 0;
    end
  public:
    proc draw()
      begin
      n = n + 1;
      super.draw();
      end
  private:
    var n : integer;

end
\end{verbatim}
No one outside {\tt DrawCount} will  ever access the value of {\tt n}.

\section{Reflective Shell Classes}
A class can be declared as\\
\verb@     reflective class Window@\\
\verb@        ...@\\
\verb@     end@\\
This means that, if a shell is attached to a {\tt Window} object, the
access to shell instance variables will be faster than if {\tt Window}
were declared as a normal class.
All subclasses of {\tt Window} must be declared as ``{\tt
reflective}'' as well.

\begin{figure}
\begin{verbatim}
class Planet
  proc init( name : String; radius, distanceSun : real )
    begin
    ...
    end

  public:
    ...
end


shell class PlanetManager( type(Planet) )
  proc init()
    begin
    numberOfPlanets = 0;
    end

  public:

    proc new( name : String; radius, distanceSun : real ) : Planet
      begin
      if numberOfPlanets >= 9
      then
        Out.writeln("Unknow solar system: too many planets");
          // ends the program
        Runtime.exit(1);
      endif
      ++numberOfPlanets;
      return super.new(name, radius, distanceSun);
      end

  private:
    var numberOfPlanets : integer;
end
\end{verbatim}
\caption{A planet manager that controls the number of planets}
\label{planet}
\end{figure}

\section{Shells and Class Objects}
A shell may be attached to an object that is a class object. An
example is shown in Figure~\ref{planet} in which a {\tt PlanetManager}
class controls the number of planets created from class {\tt
Planet}. Note:
\bi
\item there is no need to declare  a class object {\tt Planet} since
the compiler will do this;
\item class object {\tt Planet} has  methods
\begin{verbatim}
proc new( name : String; radius, distanceSun : real )
proc shallowClone() : Any
proc deepClone() : Any
...
\end{verbatim}
These method interfaces compose the type of class object {\tt Planet}
which is represented by {\tt type(Planet)}.
\ei

A shell is attached to {\tt Planet} by\\
\verb@     Meta.attachShell( Planet, PlanetManager.new() )@\\
After this,  the command \\
\verb@     saturn = Planet.new("Saturn", SaturnRadius, SaturnDistanceSun);@\\
will call {\tt PlanetManager::new()} that will then call method {\tt
new} of class object {\tt Planet} in the statement\\
\verb@     return super.new(name, radius, distanceSun);@\\
Note {\tt PlanetManager} can be used even if the source code of {\tt
Planet} is not available.

\section{Method {\tt interceptAll}}

One can declare a method\\
\verb@     proc interceptAll( mi : ObjectMethodInfo;@\\
\verb@                        vetArg : array(Any)[] )@\\
in a shell class to intercept all messages sent to the object the
shell is attached. {\tt ObjectMethodInfo} is a class of the
Green Introspective Class Library. An object of this class
 describes a specific method of an object.
{\tt vetArg} is an array containing the real parameters used to call
method {\tt mi}. As usual, basic class values are wrapped in classes
{\tt Char} and the like before they are
inserted in {\tt vetArg}.

When a message {\tt m} is sent to an object $\cal Q$ with a shell,
method {\tt m} of the shell will be executed, if one exists. If the
shell does not have a method {\tt m} but does have an {\tt
interceptAll} method, the message parameters are packed in an array
used as the argument {\tt vetArg} in a call to method {\tt
interceptAll} of the shell. The first {\tt interceptAll} parameter,
{\tt mi}, will be the object that describes method {\tt m} of object
$\cal Q$ --- there will always exist one.

Inside {\tt interceptAll},  method {\tt m} of $\cal Q$ can be called
as\\
\verb@     mi.invoke(vetArg)@\\
Method {\tt invoke} will call the method described by {\tt mi} using
parameters taken from {\tt vetArg}.

{\tt interceptAll} allows shells to have the full functionality of
metaobjects although with some overhead. See \cite{jose98} for more
details.

\section{Dynamic Extension}

Methods of a single object can be replaced by methods of a shell. Methods of a
class, which includes methods of all  objects of the class, can be
replaced by methods  of a dynamic extension class. The syntax for dynamic extensions is
identical to the syntax for shell classes. In fact, a dynamic
extension class can be used as shell class and
vice-versa. This means dynamic extension classes:
\bi
\item can inherit from other extension classes;
\item method {\tt interceptAll} also works with extension classes;
\item cannot have associate class objects. If there is a dynamic
extension class {\tt Border}, there cannot be a class object {\tt
Border}.
\ei

  Figure~\ref{ext1} shows the dynamic extension {\tt Border}
that can be attached to class {\tt Window} or its subtypes. The
programmer should specify at compile time the classes to which she
wants to attach dynamic extensions. These classes compose the
``allowed set'' for a given extension class. Different from
\cite{jose98}, it is not necessary  to declare class {\tt Window}
using ``{\tt reflective(extension)}''.

An abstract class may belong to the  ``allowed set'' of an extension
class. However, the extension class should not define any method {\tt
m} such that {\tt m} of the abstract class is declared as abstract. By
the semantic rules of Green, an abstract method will never be
called. Then the extension method that replaces {\tt m} will not be
called either. This is not an error although the compiler should issue
a warning.

To attach extension {\tt Border} to class {\tt Window} one should
write\\
\verb@     Meta.attachExtension(Window, Border)@\\
{\tt Meta} is a module and this compile-time message send will be
replaced by a message send to be executed at run time. If there is not
sufficient memory an exception {\tt OutOfMemoryException} will be thrown.
If {\tt Window}
is not in the allowed set of {\tt Border} there will be a compile
error.

\begin{figure}
\begin{verbatim}
class Window
  proc init(...)
    begin
    ...
    end

  public:
    proc draw()
      ..
    ..
end

  // same syntax as shell classes

shell class Border(Window)
  public:
    proc draw()
      begin
      self.drawBorder();
      super.draw();
      end
  private:
    proc drawBorder()
      ...
end
\end{verbatim}
\caption{Dynamic extension {\tt Border} to subtype-{\tt Window}
classes}
\label{ext1}
\end{figure}

If {\tt Border} declares instance variables, these should be attached
to all {\tt Window} objects. This is made on demand: when an object is
going to use its {\tt Border} instance variables, the run-time system
makes a test to discover if memory for these variables have been
allocated. If not, it allocates them. A hash table with pairs ({\tt
obj}, {\tt shellObj}) is used to associate {\tt Window} objects with
theirs {\tt Border} instance variables. To object {\tt obj} has been
conceptually added  the variables of object {\tt shellObj}. Object
{\tt shellObj} has only the instance variables declared in {\tt
Border}. Using the object address as key, one can retrieve the object
instance variables.

The test to discover if memory was allocated for the {\tt Border}
instance variables is added at the beginning of each {\tt Border}
method that accesses instance variables.

Suppose shell class {\tt Border} has a parameterless constructor. When
an extension {\tt Border} is attached to class {\tt Window}, the
constructor should be called on each {\tt Window} object to possibly
initiate {\tt Border} instance variables of each object. But that
would  be very expensive because all objects of {\tt Window} should be
found. An alternative solution would be to call the constructor the
first time a {\tt Window} object receives a message after the
attachement of {\tt Border} to {\tt Window}. The implementation for
this would be complex and would have a large memory overhead. A method
that calls the constructor should be created for each {\tt Window}
method.

A third solution would be to call the constructor only when the
extension instance variables are allocated. Then the constructor could
initiate these variables. This is our choice.
In the beginning of each method that accesses extension instance
variables there is code to allocate memory for the instance variables
(if this has not happened yet) and to call the extension
constructor.

Note the semantics of extension constructors is different from the
semantics of normal constructors. An extension constructor is only
called when the extension instance variables are created, which may
occur at any time during the object lifetime. Because we do not know
when the extension  constructor will be called, it should be only used
to initiate the extension instance variables.

An extension is removed from a class by the command\\
\verb@     Meta.removeExtension(Window)@\\
which will be replaced by a message send that may throw exception {\tt
NoExtensionException}. This will occur if no extension is attached to
class {\tt Window}.
More than one extension may be attached to a class. They are attached
and removed in a stack fashion.

An example of use of dynamic extensions is given next.
\begin{verbatim}
var w1, w2 : Window;

w1 = Window.new(...);
w2 = Window.new(...);
  // draw windows without borders
w1.draw();
w2.draw();

Meta.attachExtension( Window, Border );

  // draw windows with borders
w1.draw();
w2.draw();


var catch : CatchMetaException;

catch#init();
try(catch)
  Meta.removeExtension(Window);
    // draw windows without borders
  w1.draw();
  w2.draw();
end
\end{verbatim}

All message sends to all objects of a class can be intercepted by
declaring a method {\tt interceptAll} in an extension class that is
then attached to the class. This mechanism  is used with dynamic
extensions as it is with shells. The semantics is the same.

\chapter{Standard Class Objects}  \label{sco}

\begin{figure}
\setlength{\unitlength}{0.0125in}
\begin{picture}(400, 300)(0, 0)
\thicklines
\put(100.000000, 0.000000){\line(0, 1){250.000000}
}\put(300.000000, 0.000000){\line(0, 1){250.000000}}
\put(100.000000, 0.000000){\line(1, 0){200.000000}}
\put(100.000000, 250.000000){\line(1, 0){200.000000}}
\put(107.000000, 90.000000){Green}
\put(107.000000, 65.000000){Program}
\put(220.000000, 180.000000){\oval(35.000000, 20.000000)}
\put(215.000000, 177.000000){\tt Out}
\put(242.000000, 184.000000){\line(4, -1){85.000000}}
\put(220.000000, 153.000000){\oval(35.000000, 20.000000)}
\put(215.000000, 150.000000){\tt In}
\put(242.000000, 153.000000){\line(1, 0){85.000000}}
\put(372.000000, 153.000000){\oval(70.000000, 20.000000)}
\put(345.000000, 150.000000){\tt Storage}
\put(220.000000, 126.000000){\oval(70.000000, 20.000000)}
\put(193.000000, 123.000000){\tt OutError}
\put(260.000000, 126.000000){\line(3, 1){72.000000}}
\put(220.000000, 99.000000){\oval(120.000000, 20.000000)}
\put(189.000000, 96.000000){\tt Storage}
\put(284.000000, 99.000000){\line(2, 1){80.000000}}
\put(220.000000, 72.000000){\oval(65.000000, 20.000000)}
\put(201.000000, 69.000000){\tt Screen}
\put(201.000000, 39.000000){\tt souls}
\put(372.000000, 72.000000){\oval(65.000000, 20.000000)}
\put(353.000000, 69.000000){\tt Screen}
\put(260.000000, 72.000000){\line(1, 0){70.000000}}
\put(353.000000, 39.000000){\tt fleshes}
\end{picture}
\caption{Souls and fleshes in a Green Program}
\label{soul}
\end{figure}
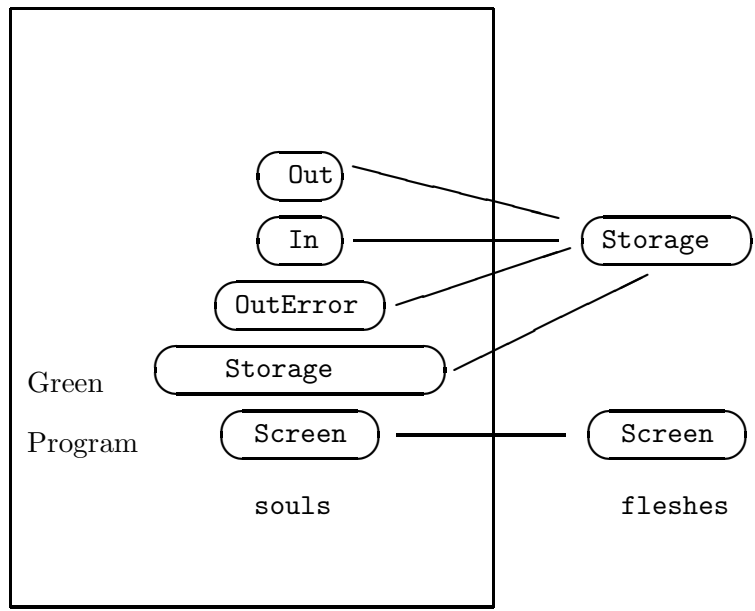


\section{Introduction}

A Green program does input, output, and controls itself through
objects {\tt In}, {\tt Out}, {\tt OutError}, {\tt Memory}, {\tt
Runtime}, {\tt Screen}, and {\tt Storage}. A brief description of each
object follows.

\bi
\item {\tt In} is used for input from the standard input.
\item {\tt Out} is used for output to the standard output.
\item {\tt OutError} is used for output to the standard error output.
\item {\tt Storage} abstracts the file system and the Internet,
although methods for working with the Internet have not been defined
yet.

\item {\tt Memory} has methods that control memory use of the whole
program. One can turn on/off the garbage collector, ask the number of
bytes of free memory available, and so on.

\item {\tt Runtime} abstracts (part of) the program run-time
system. This object has a method for finishing the program  and a
method that returns an object describing the method call stack at run
time.

\item {\tt Screen} is used for graphical handling and user interaction
through devices like keyboards and mice.  The specification of {\tt
Screen} is troublesome and beyond  the capacity of the author of this
report. Then hopefully {\tt Screen} will be defined by someone else.
\ei

Objects {\tt Memory} and {\tt Runtime} interact only with the
program. Objects {\tt In}, {\tt Out}, {\tt OutError}, {\tt Storage},
and {\tt Screen} make the linking between the program and the
Operating System. This linking is so important that it deserves to be
better  studied, which is made next.

Green assumes its programs will run in a fictitious Object-Oriented
Operating System called Rainbow. Although Rainbow does not  exist, the
features described below may be simulated by
\bi
\item a Green interpreter if the programs are translated into instructions
of some virtual machine;

\item a run-time system manager. That is, a Green program would be
compiled not to a ready-to-run executable program but to a format that
needs to be executed by other program, the run-time system manager:\\
\verb@     C:\>manager aProgram parameters@\\
So {\tt aProgram} would be controlled\footnote{To a certain extent, of
course, since {\tt aProgram} could bypass the run-time manager.} by
{\tt manager} as it is by the Operating System. The symbols
``\verb@C:\>@'' are the DOS prompt.
\ei

\section{Rainbow}

Rainbow considers every executable program in  disk or some other
storage device as a {\it class}. The loading and execution of a disk
program is carried out through the following steps:
\bi
\item the class is loaded to the main memory and one object of it is
created. This object will represent an executing process;

\item Rainbow creates objects
{\tt Storage} and {\tt Screen} that match the objects with the same
names of the Green program. These Rainbow objects are called {\it
flesh objects}. The corresponding  objects of the Green program are
called {\it souls} --- see Figure~\ref{soul}. Each {\it soul} has a reference to the {\it flesh}
object with the same name. In addition to that, soul objects {\tt In},
{\tt Out}, and {\tt OutError} have references to flesh object {\tt
Storage}. After all, input and output to standard devices are only
file manipulation. Each method of a soul object just forwards the
message to the corresponding flesh object; that is, to the flesh
object the soul object references.

\item each executable program in disk, which is just a class, may have
a set of associate shells to the program itself and to each of the
flesh objects {\tt Storage} and
{\tt Screen}. These shells are created and attached to the flesh
objects and possible to the program itself;

\item message {\tt run} is sent to the object representing the program
passing as parameter the program parameters.
\ei

This model gives us freedom to modify a program by associating shells
to it using Rainbow. Some interesting  things can be done:
\bi
\item  one can easily attach shells connecting the output of a program
(flesh object {\tt Out}) to the input (flesh object {\tt In}) of
another, thus implementing a Unix pipe;

\item a shell may be attached to flesh object {\tt Screen} such that
all graphical output  (windows, buttons, menus, etc) is done not only
to the screen associate to the program but also to the screen of
another computer in the same network;

\item a Rainbow shell may be attached to flesh object {\tt Storage} to
prevent an untrusted program from writing to or deleting disk
files. This shell would play a rôle similar to the Java Virtual
Machine when executing  applet code.
\ei
The possibilities of program control are endless.

A future feature of Green would allow any class object of the program to be
associated with a flesh object. Then the program behavior could be
easily modified by putting shells in its flesh objects. Of course, the
program can attach shells to its own soul objects ({\tt In}, {\tt
Out}, ...) thus modifying the program behavior in the same way as
attaching shells to flesh objects. Note a program has no permission to

access flesh objects (besides sending messages to them) and Rainbow
cannot access soul objects.

Objects {\tt Memory} and {\tt Runtime} may have flesh objects in the
future. Flesh objects {\tt Memory} and {\tt Runtime} would have a
subset of the methods of the corresponding  soul objects. If flesh
{\tt Memory} had all methods of soul {\tt Memory}, Rainbow could, for
example,  turn off the garbage collection, which does not make
sense. But Rainbow could furnish more memory for the program through

flesh object {\tt Memory} or inspect the program run-time stack
through flesh object {\tt Runtime}.

\section{The Class Objects}
All objects described in this chapter are detailed below. Note:
\bi
\item flesh objects have the same interface as the soul objects with
the same name;
\item object {\tt Screen} is not described --- its definition demands
knowledge about low level aspects of screen/mouse/keyboard
interactions that the author of this languages does not have;

\item methods will be added to most objects described next. Their
interfaces nowadays is minimal and need improvements. For example,
object {\tt Storage} has few methods for file handling and no for
Internet.
\ei

Here are the objects. Only the method interfaces are shown.

\begin{verbatim}
object In
  private:
      // reads from the standard input
    proc readCh()      : char
    proc readByte()    : byte
    proc readInteger() : integer
    proc readLong()    : long
    proc readReal()    : real
    proc readDouble()  : double
    proc readString()  : String
    proc readLine()    : String
end


object Out
  public:
      // print an array of objects. Any type is allowed
    proc write( v : ... array(Any)[] )
      // print a newline after each object
    proc writeln( v : ... array(Any)[] )
end


object OutError
  public:
      // print an array of objects. Any type is allowed
    proc write( v : ... array(Any)[] )
      // print a newline after each object
    proc writeln( v : ... array(Any)[] )
end



object Memory
  public:
      // returns the size in bytes of the largest memory block
    proc sizeLargestBlock() : long
      // returns the size in bytes of the free memory
    proc sizeFreeMemory() : long
      // calls the garbage collector
    proc doGarbageCollection()
      // turns on and off the garbage collection
    proc collectionOn()
    proc collectionOff()
end



object Storage
  public:
    proc removeFile( fileName : String ) : boolean
    proc renameFile( oldName, newName : String ) : boolean
      // opens a file and returns a file descriptor
    proc openFile( fileName : String ) : integer
      // closes a file with file descriptor fd
    proc closeFile( fd : integer )
      // reads n bytes from file fd to array in
    proc read( fd, n : integer; in : array(byte)[] )
      // write n bytes from out to file fd
    proc write( fd, n : integer; out : array(byte)[] )
      /* returns an error code. This method should be called after file
         or Internet operations. The values returned  are given in the
         enumerate declaration that follows. */
    proc getError() : integer

    enum( fileDoNotExist_e, cannotClose_e, smallArraySize_e,
          unknownFileDescriptor_e, readOnlyFile_e, internalError_e );

end


/* The methods of Storage are used to implement class BasicStream and
   its subclasses used for file manipulation. These classes are described
   elsewhere. */


object Runtime
  // this object represents the run-time system

  public:
    proc exit( errorCode : integer )
      /* finishes the program and returns errorCode. In an object-oriented
         operating system an object would be returned */
    proc putAtEndList( f : Function )
      /* puts f in a list of Function objects. At the end of the
         program, message exec will be sent to all objects of this
         list. Class Function belong to the Green Standard Library. */
end
\end{verbatim}
Appendix~\ref{irl} describes some methods added to {\tt Runtime} that
allows one to inspect the method call stack and the stack of exception
catch objects. See Chapter~\ref{stand} for the definition of class
{\tt Function}.

\chapter{Parameterized Classes}
\label{pclasses}

Parameterized classes are implemented using the compile-time
metaobject protocol (MOP), topic not
presented in this report. However, their use will follow nearly the
syntax described in the following paragraphs, which will be changed as
soon as the MOP is defined.


%
%
%

The language supports a simple form of parameterized classes whose
 syntax  is shown in Figure~\ref{ppc}.
Parameters are specified in a semicolon separated
list delimited by round
brackets, following the class name in the header. Each parameter can
be in one of following forms.
\bi
\item {\tt T : A}

where {\tt T} is the parameter name. Real parameters must be ext-subtypes of
{\tt A}. A class {\tt S} is an ext-subtype of class {\tt T} if {\tt S}
is subtype of {\tt T} and {\tt S} defines all {\tt init} methods {\tt
T} defines. That is, {\tt init} methods belong to a class type in the
definition of ext-subtype. Class {\tt A} must be a class. That is,
{\tt A} cannot be an expression ``\verb@type(X)@'' in which {\tt X} is
a class object.

\item {\tt T}

where {\tt T} is the parameter name. The corresponding real parameter
may belong to any type, including basic classes.

\ei

\begin{figure}
\verb@@{\tt class A( parameter-list )}\\
\verb@    // constructors@\\
\verb@  public:@\\
\verb@    // methods@\\
\verb@  private:@\\
\verb@    // methods and instance variables@\\
\verb@end@
\caption{Syntax for parameterized classes}
\label{ppc}
\end{figure}

The round brackets with the parameters
should be put in the class {\it or} in the class object,
whichever comes first. So, we can declare
\begin{verbatim}
object List( T : B )
  ...
end

class List
  ...
end
\end{verbatim}
and use {\tt List} as in\\
\verb@     var list : DS.List(Person);@\\

The compiler will check if {\tt Person} is ext-subtype of {\tt B}. If
not, an error is issued. Inside class {\tt list}, a {\tt T} object may
be allocated. The parameters to {\tt new} follow the rules for class
{\tt B}. That is, if {\tt B} defines a method\\
\verb@     init( x : integer; ch : char )@\\
then inside {\tt List} an expression\\
\verb@     T.new(3, '#A')@\\
is legal.

{\tt DS} is the module in which {\tt List} is.
To furnish the real parameters to a parameterized class, as in
``\verb@List(Person)@'',  is called ``the instantiation of class {\tt
List}''. When the compiler finds this instantiation, it replaces the
formal by the real parameters in the body of class {\tt List}, creating a
new class ``\verb@List$p$1$Person@''. Then the compiler compiles this
class, which may
have errors.

It is pretty common to design a parameterized class in which the type
parameter may be a basic class or a subclass of {\tt Any}. That is,
the type parameter may be a class with value or reference
semantics. This creates a problem because the operators (\verb@< + *@
...) of basic classes cannot be declared in normal classes therefore
dividing the world of types in two: those who have operators and those
who do not. To see why this is important, we will study the following
parameterized class.
\begin{verbatim}
class List(T)
  public:
    proc getGreatest() : T
      ...
    proc add( x : T )
    ...
end
\end{verbatim}
Method {\tt getGreatest} returns the greatest element in the list. If
{\tt T} is a basic class, this method should use operators like
\verb@<@ or \verb@>=@ to do its job. If {\tt T} is a subclass of {\tt
Any} (reference class), {\tt getGreatest} should use a method like
{\tt lessThan} or {\tt greaterEqual}. Then the {\tt List} class could
be used either with basic classes or reference classes.

    To prevent such problem, the compiler\footnote{Assume that. In
fact, parameterized classes are not supported by Green. They are
implemented using the MOP that still needs to be defined.}  will
replace the  comparison operator by message sends whenever the real
parameter {\tt T} to the parameterized class is a subclass of {\tt
Any} ({\tt T} is a normal class). Then, if {\tt x} and {\tt y} belong
to type {\tt T},\\
\verb@     a > b@\\
will be replaced by\\
\verb@     a.greaterThan(b)@\\

Methods {\tt equals}, {\tt notEqual}, {\tt lessThan}, {\tt
greaterThan}, {\tt lessEqual}, and {\tt greaterEqual} will replace
\verb@==@, \verb@<>@, \verb@<@, \verb@>@ \verb@<=@, and \verb@>=@,
respectively. The arithmetic operators will not be replaced. Their
semantics are more complex than that of comparison operators.

Parameterized classes may be implemented either with code sharing or
code duplication. In the first case there will be a single code for a
parameterized class {\tt List} regardless of the number of
instantiations of it. In the second case, the compiler will create a
different class each time {\tt List} is instantiated with a new class:
{\tt List(integer)}, {\tt List(Person)}, {\tt List(Figure)}.

The official semantics of Green parameterized classes requires code
duplication. If code sharing is used, the compiler creator should take
care of the relationship between dynamic extensions and parameterized
classes. We may describe this caveat using an example.

Class {\tt List} has method {\tt print} to print the list. Class {\tt
Slist} inherits from {\tt List} and redefines method {\tt print},
which calls {\tt List::print} through {\tt super}:
\begin{verbatim}
class SList(T) subclassOf List(T)
  public:
    proc print()
      begin
      ...
      super.print();
      end
    ...
end
\end{verbatim}
Both classes have a parameter {\tt T}, the list element type. Suppose
now an extension is attached to {\tt List(Person)} changing  its {\tt
print} method. How would {\tt SList(Figure)::print} behave since this
method has a call ``{\tt super.print()}'', which should call {\tt
List(Figure)::print} and {\tt print} was replaced in {\tt
List(Person)}~?

{\sloppy
{\tt SList(Figure)} inherits from {\tt List(Figure)} and should not be
affected by an extension on {\tt List(Person)}. But if parameterized
classes are implemented using code sharing, {\tt List(Person)::print}
and {\tt SList(Figure)::print} are the same method. And a change in
one should reflect in the other.

A solution to this problem would be to duplicate the code whenever
there is a call to super and the superclass is parameterized.
}

\chapter{The Standard Library}  \label{stand}

Green has a standard library with data structure and file handling
classes. Of course, the following classes will be improved and a lot
of new classes will be added to the standard library.

\vspace*{3ex}
\noindent {\Large \bf File Handling}
\vspace*{3ex}

\begin{verbatim}
object BasicStream
  public:
    enum (read_s, write_s, readwrite_s, append_s);
end


abstract class BasicStream
  public:

    abstract proc open( name : String; mode : integer )
                      ( exception : CatchFileException )
      /* If the name is not valid, calls
              raise(InvalidNameException) : String
         of variable exception.
         The return value should be a new file name with which open
         tries to open again. If the operation fails, raise is again called
         and the method tries to open the file. This repeats till a
         valid name is given or raise throws an exception.
             An existing file is open only for reading is mode read_s
         is used. If the file does not exist, method
                   raise(NonExistingFileException) : String
         of variable exception is called till a valid name is given
         or raise throws an exception
         In any other error, method throw(OpenFileException) is called.
      */

    abstract proc close()
                       ( exception : CatchFileException )
      // In error, calls method raise(CloseFileException) or
      // raise(FileIsClosedException)

    abstract proc getSize() : integer
                         ( exception : CatchFileException )
    abstract proc getName() : String
end


class InputStream subclassOf BasicStream
  proc init()

  public:

    proc read( v : array(char)[]; n : long )
             ( exception : CatchFileException )
      // In error, calls raise(ReadFileException)

    proc read( v : array(byte)[]; n : long )
             ( exception : CatchFileException )

    proc read( s : DynString )
             ( exception : CatchFileException )

    proc readln( s : DynString )
               ( exception : CatchFileException )
end


class OutputStream subclassOf BasicStream
  proc init()

  public:

    proc write( v : array(char)[] )
              ( exception : CatchFileException )
      // In error, calls raise(WriteFileException) or
      // raise(WriteFileInterException)

    proc write( v : array(char)[]; n : long )
              ( exception : CatchFileException )

    proc write( v : array(byte)[] )
              ( exception : CatchFileException )

    proc write( v : array(byte)[]; n : long )
              ( exception : CatchFileException )

    proc write( s : DynString )
              ( exception : CatchFileException )

    proc writeln( s : DynString )
                ( exception : CatchFileException )
end



class Stream subclassOf BasicStream
  proc init()

  public:
    // all methods of InputStream and OutputStream
    ...
end


class OpenFileException
  proc init( f : File )
  public:
    proc getFile() : File
end

// there is a class equal to the above for every every file exception.
// There is a throw method in class CatchFileException for each file
// exception

object CatchFileException
  proc init()
    begin
    singleInstance = CatchFileException.new(' ');
    end

  public:
    proc new() : CatchFileException
      begin
      return singleInstance;
      end
end


class CatchFileException
  proc init( x : char )
    begin
      // fake. Just to cause the creation of a new(char) method in
      // the class object
    end

  public:
    proc throw( exc : InvalidNameException )
    proc throw( exc : NonExistingFileException )
      // file does not exist
    proc throw( exc : OpenFileException )
      // file cannot be openend
    proc throw( exc : OpenReadOnlyFileException )
      // attempt to open for writing a read-only file
    proc throw( exc : ReadFileException )
      // attempt to read more bytes than the available or another
      // read error
    proc throw( exc : WriteSharedFileException )
      // two processes are trying to write to the same file
    proc throw( exc : WriteFileException )
      // no space for writing or other write error
    proc throw( exc : CloseFileException )
      // cannot close file
    proc throw( exc : FileIsClosedException )
      // attempt to use a non-open file

    proc raise( exc : InvalidNameException )
              ( exception : CatchFileException ) : String
    proc raise( exc : NonExistingFileException )
              ( exception : CatchFileException ) : String
    proc raise( exc : OpenFileException )
              ( exception : CatchFileException ) : String
    proc raise( exc : OpenReadOnlyFileException )
              ( exception : CatchFileException ) : String
    proc raise( exc : ReadFileException )
              ( exception : CatchFileException )
    proc raise( exc : WriteSharedFileException )
              ( exception : CatchFileException )
    proc raise( exc : WriteFileException )
              ( exception : CatchFileException )
    proc raise( exc : CloseFileException )
              ( exception : CatchFileException )
    proc raise( exc : FileIsClosedException )
              ( exception : CatchFileException )
end
\end{verbatim}
The methods of all classes for file manipulation uses methods {\tt
raise} for throwing exceptions. Methods {\tt raise} of {\tt
CatchFileException} will just throw the exception they receive as
parameters. Subclasses of {\tt CatchFileException} can take a
different approach. For example, subclass {\tt CatchFileExceptionUser}
tries to get help from the user when methods\\
\verb@     @raise(InvalidNameException)@\\
 and\\
\verb@     @raise(NonExistingFileException)@\\
are called. The user can then suggest other file name and retry the
operation. Subclass {\tt CatchFileExceptionEnd} prints an error
message and terminates the program --- a radical approach.

\vspace*{3ex}
\noindent {\Large \bf Data Structure Classes}
\vspace*{3ex}

The following classes are parameterized classes and described in
Chapter~\ref{pclasses}. They should be used with the prefix
``{\tt DS}'' as in\\
\verb@     var list  : DS.List(Person)@\\
\verb@     var queue : DS.Queue(double);@\\
Class {\tt Iter} is an iterator class. All data structure classes have
a method that return an iterator object which belong to a subclass of
{\tt Iter}.  All classes inherit from {\tt Container}. Contrary to
other object-oriented libraries, we did not do a deep inheritance
hierarchy for data structure classes. Not only this is unnecessary but
also dangerous: a class may be declared superclass of another without
being semantically subclass of it. This seems to be pretty common in
hierarchies of data structure classes.

\begin{verbatim}
abstract class Container( T : Any )
    proc init()
      begin
      index = -2;
      end
  public:
    abstract proc add( elem : T )
                     ( exception : CatchOutOfMemoryException )
      // inserts elem in the container

    abstract proc get() : T
                     ( exception : CatchNotFoundException )
      /* gets one element from the container which is not removed
        from it. */

    abstract proc remove() : T
                     ( exception : CatchNotFoundException )
      // removes one element from the container and returns it.

    abstract proc removeAll()
      // empties the container

    abstract proc getSize() : integer
      // returns the number of elements in the container

    abstract proc empty() : boolean
      // returns true if there is no element in the container

    abstract proc full() : boolean
      /* returns true if there is no more space in the container. This
         method only returns false in static containers whose number of
         elements is defined at the creation of the container */

    abstract proc getIter() : DS.Iter(T)
      // returns an iterator for the container

    abstract proc forEach( f : Function(T) )
      // calls "f.exec(x)" for each element x of the container

    abstract proc replaceBy( cmd : Command(T) )
      // replaces each element x in the container by "cmd.doIt(x)"

    abstract proc collect( f : Filter(T) ) : Container(T)
      /* collects all container elements x such that "f.test(x)"
         evaluates to true. These elements are inserted in a new
         container that is returned by this method. */

    abstract proc  remove( f : Filter(T) )
      /* removes all array elements x from the container whenever
         "f.test(x)" evaluates to true. */

    proc reset()
      /* reset, next, and endIter compose a lightWeight iterator. It
         should be used as in the example:
            list.reset();
            while (elem = list.next()) <> nil do
              Out.writeln(elem);
            list.endIter();
         This iterator can only be used when T is a reference class
         (not a basic class like char, byte, etc) and nil cannot be
         inserted in the container. Method next returns nil when there is no
         more elements.
            Method endIter should be called after the use of the
         iterator. If it is not called, the next call to reset will cause a
         run-time error:
             list.reset();
             x = list.next();
               // run-time error in the following line
             list.reset();
         This prevents one of calling reset in the middle of an use of
         the iterator:
             list.reset();
             while (x = list.next()) <> nil do
               begin
               select( list, x );
               ...
               end
         If select calls ``reset'' on list, there will be an
         error. Only one scanning of the container can be made at a time with
         this kind of iterator.

           This type of iterator does not need a separate object of
         type DS.Iter(T) thus saving memory and time of the garbage
         collector.

         This method should be implemented as
          begin
          if index <> -2
          then
             Out.writeln("Iterator error");
             Runtime.exit(1);
          else
             index = -1;
          endif
          end
      */


    abstract proc next() : T

    abstract proc endIter()
      begin
      index = -2;
      end

    proc toArray() : array(T)[]


end


abstract class Iter( T : Any )
  public:
    abstract proc more() : boolean;
    abstract proc next() : T
      assertion
        before more();
      end
    abstract proc reset()
      // begin everything again
    abstract proc toArray() : array(T)[]
      // returns all elements in an array
end


class IterFilter( T : Any )
    proc init( piter : Iter(T); pf : Filter(T) )
      begin
      iter = piter;
      end
  public:
    proc more() : boolean
      begin
      return iter.more();
      end
    proc next() : T
        assertion
          before more();
        end
      begin
      var item : T;
      loop
        if not self.more()
        then
          return nil;
        else
          item = iter.next();
          if f.test(item)
          then
            return item;
          endif
        endif
      end
      end

    proc toArray() : array(T)[]
      begin
      return iter.toArray();
      end

    proc reset()
      begin
      iter.reset();
      end

  private:
    var iter : Iter(T);
        f    : Filter(T);
end


abstract class Filter( T : Any )
  public:
    abstract proc test( x : T ) : boolean
end


abstract class Command(T)
  public:
    abstract proc doIt( x : T ) : T
end


abstract class Function(T)
  public:
    abstract proc exec( x : T )
end


/* In the classes below, a D before the class name means the class is
   a Dynamic container. That is, a class DList is better than a class
   List when the number of list elements varies greatly during the list
   lifetime. Possibly DList is implemented using a linked list and List
   using an array. So DList and the like will produce a lot of objects
   causing memory and speed overhead (because of the garbage
   collector).
*/


class List( T : Any ) subclassOf Container(T)

  /* A list of elements. There is no order to insert or remove
     elements.
     This class is better when the number of list elements does not
     vary greatly during the list lifetime. */

  public:
    // all Container(T) methods are defined but no new one is added
    ...
  private:
    ...
end


class DList( T : Any ) subclassOf Container(T)
  /* A list of elements. There is no order to insert or remove
     elements.
     This class is better when the number of elements
     varies greatly during the container lifetime. */

  public:
    // all Container(T) methods are defined but no new one is added
    ...
  private:
    ...
end



class Stack( T : Any ) subclassOf Container(T)
  /* A stack of elements.
     This class is better when the number of elements does not
     vary greatly during the container lifetime. */

  public:
    // all Container(T) methods are defined but no new one is added
    ...
  private:
    ...
end


class DStack( T : Any ) subclassOf Container(T)
  /* A stack of elements.
     This class is better when the number of elements
     varies greatly during the container lifetime. */

  public:
    // all Container(T) methods are defined but no new one is added
    ...
  private:
    ...
end



class Queue( T : Any ) subclassOf Container(T)
  /* A queue of elements.
     This class is better when the number of elements does not
     vary greatly during the container lifetime. */

  public:
    // all Container(T) methods are defined but no new one is added
    ...
  private:
    ...
end


class DQueue( T : Any ) subclassOf Container(T)
  /* A queue of elements.
     This class is better when the number of list elements
     varies greatly during the container lifetime. */

  public:
    // all Container(T) methods are defined but no new one is added
    ...
  private:
    ...
end




class DoubleQueue( T : Any ) subclassOf Container(T)
  /* A double queue of elements. Insertion and removal of elements can
     occur in both sides of the queue.
     This class is better when the number of list elements does not
     vary greatly during the container lifetime. */

  public:
    /* all Container(T) methods are defined and some ones are added.
       Method put, get, and remove are equivalente to putBack,
       getFront, and removeFront. */
    ...
    abstract proc addFront( elem : T )
                          ( exception : CatchOutOfMemoryException )
      // inserts elem in the front of the queue

    abstract proc addBack( elem : T )
                         ( exception : CatchOutOfMemoryException )
      // inserts elem in the back of the queue

    abstract proc getFront() : T
                          ( exception : CatchNotFoundException )
      /* gets one element from the front of the queue which is not removed
         from it. */

    abstract proc getBack() : T
                         ( exception : CatchNotFoundException )
      /* gets one element from the back of the queue which is not removed
         from it. */

    abstract proc removeFront() : T
                             ( exception : CatchNotFoundException )
      // removes one element from the front of the queue and returns it.

    abstract proc removeBack() : T
                            ( exception : CatchNotFoundException )
      // removes one element from the back of the queue and returns it.

  private:
    ...
end


class DDoubleQueue( T : Any ) subclassOf Container(T)
  /* A double queue of elements. Insertion and removal of elements can
     occur in both sides of the queue.
     This class is better when the number of elements
     varies greatly during the container lifetime. */

  public:
    // the same methods as DoubleQueue

    ...
  private:
    ...
end


abstract class HashFunction( T : Any )
  public:
    proc setSize( psize : integer )
      begin
      size = psize;
      end
    abstract proc hash( elem : T ) : integer
  private:
    var int size;
end


class Compose( T, U : Any )
    proc init( pt : T; pu : U )
      begin
      t = pt;
      u = pu;
      end

  public:
    proc getT() : T
      begin
      return t;
      end

    proc getU() : U
      begin
      return u;
      end
  private:
    var t : T;
        u : U;
end


class Dict( KeyType, ValueType : Any )
    proc init()
    proc init( pSize : integer )
      // pSize is the maximum estimated number of elements in the table
    proc init( pSize : integer; phashFunction : HashFunction(KeyType) )
      // phashFunction is the hash function for type KeyType

  public:
    proc add( key : KeyType; value : ValueType )
            ( exception : CatchOutOfMemoryException )
    proc get( key : KeyType ) : ValueType
            ( exception : CatchNotFoundException )
    proc remove( key : KeyType ) : ValueType
               ( exception : CatchNotFoundException )
    proc getSize() : integer
      // number of elements in the dictionary
    proc removeAll()
    proc getIter() : DS.DictIter(KeyType, ValueType)
  private:
    ...
end


class DictIter( KeyType, ValueType : Any )
    proc init( pdict : Dict(KeyType, ValueType) )

  public:
    proc more() : boolean
    proc next() : Container(KeyType, ValueType)
  private:
    ...
end


class IntegerSet
    proc init()

  public:
    proc add( n : integer )
            ( exception : CatchOutOfMemoryException )
    proc remove( n : integer )
               ( exception : CatchNotFoundException )
    proc inSet( n : integer ) : boolean
    proc empty() : boolean
    proc getIter() : DS.Iter(integer)
    proc removeAll()
  private:
    ...
end


object Vector( T : Any )
  public:
    const defaultSizeVector = 20;
end


  // since object Vector has a parameter, class Vector must have too.

class Vector( T : Any )  subclassOf  Container(T)
    proc init( p_size : integer )
      /* size of the vector. Initially the number of elements is
         zero. That is,  getSize() returns 0. */

    proc init()
      // uses defaultSizeVector to initiate the object
      begin
      init(defaultSizeVector);
      end


  /* an high-level array. Method
        get() : T
     get the last array element and
        add( elem : T )
     inserts elem after the last array element. So Vector works
     much like a stack. The maximum number of elements is given by
     getMaxSize(). However, the number of elements that have effectively
     been put in the vector is given by getSize(). If one tries to put
     an element in a position after the end of the vector (after position
     getMaxSize()), all positions between the last and this new end is
     fulfilled with nil. For example, in

         var v : Vector(Integer);
         v#init();
         v.add( Integer.new(5) );
         v.add( Integer.new(7), 2 );
         Out.writeln( v.get(1) );

     the position 1 of the vector v is initiated to nil since the
     statement
          v.add( Integer.new(7), 2 )
     bypass position 1.


        Vector is a dynamic vector. It grows and shrinks as needed.

  */
  public:


    proc add( elem : T )
            ( exception : CatchOutOfMemoryException )
      // inserts elem after the last array element

    proc add( elem : T; i : integer )
            ( exception : CatchOutOfMemoryException )
      // inserts elem in the position i. See observations above.


    proc get() : T
            ( exception : CatchNotFoundException )
      /*  gets one element from the container which is not removed
         from it. */

    proc get( i : integer ) : T
            ( exception : CatchNotFoundException )
      /* gets the i-th element from the vector which is not removed
         from it. */

    proc remove() : T
               ( exception : CatchNotFoundException )
      // removes one element from the container and returns it.

    proc remove( i : integer ) : T
               ( exception : CatchNotFoundException )
      /* removes the i-th element from the vector and returns it.  All
         other elements are left-shifted. Exception NotFoundException is
         thrown if i < 0 or i >= getNum() */

    proc getSize() : integer
      // returns the real number of vector elements
    proc getMaxSize() : integer
      // returns the maximum number of vector elements
    proc toArray() : array(T)[]
end
\end{verbatim}

\chapter{Introspective Reflection}  \label{inr}

   The  introspective reflection library is a set of objects and classes that
allow a program to access information about itself at compile or run time. For
example, the program can know the class of an object, the methods and
instance variables of that class, the parameter names and types of
each method, and so on.

There is a build-in object called {\tt GreenCompiler} whose methods are executed at
compile time. This  object has the following methods:
\bi
\item {\tt getType( id : Symbol ) : TypeNo}\\
which returns the type the identifier {\tt id} was declared with if
{\tt id} is a variable of returns the set of method signatures of {\tt
id} if it is a class object. {\tt TypeNo} is the class of the Abstract
Syntax Tree used to represent type. this method is identical to
function {\tt type} defined in Section~\ref{tycobj};

\item {\tt getCompilerName() : LiteralString}\\
 which returns the compiler name;
\item {\tt getCompilerVersion() : LiteralString}\\
which returns the compiler
version;
\item {\tt getStrDate() : LiteralString}\\
 which returns a string with the
 date of compilation in the format ``mm/dd/yyyy''.

\item {\tt getStrTime() : LiteralString}\\
 which returns a string with the
hour/minute/second of compilation in the format ``hh:mm:ss'';

\item {\tt getFileName() : LiteralString}\\
which returns a string with
the name of the file being compiled, if there is one being compiled
the moment this method is executed. The compiler may create Green code
that is not associate to any particular file. In this case {\tt
getFileName} returns an empty string;

\item {\tt getStrCurrentClass() : LiteralString}\\
 returns a string
with the name of the class being compiled. Inside either class {\tt
Person} or class object {\tt Person}, this method would return ``{\tt
Person}''. Returns an empty string if no class is being currently compiled;

\item {\tt getStrCurrentMethod() : LiteralString}\\
returns a string with the name of the method being compiled, if there
is one. Returns an empty string if no method is being currently
compiled;

\item {\tt getCurrentClass() : TypeNo}\\
returns a symbol that represents the current class. Then, inside a
method of a class-{\tt Person} (or class object {\tt Person}), the
declaration\\
\verb@     var p : GreenCompiler.getCurrentClass();@\\
is equivalent to\\
\verb@     var p : Person;@\\
If there is no current class or class object, a compile error is
issued;

\item {\tt getCurrentMethodInfo() : MethodInfoNo}\\
returns an object with information about the current method being
compiled. That is, the command\\
\verb@     mi = GreenCompiler.getCurrentMethodInfo();@\\
of a parameterless method {\tt m} may be replaced by\\
\verb@     mi = GreenCompiler.getCurrentClass().getAssociateClassInfo().getMethod("m", nil);@\\
There is an error if there is no current method;

\item {\tt getEofOfList( aClass : ClassNo ) : LiteralNo}

returns a literal value that is generally used as
end-of-list\footnote{The method pretends to return an end-of-file
mark.} (EOL) mark for list of objects of class {\tt aClass}. {\tt ClassNo}
is a class of the Abstract Syntax Tree of the compiler that represents
a class.

All subclasses of {\tt Any} use {\tt nil} as EOL. Classes {\tt
integer}, {\tt long}, {\tt real}, and {\tt double} use {\tt -1}. {\tt
byte} uses {\tt 255}. Class {\tt char} uses \verb@'\0'@. Class {\tt
boolean} cannot be parameter to this method --- that would not make
sense.

An example of use of this method is shown below in which all  array
elements are valid till {\tt -1}. Of course, one could not have put
{\tt -1} as a valid array number.

\begin{verbatim}
var v : array(integer)[];
var i : integer;
... // initialize v
i = 0;
while v[i] <> -1 do
  begin
  Out.writeln( v[i] );

  ++i;
  end
\end{verbatim}
The {\tt getEofOfList} method was devised to be used in parameterized
classes in which this code would be generalized to
\begin{verbatim}
class List(T)
  public:
    ...
    proc write()
      begin
      var i : integer;

      i = 0;
      while v[i] <> GreenCompiler.getEofOfList(T) do
        begin
        Out.writeln( v[i] );
        ++i;
        end
      end
  ...
  private:
      var v : array(T)[];
end  // List(T)
\end{verbatim}

\ei

Class {\tt LiteralString} belongs to the Abstract
Syntax Tree of the compiler and represents values of class {\tt
String}.

\begin{figure}
\begin{verbatim}
ClassInfo
    ValueClassInfo
    RefClassInfo
        AbstractClassInfo
        ConcreteClassInfo
            NormalClassInfo
            ArrayClassInfo
\end{verbatim}
\caption{Hierarchy of {\tt ClassInfo}}
\label{hcc}
\end{figure}

The hierarchy of classes that describe the program classes at run time
is shown in
Figure~\ref{hcc}. Each class in the system is described by an object
of {\tt ClassInfo} or one of its subclasses. To get the object of {\tt
ClassInfo} that describes a class {\tt Person} we do
\begin{verbatim}
var ci : ClassInfo;

ci = Person.getAssociateClassInfo();
  // print "Person"
Out.write( ci.getName() );
\end{verbatim}
The class of an object can be got by calling method {\tt
getClassObject}:
\begin{verbatim}
var p : Person;
...
var any : AnyClassObject;
any = p.getClassObject();
Out.writeln( any.getAssociateClassInfo().getName() );
\end{verbatim}
The methods of a class are got in the following way:
\begin{verbatim}
var ci : ClassInfo;
var p : Person;
...
ci = p.getClassInfo();
var v : array(ClassMethodInfo)[];
v = ci.getMethods();
for i = 1 to v.getSize() - 1 do
  Out.writeln( v[i].getName() );
\end{verbatim}

    Instead of using the {\tt ClassInfo} hierarchy to get information
about classes, one can access information about individual objects,
which includes class objects:
\begin{verbatim}
var ac : Account;
...
var objInfo : AnyObjectInfo;
objInfo = ac.getInfo();

  // list the names of all instance variables of the object
var v : array(ObjectInstanceVariableInfo)[];
v = objInfo.getInstanceVariables();
for i = 0 to v.getSize() - 1 do
  Out.writeln( v[i].getName() );

var objMethodInfo : ObjectMethodInfo;
  // get info about method "getBalance" of object ac
objMethodInfo = objInfo.getMethod("getBalance");
  // calls method "getBalance" of object ac. The balance returned is printed
Out.writeln( objMethodInfo.invoke(nil) );
\end{verbatim}
Note the exceptions this code can generate are not considered.

\begin{figure}
\begin{verbatim}
MethodInfo
    ClassMethodInfo
    ObjectMethodInfo


InstanceVariableInfo
    ClassInstanceVariableInfo
    ObjectInstanceVariableInfo


AnyObjectInfo
    ObjectInfo
    ClassObjectInfo
\end{verbatim}
\caption{Hierarchy of the introspective reflection library}
\label{aoin}
\end{figure}

  The hierarchies of {\tt MethodInfo}, {\tt InstanceVariableInfo}, and
{\tt AnyObjectInfo} are shown in Figure~\ref{aoin}.

One can create and use an array whose element type will only be known
at run time without using the reflection library. This is possible because

all arrays are subclasses of class {\tt AnyArray}. See
the following code.

\begin{verbatim}
var any : AnyClassObject = integer;
var anArray : AnyArray;
  // creates a 10-element integer array
anArray = array(any)[].new(10);
anArray.set(5, 0); // anArray[0] = 5;
anArray.set(3, 1); // anArray[1] = 3;
Out.writeln( anArray.get(0) ); // print 5
\end{verbatim}
If {\tt any} does not refer to a class object at run time, an
exception will be thrown.
All the classes of the Introspective Reflection Library are described
in Appendix~\ref{irl}. Note  a normal programmer cannot create
objects of these classes.

One should take care in using the classes of the Introspective
Reflection Library because there are objects describing these same
classes. It is easy to take information on a class {\tt Person} for
information on the class of the object that describes class {\tt
Person}. As an example, the following code will write ``{\tt
ClassInfo}'', which is the name of the class of {\tt ci} at run time.
\begin{verbatim}
var ci : ClassInfo;

ci = Person.getAssociateClassInfo();
Out.writeln( ci.getClassInfo().getName() );
\end{verbatim}
The type of a variable may be specified with {\tt type}:\\
\verb@       // Person is a class@\\
\verb@     var p : type(Person);@\\
The type of {\tt p} is the set of method signatures of class object
{\tt Person}. The compiler will create an abstract class {\tt
Type\$Person} that has the same method signatures as class
object {\tt Person}. Then this class will be set as the type of {\tt
p}. This mechanism is necessary because at run time there is an object
of class {\tt VariableInfo} that describes variable {\tt p} and this
object has a method {\tt getType} that returns an object of {\tt
ClassInfo}. Since ``{\tt type(Person)}'' is not a class, it could not
be described by this {\tt ClassInfo} object and therefore the compiler
 replaces ``{\tt type(Person)}'' by a real class, {\tt Type\$Person}.

\newpage
\appendix






\chapter{The Green Language Grammar}
\label{grammar}

This section describes the language grammar. The reserved words
and symbols of the language are shown between `` and  ''.
Anything between \{ and \} can  be repeated zero or more times and
anything between [ and ]  is optional. The prefix Un means the union of two or
more rules. The program
must be analyzed by unfolding the rule
``Program''. 

There are two kinds of comments:
\bi \item anything between \verb@/*@ and \verb@*/@. Nested comments
are allowed. 
\item anything after \verb@//@ till the end of the line.
\ei
Of course, comments are not shown in the grammar.

The rule CharConst is any  character between a single
quote '. Escape characters are allowed.
 The rule Str is a string of zero or more characters
surrounded by
double quotes ". The double quote itself can be put in a string
preceeded by the backslash character \verb@\@. The symbol \verb@+@ is
used to concatenate two strings into a larger one. For example,\\
\verb@    "ABC" + "DEF"@\\
is the same as \verb@"ABCDEF"@. If one of the strings is not literal,
the operation is made at run time. 

A literal number can have a trailing letter defining its type:\\
\verb@     35b   // byte number@\\
\verb@      2i   // integer number@\\
There should be no space between the last digit and the letter.

All words that appear between quotes in the grammar are reserved Green
keywords. Besided those, the keyword {\tt subtypeOf} is
 reserved, which  may be used in the future to state a
class is {\it subtype} of another. If a class {\tt Teacher} is
declared as subtype of {\tt Person}, the compiler would enforce that {\tt
Teacher} defines all methods defined in {\tt Person}.

\newcommand{\p}[1] { \makebox[22ex][l]{#1} }
\newcommand{\q} { \makebox[6ex]{} }
\newcommand{\rr} { \makebox[30ex]{} }

\vspace{4ex}
\p{AddExpr} ::= MultExpr \{ AddOp MultExpr \}

\p{AddOp} ::= ``$+$'' \verb@|@ ``$-$''

\p{AndExpr} ::= RelExpr \{ ``and'' RelExpr \}

\p{ArrayInit} ::= ``\#''  ``('' OrExpr \{ ``,'' OrExpr \} ``)''

\p{ArrayInitExpr} ::= ArrayInit \verb@|@ Expr

\p{ArrayInitOrExpr} ::= ArrayInit \verb@|@ OrExpr

\p{AssertClause} ::= ``assert'' [ ``before'' OrExpr ``;'' ]\\
\rr \{ StatVarDec \}\\
\rr [ ``after'' OrExpr ``;'' ]\\
\rr ``end''

\p{Assignment} ::= PostfixExpr ``$=$'' ArrayInitOrExpr

\p{BasicType} ::=  ``boolean''  \verb@|@   ``byte'' \verb@|@   
``char'' \verb@|@ ``double''   \verb@|@  ``integer'' \verb@|@ ``long''
\verb@|@ ``real'' 

\p{BitExpr} ::= ShiftExpr \{ BitOp ShiftExpr \}

\p{BitOp} ::= ``\&'' \verb@|@  ``$|$''  \verb@|@ ``\verb@^@''

\p{Block} ::= ``begin'' StatementList ``end''

\p{BooleanConst} ::= ``true'' \verb@|@ ``false''

\p{BreakStat} ::= ``break''

\p{ByteConst} ::= Digit \{ Digit \} ``b''

\p{CaseStat} ::= ``case'' Expr ``of'' EachCase \{ EachCase \} \\
\rr  [ ``otherwise'' UnStatBlock ] ``end''

\p{Class} ::= Id [ ``('' TypeList ``)'' ]

\p{ClassType} ::= Id [ ``('' TypeList ``)'' ] \verb@|@ ``type''
``(''  Id ``)''

\p{ClassDec} ::=  ShellClassDec \verb@|@\\
\rr  [ClassObjDec] [``abstract''] [ ``reflective'' ]  ``class'' Id 
\\
\rr [ ClassParamList ]\\  
\rr [ ``subclassOf'' Class ]\\
\rr  UnPubPri \\
\rr ``end''


\p{ClassObjDec} ::=  ``object'' Id [ ClassParamList ] \\
\rr UnObjPubPri \\
\rr ``end''

\p{ClassParamList} ::= ``(''   ClassParamSpecif \{ ``,'' ClassParamSpecif \}  ``)''

\p{ClassParamSpecif} ::= Id [ ``:'' ClassType ]

\p{ConstExprCase} ::= BooleanConst \verb@|@ ByteConst \verb@|@
CharConst \verb@|@ IntegerConst\\
\rr  \verb@|@ LongConst \verb@|@ Class

\p{ConstDec} ::= ``const''  ConstItem \{ ``,'' ConstItem \} ``;''

\p{ConstExpr} ::= BooleanConst  \verb@|@  ByteConst \verb@|@ 
  CharConst  \verb@|@ DoubleConst\\
  \rr \verb@|@    IntegerConst  \verb@|@  LongConst \verb@|@  RealConst

\p{ConstExprList} ::= ConstExprCase \{ ``,'' ConstExprCase \}

\p{ConstItem} ::= Id [ ``:'' ConstType ] ``='' OrExpr

\p{ConstType} ::= BasicType \verb@|@ ``String''

\p{Digit} ::= ``0'' \verb@|@ ... \verb@|@ ``9''

\p{DoubleConst} ::= Digit \{ Digit \} ``.'' \{ Digit \} [ Scale ]  
``d'' 



\p{E} ::= ``E'' \verb@|@ ``e''

\p{EachCase} ::= ConstExprList ``:'' UnStatBlock 

\p{EnumDec} ::=  ``enum'' ``('' Id [ ``='' OrExp
] \{ ``,'' Id [ ``=''  OrExp ] \} ``)''

\p{ExceptionClause} ::= ``('' ``exception'' : ClassType ``)''

\p{Expr} ::= PostfixExpr ``='' Expr \verb@|@ OrExpr

\p{ExprList} ::= Expr \{ ``,'' Expr \}

\p{FormalParamDec} ::= IdList ``:'' [ ``...'' ] Type

\p{FormalParamDecList} ::= FormalParamDec \{ ``;'' FormalParamDec \}

\p{ForStat} ::= ``for'' Id ``='' OrExpr ``to'' OrExpr ``do'' UnStatBlock

\p{Id} ::= Letter \{ Letter \verb@|@ Digit \verb@|@ ``\_'' \}

\p{IdList} ::= Id \{ ``,'' Id \}

\p{IfStat} ::= ``if'' OrExpr ``then'' StatementList [ ``else'' StatementList ] ``endif''



\p{InitMethodDec} ::= ProcHeading

\p{InitStat} ::= PostfixExpr ``\#'' ``init'' ``('' \{ Expr \} ``)''

\p{InstVarDec} ::= IdList ``:'' TypeExt ``;''

\p{InstVarDecList} ::= ``var'' InstVarDec \{ InstVarDec \}

\p{IntegerConst} ::= Digit \{ Digit \} [ ``i'' ]

\p{IntegerConstValue} ::= IntegerConst \verb@|@ Id

\p{Letter} ::= ``A'' \verb@|@ ... \verb@|@ ``Z'' \verb@|@ ``a'' \verb@|@ ... \verb@|@ ``z''

\p{LocalDec} ::= ``var'' VarDec \{ VarDec \}

\p{LongConst} ::= Digit \{ Digit \} ``L''

\p{LoopStat} ::= ``loop'' StatementList ``end''

\p{MessageReceiver} ::=  ``super'' \verb@|@ ``self'' \verb@|@ OrExpr
\verb@|@ ``exception''

\p{MessageSend} ::= [ MessageReceiver ``.'' ] Id ``('' 
ExprList ``)''

\p{MethodDec} ::= ProcHeading [ AssertClause ]  [ LocalDec ] Block

\p{MultExpr} ::= BitExpr \{ MultOp  BitExpr \}

\p{MultOp} ::= ``$/$'' \verb@|@ ``$*$'' \verb@|@ ``\%''

\p{ObjPrivatePart} ::= ObjVarDecList \verb@|@ MethodDec \verb@|@
ConstDec \verb@|@ EnumDec

\p{ObjPublicPart} ::=  MethodDec \verb@|@
ConstDec \verb@|@ EnumDec

\p{ObjVarDecList} ::= ``var'' ObjVarDec \{ ObjVarDec \}

\p{ObjVarDec} ::= IdList ``:'' Type [ ``='' ArrayInitOrExpr ]  ``;''

\p{OrExpr} ::= XorExpr \{ ``or'' XorExpr \}

\p{PostfixExpr} ::= PrimaryExpr  \verb@|@ PostfixExpr ``['' Expr ``]''
  \verb@|@\\
\rr  MessageReceiver ``.'' Id [ ``('' \{ Expr \} ``)'' ]
  
\p{PrimaryExpr} ::= Id   
\verb@|@  ConstExpr  \verb@|@ Str  \verb@|@  ``('' Expr ``)''
\verb@|@ ``self'' \verb@|@ ``result'' \verb@|@ \\
\rr ``nil'' \verb@|@ BasicType

\p{PrivatePart} ::= InstVarDecList \verb@|@ MethodDec

\p{ProcHeading} ::= [ ``abstract'' ] ``proc'' Id ``('' [
FormalParamDecList ] ``)''\\
\rr  [ ExceptionClause ]
[ ``:'' Type ]

\p{Program} ::= ClassDec \{ ClassDec \}


\p{RealConst} ::= Digit \{ Digit \} ``.'' Digit \{ Digit \} [ Scale ]  [
``r'' ]

\p{Relation} ::= ``$==$'' \verb@|@ ``$<$'' \verb@|@ ``$>$'' \verb@|@ ``$<=$''
\verb@|@ ``$>=$'' \verb@|@  ``$<>$''

\p{RelExpr} ::= AddExpr [ Relation AddExpr ]

\p{RepeatStat} ::= ``repeat'' StatementList ``until'' Expr


\p{ReturnStat} ::= ``return'' [ Expr ]

\p{Scale} ::= E [ ``$+$'' \verb@|@ ``$-$'' ] Digit \{ Digit \}

\p{ShellClassDec} ::= ``shell'' ``class'' Id ``('' ClassType ``)'' [
``subclassOf'' Id ] \\
\rr UnPubPri \\
\rr ``end''

\p{ShiftExpr} ::= UnaryExpr [ ShiftOp UnaryExpr ]

\p{ShiftOp} ::= ``\verb@<<@'' \verb@|@ ``\verb@>>@''

\p{Statement} ::= Assignment ``;'' \verb@|@
 MessageSend ``;''  \verb@|@
 ``;'' \verb@|@ InitStat ``;'' \\
 \rr  \verb@|@ ReturnStat ``;'' 
  \verb@|@ IfStat \verb@|@ WhileStat  \verb@|@
 CaseStat  \verb@|@ StatVarDec ``;''\\
 \rr  \verb@|@ ForStat  \verb@|@ TryStat \verb@|@
  RepeatStat ``;''  \verb@|@ LoopStat  \verb@|@ BreakStat ``;''

\p{StatementList} ::= \{ Statement \}

\p{StatVarDec} ::= ``var'' Id  ``:'' Type  [ ``='' ArrayInitExpr ]

\p{TryStat} ::= ``try'' ``('' Expr ``)'' StatementList ``end''

\p{Type} ::= BasicType \verb@|@ ClassType \verb@|@ 
``array'' ``('' TypeOrExpr ``)''\\
\rr  ``['' [ IntegerConstValue ] ``]'' { ``['' [ IntegerConstValue ] ``]''
} \verb@|@ MessageSend 

\p{TypeExt} ::= [ ``@'' ] Type

\p{TypeList} ::= Type \{ ``,'' Type \}

\p{TypeOrExpr} ::= TypeExt \verb@|@ OrExpr



\p{UnaryExpr} ::= PostfixExpr  \verb@|@  UnaryOp UnaryExpr

\p{UnaryOp} ::=  ``\verb@~@''  \verb@|@ ``$+$'' \verb@|@ ``$-$''
\verb@|@   ``not''  \verb@|@ ``$++$'' \verb@|@ ``$--$''

\p{UnObjPubPri} ::= [ InitMethodDec ]\\
\rr [ ``public'' ``:'' \{ ObjPublicPart \} ]\\
  \rr [ ``private'' ``:'' \{ ObjPrivatePart \} ]

\p{UnPubPri} ::=  [ InitMethodDec ] \\
\rr [ ``public'' ``:'' \{ MethodDec \} ]\\
  \rr [ ``subclass'' ``:'' \{ MethodDec \} ]\\
  \rr [ ``private'' ``:'' \{ PrivatePart \} ]

\p{UnStatBlock} ::= Statement ``;''  \verb@|@ ``begin'' StatementList ``end''

\p{VarDec} ::= IdList ``:'' TypeExt ``;''


\p{WhileStat} ::= ``while'' Expr ``do'' UnStatBlock

\p{XorExpr} ::= AndExpr \{ ``xor'' AndExpr \}


\newpage

\chapter{The Introspective Reflection Library}
\label{irl}
\label{cirl}

\begin{figure}
\begin{verbatim}
Any
  AnyClassObject
  AnyClass
    AnyArray
    AnyClassArray

    ClassInfo
        ValueClassInfo
        RefClassInfo
            AbstractClassInfo
            ConcreteClassInfo
                NormalClassInfo
                ArrayClassInfo

    MethodInfo
        ClassMethodInfo
        ClassInitMethodInfo
        ObjectMethodInfo
        ObjectInitMethodInfo



    InstanceVariableInfo
        ClassInstanceVariableInfo
        ObjectInstanceVariableInfo


    AnyObjectInfo
        ObjectInfo
        ClassObjectInfo

    MethodBodyInfo
    AssertionInfo
    VariableInfo
      ParameterInfo
    ConstantInfo
    EnumInfo
    MethodCallInfo
    LiveLocalVariableInfo
      LiveParameterInfo
    CodeAnnotation
\end{verbatim}
\caption{Classes of the Introspective Reflection Library}
\label{irlg}
\end{figure}

This appendix presents the classes of Green introspective reflection
library. The class hierarchy and some built-in Green classes are
shown in Figure~\ref{irlg}. Programmers
cannot create objects from the classes described in this Chapter.

It would be very expensive to add reflective information to all Green
programs since most of them will not need it. Therefore we assume the
methods described in this chapter will only work if the programmer has
set some compiler options so that the compiler has added reflective
information to the object files.
If a method is called and the corresponding information is not
available, it will throw the unchecked exception\\
\verb@     NoReflectiveInfoException@\\

Before describing the classes, let us remember some methods of classes
{\tt Any} and {\tt AnyClass}. Class {\tt Any} has a method

\vspace*{3ex} \noindent \verb@getInfo() : AnyObjectInfo@

returns an object that describes the object {\tt self}.

\vspace*{3ex}
\noindent and class {\tt AnyClass} defines methods

To {\tt AnyClass} are added the methods

\vspace*{3ex} \noindent \verb@getClassInfo() : ClassInfo@

  returns an object that describes the class of the object.

\vspace*{3ex} \noindent \verb@getClassObject() : AnyClassObject@

returns the class object of the object.

\vspace*{5ex}
Class {\tt AnyValue} defines methods {\tt getInfo}, {\tt
getClassInfo}, and {\tt getClassObject} exactly as described
above. {\tt AnyValue} is the superclass of the basic classes ({\tt
char}, {\tt boolean}, etc).

\vspace*{5ex}
Class {\tt AnyClassObject} defines  methods

\vspace*{3ex} \noindent \verb@getAssociateClassInfo() : ClassInfo@

returns information about the class associated to this class object.

\vspace*{3ex} \noindent \verb@getInitMethod() : ObjectInitMethodInfo@

     returns object describing the {\tt init} method of the class
object (or {\tt nil} if there is no one).

\vspace*{5ex}
{\tt AnyClassObject} is an abstract class that is {\it supertype} of
all class objects. To every class object the compiler adds all methods
defined in {\tt AnyClassObject}. therefore every class object has
methods {\tt getInfo}, {\tt getAssociateClassInfo}, and {\tt
getInitMethod} for introspective reflection.

The example below show the use of these methods.
\begin{verbatim}
var p : Person;

p = Person.new();
if p.getClassObject() == Person and
   p.getClassInfo() == Person.getAssociateClassInfo()
then
    // this is executed
  Out.writeln("Ok !");
endif
\end{verbatim}

\vspace*{8ex}

\noindent {\Large \bf Classes of the Library}

\vspace*{8ex}

\noindent {\Large Class  {\tt ClassInfo}}

 represents information about a class. Parameterized, shell, and
extension classes are not really classes and no {\tt ClassInfo} object
describes them at run time. However, these classes are used to create
real classes. For example, the declaration\\
\verb@     var list : DS.List(char);@\\
makes the compiler create a class ``\verb@List$p$1$char@''. The ``{\tt
p}'' between the dollar signals means ``parameterized''. The ``{\tt
1}'' means there is one parameter. Every class
created by the compiler has a name composed by
\bi
\item a class name which may be a parameterized, shell, or extension
class. In this example, this name is ``{\tt List}'';
\item ``\verb@$code$@'' which indicates the nature of the class or
other information. For parameterized classes, ``{\tt code}'' is ``{\tt
p}''. Shell and extensions use ``{\tt sh}'' and ``{\tt ex}'';

\item ``\verb@$n$@''  in which {\tt n} is the number of parameters;

\item  another class name or information about the class. If more than
one name is needed, \verb@$$@ is used to separate them. For example, a
declaration \\
\verb@     var dict : DS.Dict(String, Person);@\\
causes the creation of a class ``\verb@Dict$p$2$String$$Person@''.

\ei

Note \verb@$@ cannot be used in Green identifiers. Therefore the
compiler created class names do not conflict with user class names.

There are several ways of implementing parameterized classes. One
option reuses the compiled code of the parameterized code for all
instantiations of the class. For example, the code for classes
\verb@DS.List(Person)@, \verb@DS.List(Figure)@, and
\verb@DS.List(Symbol)@ would be the same. Another implementation
technique would create a new class like \verb@List$p$1$Person@ for every
instantiation. Whenever the case, there will be one {\tt ClassInfo}
object for every instantiated parameterized class. In the previous
example, there would be a {\tt ClassInfo} object describing each of
the classes \verb@DS.List(Person)@, \verb@DS.List(Figure)@, and
\verb@DS.List(Symbol)@ even if their code is shared.

If code sharing is used, each method of {\tt List} will be shared
among three classes.
Suppose variable {\tt mi} refers to a {\tt ClassMethodInfo} object
that describes a shared method of {\tt List}. Considering method\\
\verb@     getDeclaringClass@\\
 of {\tt ClassMethodInfo} returns a {\tt
ClassInfo} object describing the class in which the method {\tt m} is,
what a
call\\
\verb@     mi.getDeclaringClass()@\\
would return~? The answer is: {\tt ClassMethodInfo} has no {\tt
getDeclaringClass} method. This illustrates one more example of the
usefulness of designing the whole Green language at once. We had time
to modify everything before releasing the language. We had opportunity
to make all the parts work together without unwanted side effects.

\vspace{2ex} \noindent Methods:

\vspace*{3ex} \noindent \verb@getName() : String@

returns the name of the class.

\vspace*{3ex} \noindent \verb@toString() : String@

returns the name of the class. This method may be modified by the user
to return
more data about the class.

\vspace*{3ex} \noindent \verb@isSupertypeOf( aSubtype : ClassInfo ) : boolean@

returns {\tt true} if the class described by {\tt aSubtype} is {\it
subtype} of the class described by {\tt self}.

Example:
\begin{verbatim}
var w : Window = Window.new();
  // TextWindow is subtype of Window
var tw : TextWindow = TextWindow.new();

if w.getClassInfo().isSupertypeOf( tw.getClassInfo() )
then
  // always executed
endif
\end{verbatim}

\vspace*{3ex} \noindent \verb@isSuperclassOf( aSubclass : ClassInfo ) : boolean@

returns {\tt true} if the class described by {\tt self} is {\it
superclass} of the class described by {\tt aSubclass}.

\vspace*{3ex} \noindent \verb@getSuperclass() : ClassInfo@

returns object describing the superclass. Returns {\tt nil} if there is
no superclass; that is, the class described by {\tt self} is {\tt AnyValue} or
{\tt Any}.

\vspace*{3ex} \noindent \verb@getInstanceVariables() : DS.Iter(ClassInstanceVariableInfo)@

returns an iterator with all instance variables of the class including
the inherited ones. The first  elements are the instance
variables of the class described by {\tt self}, then those of the
superclass and so on. There may  be two variables of different classes
with the same name.

\vspace*{3ex} \noindent \verb@getMethods() : DS.Iter(ClassMethodInfo)@\\

returns an iterator with objects describing the class methods. It includes
all the inherited methods from sections {\tt public}, {\tt subclass},
and {\tt private}. It does not include  the constructors ({\tt init}
methods) which are
considered special methods. Note these constructors are not the {\tt
new} methods of the class objects. They do not create objects of the
class.

\vspace*{3ex} \noindent \verb@getPublicMethods() : DS.Iter(ClassMethodInfo)@\\

returns an iterator with objects describing the public class
methods. This includes inherited methods but not methods overridden in
subclasses. That is,   if the receiver of message {\tt getPublicMethods}
represents a class {\tt B} that overrides method {\tt m} inherited from
its superclass {\tt A}, the iterator will return
only {\tt B::m}.

\vspace*{3ex} \noindent \verb@getInitMethods() : DS.Iter(ClassInitMethodInfo)@

returns an iterator with objects describing the class constructors,
the {\tt init} methods. This is the only method of {\tt ClassInfo}
that returns information on constructors.

\vspace*{3ex} \noindent \verb@getInstanceVariable(name : String) : ClassInstanceVariableInfo@

returns  object describing the instance variable whose name is passed
as parameter. The search is made in this class, then in the
superclass, and so on. The first instance variable found is
returned. Remember  a class may have an instance variable with name
equal to another instance variable of its superclass. If there is no
variable with this name, {\tt nil} is returned.

\vspace*{3ex} \noindent
\verb@getMethod( name : String;@\\
\verb@           paramTypes : array(ClassInfo)[] ) : DS.Iter(ClassMethodInfo)@\\

returns  an iterator with objects describing the  methods with names {\tt name} whose parameter
types are in array {\tt paramTypes}. The search is
made in this class, its superclass, and so on. If there is no
 method with this name and parameter types, {\tt nil} is returned. If
{\tt paramTypes} is {\tt nil}, {\tt getMethod} assumes method {\tt
name} has no parameter.  {\tt init} methods (constructors) are not considered by this
method.

\vspace*{3ex} \noindent
\verb@getMethod_v( name : String;@\\
\verb@             paramTypes : ... array(AnyClassObject)[] ) : DS.Iter(ClassMethodInfo)@\\

Similar to method {\tt getMethod} except that here {\tt paramTypes}
represents a variable number of parameters. Each parameter should be a
class object:
\begin{verbatim}
var acc : Account;
var  ai : ClassInfo;
...
acc.deposit(100.0);
ai = acc.getClassInfo();
var mi : ClassMethodInfo;
mi = ai.getMethod_v( "deposit", real );
Out.writeln( mi.getName() );   // prints "deposit"
\end{verbatim}

Again, constructors ({\tt init} methods) are not taken into account.

\vspace*{3ex} \noindent  \verb@getPublicMethod( name : String ) : DS.Iter(ClassMethodInfo)@

returns an iterator with all public methods named {\tt
name}. Inherited methods are included if they were not overridden in
subclasses. That is, if {\tt self} represents class {\tt B} that
overrides  method {\tt m} inherited from {\tt A}, only {\tt B::m} will
be returned by the iterator. {\tt init} methods are not taken into account.

\vspace*{3ex} \noindent \verb@getThisClassInstanceVariables() : DS.Iter(ClassInstanceVariableInfo)@

returns an iterator with objects describing all the instance variables
declared in this class --- inherited ones are not considered.

\vspace*{3ex} \noindent \verb@getThisClassMethods() : DS.Iter(ClassMethodInfo)@

returns an iterator with objects describing the methods declared in this
class. Neither inherited methods nor {\tt init} methods are taken into
account.

\vspace*{3ex} \noindent \verb@getThisClassPublicMethods() : DS.Iter(ClassMethodInfo)@

returns an iterator with objects describing the public methods declared in this
class. Inherited methods are not taken in account.

\vspace*{3ex} \noindent \verb@getAssociateClassObject() : AnyClassObject@

returns the class object corresponding to the class. So the
expression\\
\verb@     p.getClassInfo().getAssociateClassObject() == Person@\\
will be {\tt true} if object {\tt p} points to an object of class {\tt
Person}.

\vspace*{3ex} \noindent \verb@isClassOf( any : Any ) : boolean@

returns {\tt true} if {\tt any} is an object of the class described by
{\tt self} (the object that received this message). Note this method
returns {\tt false} if the class of {\tt any} is a subclass or subtype of
the class described by {\tt self}. The expression\\
\verb@     p.getClassInfo().isClassOf(p)@\\
is always {\tt true}.

\vspace*{3ex} \noindent \verb@isAbstract() : boolean@

returns {\tt true} if the class is abstract.

\vspace*{3ex} \noindent \verb@isReflective() : boolean@

returns {\tt true} if the class if reflective. See
Chapter~\ref{shells} for the definition of reflective classes.

\vspace*{3ex} \noindent \verb@getNotes() : DS.Iter(CodeAnnotation)@

returns an iterator with annotations on the class. See description of
{\tt CodeAnnotation}.

\vspace*{8ex}
\noindent{\Large Class {\tt ValueClassInfo}}  subclass of {\tt
ClassInfo}

represents information about a basic class as {\tt byte} or {\tt
char}. These classes inherit from {\tt AnyValue} which does not
inherit from anyone. Basic classes do not have instance
variables\footnote{Or at least the reflective library thinks so. The
implementation {\it could} be different !} or
constructors. The operators are considered methods. So, one could
write \\
\verb@     var ci : ClassInfo;@\\
\verb@     ci = 5.getClassInfo();@\\
\verb@     Out.writeln( ci.getMethod("+").next().@\\
\verb@         getParameters().next().getType().getName() );@\\
to write the name of the first parameter type (``{\tt integer}'') of method
``{\tt +}'' of the class of ``{\tt 5}'', which is ``{\tt integer}''
too.

\vspace{2ex} \noindent Methods:

none. The goal of this class is allow one to discover if a {\tt
ClassInfo} object describes a value class. This is made by a cast:
\begin{verbatim}
var vi : ValueClassInfo;
var ci : ClassInfo;
...
try(catch)
  vi = ValueClassInfo.cast(ci);
  ...
end
\end{verbatim}
If the cast succeeds, {\tt ci} pointed to a {\tt ValueClassInfo}
object.

\vspace*{8ex}
\noindent {\Large Class {\tt RefClassInfo} } subclass of {\tt
ClassInfo}

represents information about a reference class; that is, any class
that is not a built-in class like {\tt char} or {\tt integer}.

\vspace{2ex} \noindent Methods: none.

\vspace*{8ex}
\noindent {\Large Class {\tt AbstractClassInfo} } subclass of {\tt
RefClassInfo}

represents information about an abstract class.

\vspace{2ex} \noindent Methods: none.

\vspace*{8ex}
\noindent {\Large Class {\tt ConcreteClassInfo} } subclass of {\tt
RefClassInfo}

represents information about a non-abstract class.

\vspace{2ex} \noindent Methods: none.

\vspace*{8ex}
\noindent {\Large Class {\tt NormalClassInfo} } subclass of {\tt
ConcreteClassInfo}

represents information about a non-abstract and non-array class.

\vspace{2ex} \noindent Methods: none.

\vspace*{8ex}
\noindent {\Large Class {\tt ArrayClassInfo} } subclass of {\tt ConcreteClassInfo}

represents information about array classes.

\vspace{2ex} \noindent Methods:

\vspace*{3ex} \noindent \verb@getArrayElementClass() : ClassInfo@

returns  object describing the class of the array elements.

\vspace*{3ex} \noindent \verb@getNumberOfDimensions() : integer@

returns the number of array dimensions.

\vspace*{8ex}
\noindent {\Large Class {\tt AnyArray} subclass of {\tt Any}}

this class is superclass of all array classes.

\vspace{2ex} \noindent Methods:

\vspace*{3ex} \noindent
\verb@set( v : Any; i : integer )@\\
\verb@   ( exception : CatchAnyArrayException )@

set to {\tt v} the {\tt i$^{th}$} position of the array. The {\tt
exception} variable is used to throw exceptions, described in
Chapter~\ref{exceptions}. The methods
that follow work similarly with arrays of two or more dimensions.
These methods may throw the following exceptions:
\bi
\item {\tt TypeErrorException}. This exception is thrown if
there is a type error; that is, if
the array element type is {\tt T} and  the run-time type of
object {\tt v} is not a subtype of {\tt T};

\item {\tt TooManyDimensionsException}, which is thrown  if the array has
less dimensions than those specified. For example, if one uses the
next method {\tt set} with a one-dimensional array. There should have been
specified only only index, {\tt i} or {\tt j}.

\ei

\vspace*{3ex} \noindent
\verb@set( v : Any; i, j : integer )@\\
\verb@   ( exception : CatchAnyArrayException )@

\vspace*{3ex} \noindent \verb@set( v : Any; i, j, k : integer; others : ... array(integer)[] )@\\
\verb@   ( exception : CatchAnyArrayException )@

\vspace*{3ex} \noindent \verb@get( i : integer ) : Any@\\
\verb@   ( exception : CatchAnyArrayException )@

returns the element of the array index {\tt i}. The methods that
follow work similarly with arrays of two or more dimensions. The {\tt
get} methods may throw exception {\tt TooManyDimensionsException}.

\vspace*{3ex} \noindent \verb@get( i, j : integer ) : Any@\\
\verb@   ( exception : CatchAnyArrayException )@

\vspace*{3ex} \noindent \verb@get( i, j, k : integer, others : ... array(integer)[] )@\\
\verb@   ( exception : CatchAnyArrayException )@

\vspace*{3ex} \noindent \verb@getSize() : integer@

returns the array size.

\vspace*{3ex} \noindent \verb@toString() : String@

returns a string like ``{\tt array(T)[][]}'' if the array has two
dimensions and {\tt T} as the element type.

\vspace*{8ex}
\noindent {\Large Class {\tt MethodInfo}}

represents information about methods.

\vspace{2ex} \noindent Methods:

\vspace*{3ex} \noindent \verb@getName() : String@

returns the method name.

\vspace*{3ex} \noindent \verb@toString() : String@

returns a string with the method name and the parameter names and
types. If mi represents a method\\
\verb@     proc get( i, j : integer ) : Person@\\
then ``{\tt mi.toString()}'' would return the string\\
 ``\verb@get( i : integer; j : integer ) : Person@''\\

\vspace*{3ex} \noindent
\verb@getBodyInfo() : MethodBodyInfo@\\
\verb@           ( exception : CatchNoReflectiveBodyInfoException )@

returns  object describing the method local variables and
statements. Throws exception\\
 {\tt NoReflectiveBodyInfoException} if no
reflective information about the method body is available.

\vspace*{3ex} \noindent \verb@getVisibility() : integer@

return an integer \verb@MethodInfo.constructor_v@, \verb@MethodInfo.public_v@,
\verb@MethodInfo.subclass_v@, or \verb@MethodInfo.private_v@ telling
the method visibility.

\vspace*{3ex} \noindent \verb@getParameterTypes() : DS.Iter(ClassInfo)@

returns an iterator with all the parameter types.

\vspace*{3ex} \noindent \verb@getReturnType() : ClassInfo@

returns object describing  the method return type. If there is no
return type, nil is returned.

\vspace*{3ex} \noindent \verb@getParameters() : DS.Iter(ParameterInfo)@

returns an iterator with objects describing the method parameters in the
order they were declared.

\vspace*{3ex} \noindent \verb@getExceptionClass() : ClassInfo@

returns  object describing the class of the parameter {\tt exception}
of the method. If there is no
{\tt exception} parameter, {\tt nil} is returned.

\vspace*{3ex} \noindent \verb@getNotes() : DS.Iter(CodeAnnotation)@

returns an iterator with annotations on the method.

\vspace*{8ex}
\noindent {\Large Class {\tt ClassMethodInfo}} subclass of {\tt MethodInfo}

represents a method of a class. The  class in which the method is
cannot be discovered using the methods of {\tt ClassMethodInfo},
including the inherited ones from {\tt MethodInfo}. This was made on
purpose to allow software tools (or the compiler itself) to plug a
method to more than one class.

\vspace{2ex} \noindent Methods:

\vspace*{3ex} \noindent \verb@isAbstract() : boolean@

returns {\tt true} if this method is abstract.

\vspace*{3ex} \noindent
\verb@invoke( obj : Any; args : array(Any)[] ) : Any@\\
\verb@      ( exception : CatchInvokePackedException )@

if {\tt self} describes a method {\tt m}, {\tt invoke} would be
equivalent to\\
\verb@     obj.m( args[0], args[1], ... args[n] )@\\
in which {\tt n} is the index of the last {\tt args} object. This
method will throw exception {\tt PackedException} if  method {\tt m} has
thrown an exception. The exception thrown by {\tt m} is packed in
an exception object of class {\tt PackedException}.
If the arguments cannot be {\it cast} to the method arguments, or the
number of arguments is wrong, an exception {\tt
WrongParametersException} is thrown. If {\tt obj} does not have the
method described by {\tt self}, exception {\tt TypeErrorException} is
thrown.

Note method {\tt invoke} may be used to call a method of {\tt public},
{\tt subclass}, or {\tt private} sections. It will never call
constructor since constructors are described by class {\tt
ClassInitMethodInfo}.

\vspace*{3ex} \noindent
\verb@invoke_v( obj : Any; args : ... array(Any)[] ) : Any@\\
\verb@      ( exception : CatchInvokePackedException )@\\
the same as {\tt invoke} but with a variable number of parameters.

\vspace*{8ex}
\noindent {\Large Class {\tt ClassInitMethodInfo}} subclass of {\tt
MethodInfo}

represents an {\tt init} method of a class, a constructor. A
constructor cannot be called by the user outside its class. So, there is no {\tt
invoke} method here.

\vspace{2ex} \noindent Methods: none.

\vspace*{8ex}
\noindent {\Large Class {\tt ObjectMethodInfo}} subclass of {\tt MethodInfo}

represents a method of an object which may have a class or may be a
class object.

\vspace{2ex} \noindent Methods:

\vspace*{3ex} \noindent
\verb@invoke( args : array(Any)[] ) : Any@\\
\verb@      ( exception : CatchInvokePackedException )@

if {\tt self} describes a method {\tt m} of object {\tt obj}, {\tt
invoke} is equivalent to call {\tt m} on {\tt obj} using {\tt args} as
parameters. This
method will throw exception {\tt PackedException} if  method {\tt m} has
thrown an exception. The exception thrown by {\tt m} is packed in
an exception object of class {\tt PackedException}.

If the arguments cannot be cast to the method arguments or the number
of arguments is wrong, exception {\tt WrongParametersException} is thrown.

Then the following
situation may occur: class {\tt A} defines method {\tt m()} which is
overridden in a subclass {\tt B} of {\tt A}. Object {\tt objmi}
describes method {\tt A::m} but  {\tt objmi}
belongs to class {\tt B}. Then\\
\verb@     objmi.invoke(args)@\\
will call {\tt A::m}. Objects like {\tt objmi} will never be returned
by method {\tt getMethods} of {\tt AnyObjectInfo} (see ahead) which would return
object describing {\tt B::m} since {\tt objmi} is linked to a {\tt B}
object. However, the shell method {\tt interceptAll} uses objects like
{\tt objmi} in which the {\it method} and the {\it object} described
do not match.

\vspace*{3ex} \noindent
\verb@invoke_v( args : ...array(Any)[] ) : Any@\\
\verb@      ( exception : CatchInvokePackedException )@

the same as the previous {\tt invoke} but with a variable number of
parameters.

\vspace*{8ex}
\noindent {\Large Class {\tt ObjectInitMethodInfo}} subclass of {\tt
MethodInfo}

represents an {\tt init} method of a class object. A class-object
constructor cannot be called outside the class object. So, there is
not {\tt invoke} method here.

\vspace{2ex} \noindent Methods: none.

\vspace*{8ex}
\noindent {\Large Class {\tt InstanceVariableInfo}}

describes an instance variable.

\vspace{2ex} \noindent Methods:

\vspace*{3ex} \noindent \verb@getName() : String@

returns the name of the instance variable.

\vspace*{3ex} \noindent \verb@getType() : ClassInfo@

returns the instance variable type.

\vspace*{3ex} \noindent \verb@isExpanded() : boolean@

returns {\tt true} if the variable is expanded. That is, the variable
is declared with \verb!@! and obeys value semantics.

\vspace*{3ex} \noindent \verb@getNotes() : DS.Iter(CodeAnnotation)@

returns an iterator with annotations on the instance variable.

\vspace*{8ex}
\noindent {\Large Class {\tt ClassInstanceVariableInfo}} subclass of
{\tt InstanceVariableInfo}

describes an instance variable of a class.

\vspace{2ex} \noindent Methods:

\vspace*{3ex} \noindent
\verb@set( obj : Any; v : Any )@\\
\verb@   ( exception : CatchTypeErrorException )@

set the variable described by {\tt self} of object {\tt obj} to {\tt
v}.   Exception {\tt TypeErrorException} is thrown if
there is a type error; that is, if
the instance variable type is {\tt T} and  the run-time type of
object {\tt v} is not a subtype of {\tt T}. Or if {\tt obj} does not
have the instance variable described by {\tt self}.

\vspace*{3ex} \noindent
\verb@get( obj : Any ) : Any@\\
\verb@   ( exception : CatchTypeErrorException )@\\

return the value of the variable described by {\tt self} of the object
{\tt obj}. Exception {\tt TypeErrorException} is thrown if the object
{\tt obj} does not have the instance variable described by {\tt
self}.

\vspace*{8ex}
\noindent {\Large Class {\tt ObjectInstanceVariableInfo}} subclass of
{\tt InstanceVariableInfo}

describes a variable of an object.

\vspace{2ex} \noindent Methods:

\vspace*{3ex} \noindent \verb@set( v : Any )@\\
\verb@   ( exception : CatchTypeErrorException )@

sets the variable to {\tt v}. Exception {\tt TypeErrorException} is thrown if
there is a type error; that is, if
the  variable type is {\tt T} and  the run-time type of
object {\tt v} is not a subtype of {\tt T}.

\vspace*{3ex} \noindent \verb@get() : Any@

returns the variable value.

\vspace*{8ex}
\noindent {\Large Class {\tt AnyObjectInfo}}

describes an object.

\vspace{2ex} \noindent Methods:

\vspace*{3ex} \noindent \verb@getObject() : Any@

returns the object that {\tt self} describes.

\vspace*{3ex} \noindent
\verb@getInstanceVariables() : DS.Iter(ObjectInstanceVariableInfo)[]@

returns an iterator with objects describing all variables of the object
described by {\tt self}.

\vspace*{3ex} \noindent \verb@getMethods() : DS.Iter(ObjectMethodInfo)@

returns an iterator with objects describing all methods of the object
described by {\tt self}. If the object has a class, this includes all
inherited methods from {\tt public}, {\tt subclass}, and {\tt private}
sections. This does not include  the constructors.

If the object does not have a class (it is a class object), the
iterator will return objects for all public and private methods of the
class object. This excludes any methods {\tt init} of the class object.

\vspace*{3ex} \noindent
\verb@getPublicMethods() : DS.Iter(ObjectMethodInfo)@

returns an iterator with objects describing the public object
methods. If the object has a class, this includes inherited methods
but not overridden methods. That is, if {\tt self} represents an
object of class {\tt B} that overrides method {\tt m} inherited from
superclass {\tt A}, the iterator will return only {\tt B::m}.

If the object is a class object, the iterator yields all public
methods of the class object.

\vspace*{3ex} \noindent
\verb@getPublicMethod( name : String ) : DS.Iter(ObjectMethodInfo)@

returns an iterator with all public object methods named {\tt
name}. If the object has a class, inherited methods are included. But
only if they were not overridden in subclasses. That is, if {\tt self}
represents an object of class {\tt B} that overrides method {\tt m} inherited from
superclass {\tt A}, the iterator will return only {\tt
B::m}. Constructors are not taken into account.

\vspace*{3ex} \noindent \verb@getMethod( name : String; paramTypes : array(ClassInfo)[] ) : ObjectMethodInfo@

returns object describing the object method {\tt name} whose parameter
types are in array {\tt paramTypes}. If the object has a class, the
search for the method is made in this class, its superclass, and so
on. Public, subclass, and private methods are considered.

If the object is a class object, the search is made in the public and
private sections of the object. The constructor {\tt init()} is
not taken into account.

If not method is found, {\tt nil} is returned. If {\tt paramTypes} is
{\tt nil}, {\tt getMethod} assumes the method to be searched has no
parameters.

\vspace*{3ex} \noindent \verb@getMethod( name : String ) : ObjectMethodInfo@

returns  object describing  method {\tt name} of {\tt self}. If there is not
any method with this name, {\tt nil} is returned. The normal method
look-up is made to find the method. The first method found is returned.

\vspace*{3ex} \noindent \verb@getInstanceVariable( name : String ) : ObjectInstanceVariableInfo@

returns  object describing variable called {\tt name}. It is returned the
first variable found in a search from the object class to its
superclass, if the object has a class. If there is no variable with
this name, {\tt nil} is returned.

\vspace*{3ex} \noindent \verb@getTypeInfo() : ClassInfo@

returns  object describing a class that has the same type as the
object. The type of an object has the same definition as the type of a
class. This method returns an object describing
\bi
\item the object class if the object has a class;
\item an abstract class if the object is a class object. This abstract
class has the same type as the object.
\ei

\vspace*{8ex}
\noindent {\Large Class {\tt ObjectInfo}} subclass of {\tt AnyObjectInfo}

describes an object that has a class

\vspace{2ex} \noindent Methods:

No methods in this class. Used only to discriminate objects
that were created from a class from class objects that are the
classes.

\vspace*{8ex}
\noindent {\Large Class {\tt ClassObjectInfo}} subclass of {\tt AnyObjectInfo}

describes a class object.

\vspace{2ex} \noindent Methods:

\vspace*{3ex} \noindent {\tt getInitMethod() : ObjectInitMethodInfo}

returns an object describing the {\tt init} method of the class
object, if one exists.

\vspace*{3ex} \noindent {\tt getPublicConstants() :
DS.Iter(ConstantInfo)}

returns an iterator with objects describing  the public constants of
the class object, which does not include enumerated constants.

\vspace*{3ex} \noindent {\tt getConstants() : DS.Iter(ConstantInfo)}

returns an iterator with objects describing  all  constants of
the class object, which does not include enumerated constants.

\vspace*{3ex} \noindent {\tt getPublicEnumConstants() :
DS.Iter(EnumInfo)}

returns an iterator with objects describing the public enumerate
declarations of the class object.

\vspace*{3ex} \noindent {\tt getEnumConstants() : DS.Iter(EnumInfo)}

returns an iterator with objects describing all enumerate
declarations of the class object.

\vspace*{3ex} \noindent \verb@getNotes() : DS.Iter(CodeAnnotation)@

returns an iterator with annotations on the class object.

\vspace*{3ex} \noindent
\verb@new( args : array(Any)[] )@\\
\verb@   ( exception : CatchNewException ) : Any@

if {\tt coi} describes class {\tt Person}, a call  {\tt coi.new(args)} is
equivalent to \\
\verb@     Person.new( args[0], args[1], ... args[n] )@\\
in which {\tt n} is the last valid index of {\tt args}. The first new method that accepts the parameters is called.
This
method will throw exception {\tt PackedException} if  the class method
{\tt new} has
thrown an exception. The exception  is packed in
an exception object of class {\tt PackedException}.

Exception {\tt WrongParametersException}  will be thrown if no {\tt new} method of the
class object matches the parameters passed to the above method {\tt
new}. That is, the types or number of parameters is wrong.

Exception {\tt OutOfMemoryException} will be thrown if there is not
sufficient memory for creating the object.

\vspace*{3ex} \noindent
\verb@new_v( args : ... array(Any)[] )@\\
\verb@     ( exception : CatchNewException ) : Any@

the same as the previous method {\tt new} but with variable number of parameters.

\vspace*{8ex}
\noindent {\Large Class {\tt MethodBodyInfo}}

describes the assertions, local variables,   and the statements of a method.

\vspace{2ex} \noindent Methods:

\vspace*{3ex} \noindent \verb@getMethodInfo() : MethodInfo@

returns  object describing the method.

\vspace*{3ex} \noindent \verb@getAssertionInfo() : AssertionInfo@

returns  object describing the method assertions. If there is no
assertion, {\tt nil} is returned.

\vspace*{3ex} \noindent \verb@getLocalVariables() : DS.Iter(VariableInfo)@

returns an iterator with objects describing the method local variables.
This includes all local variables, including those declared inside the
{\tt begin}-{\tt end} block.

It is worth noting there would be problems if a block inside the {\tt
begin}-{\tt end} of a method introduced a new scope. Then a method
\begin{verbatim}
proc search() : boolean
  var i : integer;
  begin
  ...
  if i > 0
  then
    var i : integer;
    ...
  endif
  end
\end{verbatim}
would have two different variables with the
same name. At least confusing.  This example enforces what we have
said: it is better to design the whole language at once so
incompatibilities among the language elements can be elegantly avoided.

\vspace*{3ex} \noindent \verb@getStatements() : DS.Iter(StatementNo)@

returns an iterator with objects describing the method statements in the
order they appear in the method declaration. {\tt StatementNo} and
other classes related to the Abstract Syntax Tree of the compiler
will not be described in this report.

\vspace*{8ex}
\noindent {\Large Class {\tt AssertionInfo}}

describes the assertion section of a method. None of the following
methods reveals the method in which the assertion is.

\vspace*{3ex} \noindent \verb@getBeforeExpression() : ExprAST@

returns  object describing the expression following keyword {\tt
before} in the assertion clause. Class {\tt ExprAST} belongs
to the Abstract Syntax Tree of the compiler and is not described in
this report. If there is no {\tt before} expression, {\tt nil} is returned.

\vspace*{3ex} \noindent \verb@getAfterExpression() : ExprAST@

returns  object describing the expression following keyword {\tt
after} in the assertion clause. Class {\tt ExprAST} belongs
to the Abstract Syntax Tree of the compiler and is not described in
this report. If there is no {\tt after} expression, {\tt nil} is returned.

\vspace*{3ex} \noindent \verb@getStatVarDeclarations() : DS.Iter(StatVarDecNo)@

returns an iterator with objects describing the initialization of
variables after the {\tt before} expression. As an example, if the
assertion clause is
\begin{verbatim}
assertion
  before not full();
  var oldSize : integer = getSize();
  after oldSize == getSize() - 1;
end
\end{verbatim}
This method would return a one-element iterator with an object describing
variable {\tt oldSize} and the expression that is assigned to it,
``{\tt getSize()}''.

\vspace*{8ex}
\noindent {\Large Class {\tt VariableInfo}}

describes a local variable.

\vspace{2ex} \noindent Methods:

\vspace*{3ex} \noindent \verb@getName() : String@

returns the variable name.

\vspace*{3ex} \noindent \verb@getType() : ClassInfo@

returns the variable type.

\vspace*{4ex}

\vspace*{3ex} \noindent \verb@getDeclaringMethod() : MethodInfo@\\

returns object describing the method in which this local variable was
declared.

\vspace*{3ex} \noindent \verb@isExpanded() : boolean@

returns {\tt true} if the variable is expanded. That is, the variable
is declared with \verb!@! and obeys value semantics. Note parameters are
never expanded.

\vspace*{3ex} \noindent \verb@getNotes() : DS.Iter(CodeAnnotation)@

returns an iterator with annotations on the variable.

\vspace*{8ex}
\noindent {\Large Class {\tt ParameterInfo}} subclass of {\tt VariableInfo}

describes a method parameter.

\vspace{2ex} \noindent Methods:

\vspace*{3ex} \noindent \verb@setDeclaringMethod( MethodInfo declaringMethod )@

sets the method declaring this parameter.

\vspace*{3ex} \noindent \verb@isVariableNumber() : boolean@

returns {\tt true} if this parameter is an array used to make the
method accept a variable number of parameters, as {\tt v} in\\
\verb@     proc print( f : Font; v : ... array(Any)[] )@\\
Then if {\tt pi} describes parameter {\tt v} of this method, \\
\verb@     pi.isVariableNumber()@\\
would return {\tt true}.

\vspace*{8ex}
\noindent {\Large Class {\tt ConstantInfo}}

describes a constant.

\vspace{2ex} \noindent Methods:

\vspace*{3ex} \noindent \verb@getName() : String@

returns the variable name.

\vspace*{3ex} \noindent \verb@getType() : ClassInfo@

returns the variable type.

\vspace*{3ex} \noindent \verb@getVisibility() : integer@

return an integer  \verb@MethodInfo.public_v@,
or \verb@MethodInfo.private_v@ telling
the constant visibility.

\vspace*{3ex} \noindent \verb@getValue() : Any@

returns the constant value.

\vspace*{3ex} \noindent \verb@getNotes() : DS.Iter(CodeAnnotation)@

returns an iterator with annotations on the constant.

\vspace*{8ex}
\noindent {\Large Class {\tt EnumInfo}}

describes a declaration of enumerated constants.

\vspace{2ex} \noindent Methods:

\vspace*{3ex} \noindent \verb@getConstants() : DS.Iter(ConstantInfo)@

returns objects describing each of the enumerated constants.

\vspace*{3ex} \noindent \verb@getVisibility() : integer@

return an integer  \verb@MethodInfo.public_v@,
or \verb@MethodInfo.private_v@ telling
the enumeration visibility.

\vspace*{3ex} \noindent \verb@getNotes() : DS.Iter(CodeAnnotation)@

returns an iterator with annotations on the enumerate.

\vspace*{8ex}

\noindent To object {\tt Runtime} are added the following methods:

\vspace*{3ex} \noindent \verb@getClasses() : DS.Iter(ClassInfo)@

returns an iterator with objects describing all program classes.
Note shell, extension, and parameterized
classes are not real classes and no object describes them. However,
the compiler creates some classes based on these classes
that are returned by this method and can be searched using the following method.

\vspace*{3ex} \noindent \verb@searchForClass( name : String ) : ClassInfo@

returns  object describing class {\tt name}. The search is made among
all program classes. If no class is found, {\tt nil} is returned.

\vspace*{3ex} \noindent
\verb@getMethodCallStack() : DS.Stack(MethodCallInfo)@\\
\verb@                  ( exception : CatchNoReflectiveCallInfoException )@

returns a stack with one object for each method in the method call
stack. There
will always be at least one method in the stack. Exception {\tt
NoReflectiveCallInfoException} is thrown if the program was not
compiled with information about the run-time stack.

\vspace*{3ex} \noindent \verb@getCatchObjectStack() : DS.Stack(Catch)@

returns a stack with catch objects.

The following code
prints the name of all exceptions that can be caught by the active
catch objects, which are those returned by {\tt
getCatchObjectStack}. This method will always return a correct stack
at run time, even if the program was not compiled with reflective
information. An example of use of this method is given below.

\begin{verbatim}
var cmi : ClassMethodInfo;
var iterMethod : DS.Iter(ClassMethodInfo);
var iterStack  : DS.Iter(Catch) =
Runtime.getCatchObjectStack().getIter();

while iterStack.more() do
  begin
  iterMethod = iterStack.next().getClassInfo().getPublicMethods();
  while (cmi = iterMethod.next()) <> nil do
    if cmi.getName().equals("throw")
    then
      Out.writeln( cmi.getParameters().next().getType().getName() );
    endif
  end
\end{verbatim}

Note we used two different techniques to scan the elements of the
iterators.

  It is not possible to add or remove catch objects from the
stack. But one may attach a shell to a catch object of the stack. Thus
a {\tt throw} method of a shell may be called when an exception is
thrown.

\vspace*{3ex} \noindent
\verb@getMethodThrownException()@\\
\verb@                       (exception : CatchNoExceptionInfoException) : MethodInfo@


return information on the method that has thrown the last
exception. Exception\\
\verb@     NoExceptionInfoException@\\
 is thrown if the
program was not compiled to keep the information requested by this
method.

\vspace*{3ex} \noindent {\tt setCatchUnchecked( myCatch :
CatchUncheckedException )}

set the first reference  of the stack of catch objects to {\tt
myCatch}. Since the type of this object is {\tt
CatchUncheckedException}, it is able to treat all unchecked
exceptions. This works as follows. Before the program starts, a
default catch object is pushed into th stack of catch objects. This
object will be used to treat the unchecked exceptions of th program
(unless the program catches them itself). The {\tt setCatchUnchecked}
method just allow the user to change this default catch object.

\vspace*{3ex} \noindent {\tt getCatchUnchecked() :
CatchUncheckedException}

get the first object of the stack of catch objects.

\vspace*{8ex}
\noindent {\Large Class {\tt MethodCallInfo}}

describes a method call. The method is in the method call stack and
therefore its local variables and parameters are live.

\vspace{2ex} \noindent Methods:

\vspace*{3ex} \noindent \verb@getMethodInfo() : MethodInfo@

returns  object describing the method.

\vspace*{3ex} \noindent \verb@getLiveLocalVariables() : DS.Iter(LiveLocalVariableInfo)@

returns an iterator with objects describing the local variables of the
method, which are, of course, live.

\vspace*{3ex} \noindent \verb@getLiveParameters() : DS.Iter(LiveParameterInfo)@

returns an iterator with objects describing the method parameters, which
are, of course, live.

\vspace*{8ex}
\noindent {\Large Class {\tt LiveLocalVariableInfo}}

describes a local variable of a method that is currently active; that
is, the method is in the stack of called methods.

\vspace{2ex} \noindent Methods:

\vspace*{3ex} \noindent \verb@getVariableInfo() : VariableInfo@

returns information about the variable.

\vspace*{3ex} \noindent
\verb@set( v : Any )@\\
\verb@   ( exception : CatchTypeErrorException )@

sets the variable to {\tt v}. If the run time type of {\tt v} is not a
subtype of the type of the variable, exception {\tt
TypeErrorException} is thrown.

\vspace*{3ex} \noindent \verb@get() : Any@

returns the variable value.

\vspace*{8ex}
\noindent {\Large Class {\tt LiveParameterInfo}}  subclass of {\tt LiveLocalVariableInfo}

\vspace{2ex} \noindent Methods: none.

\vspace*{8ex}
\noindent {\Large Abstract Class {\tt CodeAnnotation}}

an annotation about a class, method, or variable. The object returned
by method {\tt getDescription} of this class should be {\it cast} to
some useful
type. An annotation gives further information which is not related to
the structure of the class, method, or variable. For example, an
object of {\tt CodeAnnotation} attached to a method could tell if the
method changes the object instance variables or calls other object
methods. If it does not, a debugger could allow the programmer to call
this read-only method in an Inspect window. This information is
available to the compiler and given to the programmer through a {\tt
CodeAnnotation} object.

A {\it final} class cannot be subclassed and has no {\it
subtype}. A {\it final} method cannot be overridden in a
subclass. Green has no {\tt final} keyword  to mark final classes and
methods. These classes and methods should be marked final by the
programming environment or through a special statement in the module
system.\footnote{Yet to be designed.} There should be a ``final'' code
annotation to each final class or method.

Objects of {\tt CodeAnnotation} can provide other information that
would only be available in the program documentation. For example, a
{\tt CodeAnnotation} object linked to a class {\tt Button} could tell
which methods are used to add and remove a listener to the button. A
button is a graphical component that may be pressed with the
mouse. When this occurs, a message ``{\tt
actionPerformed}''\footnote{We are using the names employed in Java
AWT.} is sent to each of the button listeners which are objects with a
{\tt actionPerformed} method.

Figures~\ref{ca1} and \ref{ca2} show examples of use of {\tt
CodeAnnotation} objects. In Figure~\ref{ca1}
 a test is made to discover if a method changes the
object instance variables or call other methods. The second example in
Figure~\ref{ca2}
prints the names of  methods of a class that are used to add and
remove listeners. It uses a class {\tt EventSourceInfo} which keeps
information on the methods.

\begin{figure}
\begin{verbatim}
var change : boolean;
var an     : CodeAnnotation;
var mi     : ObjectMethodInfo;
var iter   : DS.Iter(CodeAnnotation);


  //  mi will hold data on method get of object x
mi = x.getInfo().getMethod("get", nil);
iter = mi.getNotes();

if (an = iter.next()) <> nil  and
   an.getName().equals("changeState")
then
    // the name of the annotation is changeState
  try(HCatchAll)
    change = boolean.cast(an.getDescription());
    if not change
    then
      Out.writeln("Read only method");
    endif
  end
endif
\end{verbatim}
\caption{Test to discover if a method is read-only}
\label{ca1}
\end{figure}

\begin{figure}
\begin{verbatim}
class EventSourceInfo
  public:
      // only the method headers are shown
    proc getAddMethod() : ClassMethodInfo
    proc getRemoveMethod() : ClassMethodInfo
    ...
end

...

var an     : CodeAnnotation;
var aClass : AnyClassObject;
var esi    : EventSourceInfo
var ci     : ClassInfo;
var iter   : DS.Iter(CodeAnnotation);

ci = aClass.getAssociateClassInfo();
iter = ci.getNotes();

if (an = iter.next()) <> nil   and
   an.getName().equals("eventSource")
then
  try(HCatchAll)
    esi = EventSourceInfo.cast( an.getDescription() );
  end
  Out.writeln( "add method : "   , esi.getAddMethod().getName(), "\n",
               "remove method : ", esi.getRemoveMethod().getName() );
endif
\end{verbatim}
\caption{Prints the names of the methods to add and remove listeners}
\label{ca2}
\end{figure}

We hope the compile-time objects created by the programmer will be
able to add {\tt CodeAnnotation} objects to classes, methods, and
variables as the compiler itself.

\vspace{2ex} \noindent Methods:

\vspace*{3ex} \noindent {\tt getDescription() : Any}

returns an object describing the annotation.

\vspace*{3ex} \noindent {\tt getName() : String}

returns the annotation name.

\chapter{The Exception Library}

The predefined Green exceptions are defined below. The class hierarchy
is shown in Figure~\ref{hexc}. All exception classes redefine method
{\tt toString} to return a message explaining the exception.

\begin{figure}
\begin{verbatim}
Exception
  TypeErrorException
  WrongParametersException
  NotFoundException
  PackedException
  TooManyDimensionsException

  MetaException
    ClassNotInAllowedSetException
    NoShellException
    NoExtensionException

  UncheckedException
    StackOverflowException
    IllegalArrayIndexException
    OutOfMemoryException
    InternalErrorException
    MessageSendToNilException

    NoReflectiveInfoException
      NoReflectiveBodyInfoException
      NoReflectiveCallInfoException

    ArithmeticException
      DivisionByZeroException
      RealOverflowException
      RealUnderflowException

    AssertionException
      AssertionAfterException
      AssertionBeforeException
      AssertionCastCharException
      AssertionCastBooleanException
      AssertionCastByteException
      AssertionCastIntegerException
      AssertionCastLongException
      AssertionCastRealException
      AssertionCastDoubleException
\end{verbatim}
\caption{Hierarchy of predefined Green exceptions}
\label{hexc}
\end{figure}

\vspace*{8ex}
\noindent {\Large Class {\tt Exception}}

an abstract class that is superclass of all predefined exception
classes.

\vspace{2ex} \noindent Methods: none.

\vspace*{8ex}
\noindent {\Large Class {\tt TypeErrorException}} subclass of
{\tt Exception}

exception thrown by methods {\tt set} of class {\tt AnyArray}, {\tt
ClassInstanceVariableInfo},\\
 {\tt ObjectInstanceVariableInfo}, and {\tt
LiveLocalVariableInfo} if there is a type error.

Also thrown if a cast
to the type of a class failed. Then, if the cast \\
\verb@     rectangle = Rectangle.cast(figure);@\\
fails, exception {\tt TypeErrorException} will be thrown.

\vspace{2ex} \noindent Constructors:

\vspace*{3ex} \noindent \verb@init()@

\vspace{2ex} \noindent Methods: none.

\vspace*{8ex}
\noindent {\Large Class {\tt WrongParametersException}} subclass of
{\tt Exception}

this exception will be thrown by method {\tt new} of {\tt
ClassObjectInfo}  if no {\tt new} method of the
class object matches the parameters passed to the above method {\tt
new}. Other methods like {\tt invoke} of {\tt ClassMethodInfo} may
throw this exception for similar reasons.

\vspace*{8ex}
\vspace{2ex} \noindent Constructors:

\vspace*{3ex} \noindent \verb@init()@

\vspace{2ex} \noindent Methods: none

\vspace*{8ex}
\noindent {\Large Class {\tt NotFoundException}} subclass of
{\tt Exception}

usually this exception is thrown by search methods when the item
searched is not found.

\vspace{2ex} \noindent Constructors:

\vspace*{3ex} \noindent \verb@init()@

\vspace{2ex} \noindent Methods: none

\vspace*{8ex}
\noindent {\Large Class {\tt PackedException}} subclass of
{\tt Exception}

thrown by methods {\tt invoke} of classes {\tt
ObjectMethodInfo} and {\tt ClassMethodInfo} when the executed method
throws an exception that is packed by an object of {\tt
PackedException}.

\vspace{2ex} \noindent Constructors:

\vspace*{3ex} \noindent \verb@init( exc : Exception )@

packs the exception passed as parameter.

\vspace{2ex} \noindent Methods:

\vspace*{3ex} \noindent \verb@getException() : Exception@

returns the exception packed by the receiver object.

\vspace*{8ex}
\noindent {\Large Class {\tt TooManyDimensionsException}} subclass of
{\tt Exception}

used to signal that an array has less dimensions than those specified
in the parameters of method {\tt set} of {\tt AnyArray}.

\vspace{2ex} \noindent Constructors:

\vspace*{3ex} \noindent \verb@init( givenNumber, numDimensions : integer )@

the parameters specify the number of parameters given in the method
and the array dimension.

\vspace{2ex} \noindent Methods:

\vspace*{3ex} \noindent \verb@getArrayDimension() : integer@

returns the number of array dimensions.

\vspace*{3ex} \noindent \verb@getGivenNumber() : integer@

returns the number of elements used to index the array.

\vspace*{8ex}
\noindent {\Large Class {\tt UncheckedException}}  subclass of
{\tt Exception}

this abstract class is superclass of all unchecked Green
exceptions. These exceptions  need not to be caught by the program when thrown by the run-time system or
compile-created code.

\vspace{2ex} \noindent Constructors: none.

\vspace{2ex} \noindent Methods: none.

\vspace*{8ex}
\noindent {\Large Class {\tt StackOverflowException}} subclass of {\tt
UncheckedException}

thrown when there is no more  space for the run-time call stack.

\vspace{2ex} \noindent Constructors:

\vspace*{3ex} \noindent \verb@init()@

\vspace{2ex} \noindent Methods: none.

\vspace*{8ex}
\noindent {\Large Class {\tt IllegalArrayIndexException}} subclass
of {\tt UncheckedException}

thrown when an illegal index is used in an array.

\vspace{2ex} \noindent Constructors:

\vspace*{3ex} \noindent \verb@init( index : integer; theArray : AnyArray )@

{\tt index} was used to index array {\tt theArray}.

\vspace{2ex} \noindent Methods:

\vspace*{3ex} \noindent \verb@getIndex() : integer@

returns the index.

\vspace*{3ex} \noindent \verb@getArray() : AnyArray@

returns the array.

\vspace*{8ex}
\noindent {\Large Class {\tt OutOfMemoryException}} subclass of {\tt
UncheckedException}

thrown when there is no more dynamic memory.

\vspace{2ex} \noindent Constructors:

\vspace*{3ex} \noindent \verb@init()@

\vspace{2ex} \noindent Methods: none.

\vspace*{8ex}
\noindent {\Large Class {\tt InternalErrorException}}  subclass of {\tt
UncheckedException}

thrown when there is an internal error in the run-time system.

\vspace{2ex} \noindent Constructors:

\vspace*{3ex} \noindent \verb@init( s : String )@

{\tt s} is a description on the error.

\vspace{2ex} \noindent Methods:

\vspace*{3ex} \noindent \verb@getErrorString() : String@

returns the error string passed as parameter to the constructor.

\vspace*{8ex}
\noindent {\Large Class {\tt MessageSendToNilException}} subclass of {\tt
UncheckedException}

thrown when a message is sent to {\tt nil}.

\vspace{2ex} \noindent Constructors:

\vspace*{3ex} \noindent \verb@init()@

\vspace{2ex} \noindent Methods: none.

\vspace*{8ex}
\noindent {\Large Class {\tt NoReflectiveInfoException}} subclass of {\tt
UncheckedException}

abstract class used as a superclass to the exception classes used to
signal errors in the reflective system of Green.

\vspace{2ex} \noindent Constructors: none.

\vspace{2ex} \noindent Methods: none.

\vspace*{8ex}
\noindent {\Large Class {\tt NoReflectiveBodyInfoException}}
subclass of {\tt NoReflectiveInfoException}

thrown if there is no reflective information about the body of a
method.

\vspace{2ex} \noindent Constructors:

\vspace*{3ex} \noindent \verb@init()@

\vspace{2ex} \noindent Methods: none.

\vspace*{8ex}
\noindent {\Large Class {\tt NoReflectiveCallInfoException}} subclass
of {\tt NoReflectiveInfoException}

exception used to signal that there is no reflective information about
the run-time method call stack.

\vspace{2ex} \noindent Constructors:

\vspace*{3ex} \noindent \verb@init()@

\vspace{2ex} \noindent Methods: none.

\vspace*{8ex}
\noindent {\Large Class {\tt ArithmeticException}} subclass of {\tt
UncheckedException}

abstract class of all arithmetic exception classes.

\vspace{2ex} \noindent Constructors:

\vspace*{3ex} \noindent \verb@init()@

\vspace{2ex} \noindent Methods: none.

\vspace*{8ex}
\noindent {\Large Class {\tt DivisionByZeroException}} subclass of
{\tt ArithmeticException}

\vspace{2ex} \noindent Constructors:

\vspace*{3ex} \noindent \verb@init()@

\vspace{2ex} \noindent Methods: none

\vspace*{8ex}
\noindent {\Large Class {\tt RealOverflowException}} subclass of
{\tt ArithmeticException}

thrown by the run-time system in overflow  of real numbers.

\vspace{2ex} \noindent Constructors:

\vspace*{3ex} \noindent \verb@init()@

\vspace{2ex} \noindent Methods: none.

\vspace*{8ex}
\noindent {\Large Class {\tt RealUnderflowException}} subclass of
{\tt ArithmeticException}

thrown by the run-time system in underflow of real numbers.

\vspace{2ex} \noindent Constructors:

\vspace*{3ex} \noindent \verb@init()@

\vspace{2ex} \noindent Methods: none.

\vspace*{8ex}
\noindent {\Large Class {\tt AssertionException}} subclass of {\tt
1UncheckedException}

abstract superclass of the exception classes for assertions.

\vspace{2ex} \noindent Constructors: none

\vspace{2ex} \noindent Methods:

\vspace*{3ex} \noindent \verb@getMethodInfo() : MethodInfo@

returns an object describing the method in which the assertion is.

\vspace*{8ex}
\noindent {\Large Class {\tt   AssertionBeforeException}} subclass of
{\tt AssertionException}

\vspace{2ex} \noindent Constructors:

\vspace*{3ex} \noindent \verb@init( mi : MethodInfo )@

the {\tt before} part of the assertion clause of the method described by {\tt mi}
evaluated to {\tt false}.

\vspace{2ex} \noindent Methods: none.

\vspace*{8ex}
\noindent {\Large Class {\tt   AssertionAfterException}} subclass of
{\tt AssertionException}

\vspace{2ex} \noindent Constructors:

\vspace*{3ex} \noindent \verb@init( mi : MethodInfo )@

the {\tt after} part of the assertion clause of the method described by {\tt mi}
evaluated to {\tt false}.

\vspace{2ex} \noindent Methods: none.

\vspace*{8ex}

the classes that follow are used in casts among the basic classes. All
of them are subclasses of {\tt AssertionException}. An
object of, say, {\tt AssertionCastCharException} is thrown if a cast
from something to {\tt char} failed. The compiler should have some
option to disable the assertions of methods {\tt cast} of basic
classes.

\begin{verbatim}
AssertionCastCharException
AssertionCastBooleanException
AssertionCastByteException
AssertionCastIntegerException
AssertionCastLongException
AssertionCastRealException
AssertionCastDoubleException
\end{verbatim}
All of these classes are similar. As an example we describe class {\tt
AssertionCastCharException}.

\begin{verbatim}
class AssertionCastCharException subclassOf AssertionException
    proc init( p_originalValueClass : AnyClassObject; p_value : Any )
      begin
      originalValueClass = p_originalValueClass;
      value = p_value;
      end
  public:
    proc getOriginalValueClass() : AnyClassObject
      begin
      return originalValueClass;
      end
    proc getOriginalValue() : Any
      begin
      return value;
      end
  private:
    var originalValueClass : AnyClassObject;
        value : Any;
end
\end{verbatim}
{\tt originalValueClass} is the class of the object that could not
have been converted to {\tt char}. The value that could not have been
converted to {\tt char} was transformed into an
object of a wrapper class which is {\tt p\_value}.

\vspace*{8ex}
\noindent {\Large Class {\tt MetaException}} subclass of
{\tt Exception}

abstract superclass of all classes related to the meta level.

\vspace{2ex} \noindent Constructors: none.

\vspace{2ex} \noindent Methods: none.

\vspace*{8ex}
\noindent {\Large Class {\tt ClassNotInAllowedSetException}}
subclass of {\tt MetaException}

an object of this class is thrown if the program tries to attach a
shell to an object of a class not prepared for being used with
shells. That is, the class is not in the ``allowed set'' of the shell
class.
See Chapter~\ref{shells} for more information.

\vspace{2ex} \noindent Constructors:

\vspace*{3ex} \noindent \verb@init( value : Any )@

{\tt value} is the object one tried to attach a shell.

\vspace{2ex} \noindent Methods:

\vspace*{3ex} \noindent \verb@getValue() : Any@

returns the object one tried to attach to a shell.

\vspace*{8ex}
\noindent {\Large Class {\tt NoShellException}} subclass of {\tt
MetaException}

an object of this class is thrown if the program tries to remove a
shell from an object without a shell.

\vspace{2ex} \noindent Constructors:

\vspace*{3ex} \noindent \verb@init( any : Any )@

{\tt any} is the object from which the program tried to remove the shell.

\vspace{2ex} \noindent Methods:

\vspace*{3ex} \noindent \verb@getObject() : Any@

returns the object from which the program tried to remove a shell.

\vspace*{8ex}
\noindent {\Large Class {\tt NoExtensionException}}

an object of this class is thrown if the program tries to remove an
extension from a class without an attached extension.

\vspace{2ex} \noindent Constructors:

\vspace*{3ex} \noindent \verb@init( aClass : AnyClassObject )@

{\tt aClass} is the class from which the program tried to remove the extension.

\vspace{2ex} \noindent Methods:

\vspace*{3ex} \noindent \verb@getTheClassObject() : AnyClassObject@

returns the class object from which the program tried to remove an extension.

\chapter{The Main Green Classes}

This Chapter presents the main Green classes and objects along their
methods. Figure~\ref{greenh} shows a hierarchy of Green classes. In
this figure are also shown the array class ``\verb@array(char)[]@''
and some class objects like {\tt Runtime} and {\tt Out}. These objects
are preceded by \verb@*@ to indicate they are not classes. Each class
object is {\it subtype} of {\tt AnyClassObject}.
The container and stream hierarchies are shown in
Figure~\ref{cont2}. The hierarchy of the Introspective Reflection
Library is shown in Figure~\ref{irlg}.

This appendix have not been updated with the last modifications in
this report. Refer to the report itself to get precise information.

\begin{figure}
\begin{verbatim}
AnyValue
  char
  boolean
  byte
  integer
  long
  real
  double
Any
  AnyClassObject
    *In
    *Out
    *OutError
    *Screen
    *Storage
    *Runtime
    *Memory
    *char
    *boolean
    *byte
    *integer
    *long
    *real
    *double
  AnyClass
    Char
    Boolean
    Byte
    Integer
    Long
    Real
    Double
    AnyArray
      array(char)[]
    Nil
    String
    DynString
\end{verbatim}
\caption{The Green class hierarchy}
\label{greenh}
\end{figure}

\begin{figure}
\begin{verbatim}
Any
  AnyClass
    BasicStream
      InputStream
      OutputStream
      Stream
    CatchFileException
    Iter( T : Any )
    IterFilter( T : Any )
    Filter( T : Any )
    Command(T)
    Function(T)
    Container( T : Any )
      List( T : Any )
      DList( T : Any )
      Stack( T : Any )
      DStack( T : Any )
      Queue( T : Any )
      DQueue( T : Any )
      DoubleQueue( T : Any )
      DDoubleQueue( T : Any )
      Vector( T : Any )
    HashFunction( T : Any )
    Compose( T, U : Any )
    Dict( T, U : Any )
    DictIter( T, U : Any )
    IntegerSet
\end{verbatim}
\caption{Container and stream class hierarchy}
\label{cont2}
\end{figure}

\newpage

\vspace*{5ex}
\begin{tabular}{| |l ||} \hline
{\tt AnyValue} \\ \hline \hline
\verb@toString() : String@\\ \hline
\verb@getInfo() : AnyObjectInfo@\\ \hline
\verb@getClassInfo() : ClassInfo@\\ \hline
\verb@getClassObject() : AnyClassObject@\\ \hline

\end{tabular}

 \vspace*{5ex} \begin{tabular}{| |l ||} \hline

{\tt Any}\\ \hline \hline
\verb@toString() : String@\\ \hline
\verb@isObjectOf( aClass : AnyClassObject ) : boolean@\\ \hline
\verb@shallowClone() : Any@\\
\verb@            ( exception : CatchOutOfMemoryException )@\\ \hline
\verb@deepClone() : Any@\\
\verb@         ( exception : CatchOutOfMemoryException )@\\ \hline
\verb@shallowCopy( other : Any ) : boolean@\\ \hline
\verb@shallowEqual( other : Any ) : boolean@\\ \hline
\verb@deepEqual( other : Any ) : boolean@\\ \hline
\verb@getInfo() : AnyObjectInfo@\\ \hline
\verb@equals( other : Any ) : boolean@\\  \hline

\end{tabular}

\vspace*{5ex} \begin{tabular}{| |l ||} \hline

{\tt AnyClassObject subclassOf Any}\\ \hline \hline
\verb@toString() : String@\\ \hline
\verb@isObjectOf( aClass : AnyClassObject ) : boolean@\\ \hline
\verb@shallowClone() : Any@\\
\verb@            ( exception : CatchOutOfMemoryException )@\\ \hline
\verb@deepClone() : Any@\\
\verb@         ( exception : CatchOutOfMemoryException )@\\ \hline
\verb@shallowCopy( other : Any ) : boolean@\\ \hline
\verb@shallowEqual( other : Any ) : boolean@\\ \hline
\verb@deepEqual( other : Any ) : boolean@\\ \hline
\verb@equals( other : Any ) : boolean@\\ \hline
\verb@getInfo() : AnyObjectInfo@\\ \hline
\verb@getAssociateClassInfo() : ClassInfo@\\ \hline
\verb@getInitMethod() : ObjectMethodInfo@\\ \hline

\end{tabular}

\vspace*{5ex}

\begin{tabular}{| |l ||} \hline
{\tt AnyClass subclassOf Any}\\ \hline \hline
\verb@getClassInfo() : ClassInfo@\\ \hline
\verb@getClassObject() : AnyClassObject@\\ \hline

\end{tabular}

\newpage

\vspace*{5ex} \begin{tabular}{| |l ||} \hline
{\tt object In } \\ \hline \hline
\verb@readCh()      : char @ \\ \hline
\verb@readByte()    : byte @ \\ \hline
\verb@readInteger() : integer @ \\ \hline
\verb@readLong()    : long @ \\ \hline
\verb@readReal()    : real @ \\ \hline
\verb@readDouble()  : double @ \\ \hline
\verb@readString()  : String @ \\ \hline
\verb@readLine()    : String @ \\ \hline
\end{tabular}

  \vspace*{5ex} \begin{tabular}{| |l ||} \hline
{\tt object Out } \\ \hline \hline
\verb@write( v : ... array(Any)[] ) @ \\ \hline
\verb@writeln( v : ... array(Any)[] ) @ \\ \hline
\end{tabular}

  \vspace*{5ex} \begin{tabular}{| |l ||} \hline
{\tt object OutError } \\ \hline \hline
\verb@write( v : ... array(Any)[] ) @ \\ \hline
\verb@writeln( v : ... array(Any)[] ) @ \\ \hline
\end{tabular}

\vspace*{5ex} \begin{tabular}{| |l ||} \hline
{\tt object Storage } \\ \hline \hline
\verb@removeFile( fileName : String ) : boolean @ \\ \hline
\verb@renameFile( oldName, newName : String ) : boolean @ \\ \hline
\verb@openFile( fileName : String ) : integer @ \\ \hline
\verb@closeFile( fd : integer ) @ \\ \hline
\verb@read( fd, n : integer; in : array(byte)[] ) @ \\ \hline
\verb@write( fd, n : integer; out : array(byte)[] ) @ \\ \hline
\verb@getError() : integer @ \\ \hline
\end{tabular}

\vspace*{5ex} \begin{tabular}{| |l ||} \hline
{\tt object Runtime} \\ \hline \hline
\verb@exit( errorCode : integer )@\\ \hline
\verb@putAtEndList( f : Function )@\\ \hline
\verb@getClasses() : DS.Iter(ClassInfo)@\\ \hline
\verb@searchForClass( name : String ) : ClassInfo@\\ \hline
\verb@getMethodCallStack() : DS.Stack(MethodCallInfo)@\\
\verb@                  ( exception : CatchNoReflectiveCallInfoException )@\\ \hline
\verb@getCatchObjectStack() : DS.Stack(Catch)@\\ \hline

\end{tabular}

\newpage

\vspace*{5ex} \begin{tabular}{| |l ||} \hline
{\tt object Memory } \\ \hline \hline
\verb@sizeLargestBlock() : long @ \\ \hline
\verb@sizeFreeMemory() : long @ \\ \hline
\verb@doGarbageCollection() @ \\ \hline
\verb@collectionOn() @ \\ \hline
\verb@collectionOff() @ \\ \hline
\end{tabular}

\vspace*{5ex} \begin{tabular}{| |l ||} \hline
{\tt object char } \\ \hline \hline
\verb@getSizeInBits() : integer @ \\ \hline
\verb@getSize() : integer @ \\ \hline
\verb@cast( any : AnyClass ) @ \\
\verb@    ( exception : CatchTypeErrorException ) : char@\\ \hline
\verb@cast( value : byte ) : char @ \\ \hline
\verb@cast( value : integer ) : char @ \\ \hline
\verb@castOk( any : AnyClass ) : boolean @ \\ \hline
\verb@castOk( value : byte ) : boolean @ \\ \hline
\verb@castOk( value : integer ) : boolean @ \\ \hline
\verb@getMinValue() : char @ \\ \hline
\verb@getMaxValue() : char  @ \\ \hline
\verb@getMaxIntegerChar() : integer @ \\ \hline
\verb@getMinIntegerChar() : integer @ \\ \hline
\end{tabular}

 \vspace*{5ex} \begin{tabular}{| |l ||} \hline
{\tt object boolean } \\ \hline \hline
\verb@getSizeInBits() : integer @ \\ \hline
\verb@getSize() : integer @ \\ \hline
\verb@cast( any : AnyClass ) @ \\
\verb@    ( exception : CatchTypeErrorException ) : boolean@\\ \hline
\verb@castOk( any : AnyClass ) : boolean @ \\ \hline
\verb@cast( value : integer ) : boolean @ \\ \hline
\verb@cast( value : byte ) : boolean @ \\ \hline
\verb@getMinValue() : boolean @ \\ \hline
\verb@getMaxValue() : boolean @ \\ \hline
\end{tabular}

\newpage

 \vspace*{5ex} \begin{tabular}{| |l ||} \hline
{\tt object byte } \\ \hline \hline
\verb@getSizeInBits() : integer @ \\ \hline
\verb@getSize() : integer @ \\ \hline
\verb@cast( any : AnyClass ) @ \\
\verb@    ( exception : CatchTypeErrorException ) : byte@\\ \hline
\verb@cast( value : boolean ) : byte @ \\ \hline
\verb@cast( value : char ) : byte @ \\ \hline
\verb@cast( value : integer ) : byte @ \\ \hline
\verb@cast( value : long ) : byte @ \\ \hline
\verb@cast( value : real ) : byte @ \\ \hline
\verb@cast( value : double ) : byte @ \\ \hline
\verb@castOk( any : AnyClass ) : boolean @ \\ \hline
\verb@castOk( value : integer ) : boolean @ \\ \hline
\verb@castOk( value : long ) : boolean @ \\ \hline
\verb@castOk( value : real ) : boolean @ \\ \hline
\verb@castOk( value : double ) : boolean @ \\ \hline
\verb@getMinValue() : byte @ \\ \hline
\verb@getMaxValue() : byte @ \\ \hline
\end{tabular}

\vspace*{5ex} \begin{tabular}{| |l ||} \hline
{\tt object integer } \\ \hline \hline
\verb@getSizeInBits() : integer @ \\ \hline
\verb@getSize() : integer @ \\ \hline
\verb@cast( any : AnyClass ) @ \\
\verb@    ( exception : CatchTypeErrorException ) : integer@\\ \hline

\verb@cast( value : char ) : integer @ \\ \hline
\verb@cast( value : boolean ) : integer @ \\ \hline
\verb@cast( value : byte ) : integer @ \\ \hline
\verb@cast( value : long ) : integer @ \\ \hline
\verb@cast( value : real ) : integer @ \\ \hline
\verb@cast( value : double ) : integer @ \\ \hline
\verb@castOk( any : AnyClass ) : integer @ \\ \hline
\verb@castOk( value : double ) : boolean @ \\ \hline
\verb@castOk( value : real ) : boolean @ \\ \hline
\verb@castOk( value : long ) : boolean @ \\ \hline
\verb@getMinValue() : integer @ \\ \hline
\verb@getMaxValue() : integer @ \\ \hline
\end{tabular}

\newpage

\vspace*{5ex} \begin{tabular}{| |l ||} \hline
{\tt object long } \\ \hline \hline
\verb@getSizeInBits() : integer @ \\ \hline
\verb@getSize() : integer @ \\ \hline
\verb@cast( any : AnyClass ) @ \\
\verb@    ( exception : CatchTypeErrorException ) : long@\\ \hline
\verb@cast( value : byte ) : long @ \\ \hline
\verb@cast( value : integer ) : long @ \\ \hline
\verb@cast( value : real ) : long @ \\ \hline
\verb@cast( value : double ) : long @ \\ \hline
\verb@castOk( any : AnyClass ) : boolean @ \\ \hline
\verb@castOk( value : double ) : boolean @ \\ \hline
\verb@castOk( value : real ) : boolean @ \\ \hline
\verb@getMinValue() : long @ \\ \hline
\verb@getMaxValue() : long @ \\ \hline
\end{tabular}

 \vspace*{5ex} \begin{tabular}{| |l ||} \hline
{\tt object real } \\ \hline \hline
\verb@getSizeInBits() : integer @ \\ \hline
\verb@getSize() : integer @ \\ \hline
\verb@cast( any : AnyClass ) @ \\ \hline
\verb@    ( exception : CatchTypeErrorException ) : real@\\ \hline
\verb@cast( value : byte ) : real @ \\ \hline
\verb@cast( value : integer ) : real @ \\ \hline
\verb@cast( value : long ) : real @ \\ \hline
\verb@cast( value : double ) : real @ \\ \hline
\verb@castOk( any : AnyClass ) : boolean @ \\ \hline
\verb@castOk( value : double ) : boolean @ \\ \hline
\verb@getRadix() : integer @ \\ \hline
\verb@getRounds() : integer @ \\ \hline
\verb@getPrecision() : integer @ \\ \hline
\verb@getEpsilon() : real @ \\ \hline
\verb@getMantDig() : integer @ \\ \hline
\verb@getMinValue() : real @ \\ \hline
\verb@getMaxValue() : real @ \\ \hline
\verb@getMaxExp() : real  @ \\ \hline
\verb@getMinExp() : real @ \\ \hline

\end{tabular}

\newpage

\vspace*{5ex} \begin{tabular}{| |l ||} \hline
{\tt object double } \\ \hline \hline
\verb@getSizeInBits() : integer @ \\ \hline
\verb@getSize() : integer @ \\ \hline
\verb@cast( any : AnyClass ) @ \\ \hline
\verb@    ( exception : CatchTypeErrorException ) : double@\\ \hline
\verb@cast( value : byte ) : double @ \\ \hline
\verb@cast( value : integer ) : double @ \\ \hline
\verb@cast( value : long ) : double @ \\ \hline
\verb@cast( value : real ) : double @ \\ \hline
\verb@castOk( any : AnyClass ) : double @ \\ \hline
\verb@getRadix() : integer @ \\ \hline
\verb@getRounds() : integer @ \\ \hline
\verb@getPrecision() : integer @ \\ \hline
\verb@getEpsilon() : double @ \\ \hline
\verb@getMantDig() : integer @ \\ \hline
\verb@getMinValue() : double @ \\ \hline
\verb@getMaxValue() : double @ \\ \hline
\verb@getMaxExp() : double  @ \\ \hline
\verb@getMinExp() : double @ \\ \hline
\end{tabular}

 \vspace*{5ex} \begin{tabular}{| |l ||} \hline
{\tt AnyClass} \\ \hline \hline
\verb@getClassInfo() : ClassInfo@\\ \hline
\verb@getClassObject() : AnyClassObject@\\ \hline
\end{tabular}

\vspace*{5ex} \begin{tabular}{| |l ||} \hline
{\tt Char} \\ \hline \hline
\verb@init( value : char ) @ \\ \hline
\verb@get() : char @ \\ \hline
\end{tabular}

\vspace*{5ex} \begin{tabular}{| |l ||} \hline
{\tt Boolean} \\ \hline \hline
\verb@init( value : boolean ) @ \\ \hline
\verb@get() : boolean @ \\ \hline
\end{tabular}

\newpage

 \vspace*{5ex} \begin{tabular}{| |l ||} \hline
{\tt Byte} \\ \hline \hline
\verb@init( value : byte ) @ \\ \hline
\verb@get() : byte @ \\ \hline
\end{tabular}

\vspace*{5ex} \begin{tabular}{| |l ||} \hline
{\tt Integer} \\ \hline \hline
\verb@init( value : integer ) @ \\ \hline
\verb@get() : integer @ \\ \hline
\end{tabular}

\vspace*{5ex} \begin{tabular}{| |l ||} \hline
{\tt Long} \\ \hline \hline
\verb@init( value : long ) @ \\ \hline
\verb@get() : long @ \\ \hline
\end{tabular}

\vspace*{5ex} \begin{tabular}{| |l ||} \hline
{\tt Real} \\ \hline \hline
\verb@init( value : real ) @ \\ \hline
\verb@get() : real @ \\ \hline
\end{tabular}

\vspace*{5ex} \begin{tabular}{| |l ||} \hline
{\tt Double} \\ \hline \hline
\verb@init( value : double ) @ \\ \hline
\verb@get() : double @ \\ \hline
\end{tabular}

\vspace*{5ex} \begin{tabular}{| |l ||} \hline

{\tt AnyArray}\\ \hline \hline
\verb@set( v : Any; i : integer )@\\
\verb@   ( exception : CatchAnyArrayException )@\\ \hline
\verb@set( v : Any; i, j : integer )@\\
\verb@   ( exception : CatchAnyArrayException )@\\ \hline
\verb@set( v : Any; i, j, k : integer; others : ... array(integer)[] )@\\
\verb@   ( exception : CatchAnyArrayException )@\\ \hline
\verb@get( i : integer ) : Any@\\
\verb@   ( exception : CatchAnyArrayException )@\\ \hline
\verb@get( i, j : integer ) : Any@\\
\verb@   ( exception : CatchAnyArrayException )@\\ \hline
\verb@get( i, j, k : integer, others : ... array(integer)[] )@\\
\verb@   ( exception : CatchAnyArrayException )@\\ \hline
\verb@getSize() : integer@\\ \hline
\verb@toString() : String@\\ \hline

\end{tabular}

\newpage

\vspace*{5ex} \begin{tabular}{| |l ||} \hline
{\tt array(char)[]} \\ \hline \hline

\verb@getSize() : integer@\\ \hline
\verb@init( first, second, third, ... : integer )@\\ \hline
\verb@getIter() : DS.Iter(T) @\\ \hline
\verb@forEach( f : Function(T) )@\\ \hline
\verb@replaceBy( cmd : Command(T) )@\\ \hline
\verb@collect( f : Filter(T) ) : array(T)[]@\\ \hline
\verb@remove( f : Filter(T) )@\\ \hline
\verb@reset( up : boolean )@\\ \hline
\verb@more() : boolean@\\ \hline
\verb@next() : T@\\ \hline

\end{tabular}

\vspace*{5ex} \begin{tabular}{| |l ||} \hline
{\tt Nil}\\ \hline \hline
no methods defined in this class \\ \hline
\end{tabular}

\newpage

\vspace*{5ex} \begin{tabular}{| |l ||} \hline
{\tt String} \\ \hline \hline
\verb@get( i : integer ) : char @ \\ \hline
\verb@getIter() : DS.Iter(char) @ \\ \hline
\verb@cmp( other : String ) : integer @ \\ \hline
\verb@cmpIgnoreCase( other : String ) : integer @ \\ \hline
\verb@newConcat( other : String )@ \\
\verb@         ( exception : CatchOutOfMemoryException ) : String@\\ \hline
\verb@tocharArray( copyto : array(char)[] )   @ \\
\verb@           ( exception : CatchOutOfMemoryException )@\\ \hline
\verb@tocharArray( copyto : array(char)[]; i : integer )  @ \\
\verb@           ( exception : CatchOutOfMemoryException )@\\ \hline
\verb@tobyteArray( copyto : array(byte)[] )  @ \\
\verb@           ( exception : CatchOutOfMemoryException )@\\ \hline
\verb@tobyteArray( copyto : array(byte)[]; i : integer )  @ \\
\verb@           ( exception : CatchOutOfMemoryException )@\\ \hline
\verb@getSize() : integer @ \\ \hline
\verb@newToLowerCase()  @ \\
\verb@              ( exception : CatchOutOfMemoryException ) : String@\\ \hline
\verb@newToUpperCase()  @ \\
\verb@              ( exception : CatchOutOfMemoryException ) : String@\\ \hline
\verb@getSubset( from, to2 : integer )  @ \\
\verb@         ( exception : CatchOutOfMemoryException ) : String@\\ \hline
\verb@search( s : String ) : integer @ \\ \hline
\verb@init( s : String ) @ \\ \hline
\verb@hashCode() : integer @ \\ \hline
\verb@tobyte() : byte @ \\ \hline
\verb@tointeger() : integer @ \\ \hline
\verb@tolong() : long @ \\ \hline
\verb@toreal() : real @ \\ \hline
\verb@todouble() : double @ \\ \hline
\verb@toDynString() @ \\
\verb@           ( exception : CatchOutOfMemoryException ) : DynString@\\ \hline
\end{tabular}

\newpage

\vspace*{5ex} \begin{tabular}{| |l ||} \hline
{\tt DynString} \\ \hline \hline
\verb@init( s : String ) @ \\ \hline
\verb@init( s : DynString ) @ \\ \hline
\verb@get( i : integer ) : char @ \\ \hline
\verb@getIter() : DS.Iter(char) @ \\ \hline
\verb@cmp( other : DynString ) : integer @ \\ \hline
\verb@cmpIgnoreCase( other : DynString ) : integer @ \\ \hline
\verb@concat( other : String )  @ \\
\verb@      ( exception : CatchOutOfMemoryException )@\\ \hline
\verb@tocharArray( copyto : array(char)[] )   @ \\
\verb@      ( exception : CatchOutOfMemoryException )@\\ \hline
\verb@tocharArray( copyto : array(char)[]; i : integer )  @ \\
\verb@      ( exception : CatchOutOfMemoryException )@\\ \hline
\verb@tobyteArray( copyto : array(byte)[] )  @ \\
\verb@      ( exception : CatchOutOfMemoryException )@\\ \hline
\verb@tobyteArray( copyto : array(byte)[]; i : integer )  @ \\
\verb@      ( exception : CatchOutOfMemoryException )@\\ \hline
\verb@getSize() : integer @ \\ \hline
\verb@toLowerCase()  @ \\ \hline
\verb@toUpperCase()  @ \\ \hline
\verb@getSubset( from, to2 : integer )  @ \\ \hline
\verb@         ( exception : CatchOutOfMemoryException ) : DynString@\\ \hline
\verb@search( s : DynString ) : integer @ \\ \hline
\verb@hashCode() : integer @ \\ \hline
\verb@tobyte() : byte @ \\ \hline
\verb@tointeger() : integer @ \\ \hline
\verb@tolong() : long @ \\ \hline
\verb@toreal() : real @ \\ \hline
\verb@todouble() : double @ \\ \hline
\verb@removeSpaceBegin() @ \\ \hline
\verb@removeSpaceEnd() @ \\ \hline
\verb@toString() @ \\
\verb@      ( exception : CatchOutOfMemoryException ) : String@\\ \hline
\verb@prepend( toadd : DynString ) @ \\
\verb@       ( exception : CatchOutOfMemoryException )@\\ \hline
\verb@removeAllCh( ch : char ) : boolean @ \\ \hline
\verb@remove( i : integer ) @ \\ \hline
\verb@insert( i : integer; ch : char )  @ \\
\verb@      ( exception : CatchOutOfMemoryException )@\\ \hline
\verb@add( i : integer; ch : char ) @ \\
\verb@      ( exception : CatchOutOfMemoryException )@\\ \hline
\verb@add( ch : char ) @ \\
\verb@      ( exception : CatchOutOfMemoryException )@\\ \hline
\end{tabular}

\newpage

\vspace*{5ex} \begin{tabular}{| |l ||} \hline
{\tt object Meta} \\ \hline \hline
\verb@attachShell( any : IdentAST; exp : ExprAST ) @ \\ \hline
\verb@removeShell( any : IdentAST ) @ \\ \hline
\verb@attachExtension( aClass : ClassNo; dynExt : ExtensionClassNo ) @ \\ \hline
\verb@removeExtension( aClass : ClassNo ) @ \\ \hline
\end{tabular}

\vspace*{5ex} \begin{tabular}{| |l ||} \hline
{\tt BasicStream} \\ \hline \hline
\verb@open( name : String; mode : integer ) @ \\
\verb@    ( exception : CatchFileException )@\\ \hline
\verb@close() @ \\
\verb@    ( exception : CatchFileException )@\\ \hline
\verb@getSize() : integer @ \\
\verb@    ( exception : CatchFileException )@\\ \hline
\end{tabular}

\vspace*{5ex} \begin{tabular}{| |l ||} \hline
{\tt InputStream subclassOf BasicStream} \\ \hline \hline
\verb@init( name : String ) @ \\
\verb@    ( exception : CatchFileException )@\\ \hline
\verb@read( v : array(char)[]; n : long ) @ \\
\verb@    ( exception : CatchFileException )@\\ \hline
\verb@read( v : array(byte)[]; n : long ) @ \\
\verb@    ( exception : CatchFileException )@\\ \hline
\verb@read( s : DynString ) @ \\
\verb@    ( exception : CatchFileException )@\\ \hline
\verb@readln( s : DynString ) @ \\
\verb@    ( exception : CatchFileException )@\\ \hline
\end{tabular}

\vspace*{5ex} \begin{tabular}{| |l ||} \hline
{\tt OutputStream subclassOf BasicStream} \\ \hline \hline
\verb@init( name : String ) @ \\
\verb@    ( exception : CatchFileException )@\\ \hline
\verb@write( v : array(char)[] ) @ \\
\verb@    ( exception : CatchFileException )@\\ \hline
\verb@write( v : array(char)[]; n : long ) @ \\
\verb@    ( exception : CatchFileException )@\\ \hline
\verb@write( v : array(byte)[] ) @ \\
\verb@    ( exception : CatchFileException )@\\ \hline
\verb@write( v : array(byte)[]; n : long ) @ \\
\verb@    ( exception : CatchFileException )@\\ \hline
\verb@write( s : DynString ) @ \\
\verb@    ( exception : CatchFileException )@\\ \hline
\verb@writeln( s : DynString ) @ \\
\verb@    ( exception : CatchFileException )@\\ \hline
\end{tabular}

\newpage

\vspace*{5ex} \begin{tabular}{| |l ||} \hline
{\tt Stream subclassOf BasicStream}  \\ \hline \hline
\verb@open( name : String; mode : integer ) @ \\
\verb@    ( exception : CatchFileException )@\\ \hline
 This class has all  methods of {\tt
InputStream} and {\tt OutputStream}. \\ \hline

\end{tabular}

\vspace*{5ex} \begin{tabular}{| |l ||} \hline
{\tt CatchFileException} \\ \hline \hline
\verb@init()@\\ \hline
\verb@throw( exc : OpenFileException ) @ \\ \hline
\verb@throw( exc : CloseFileException ) @ \\ \hline
\verb@throw( exc : ReadFileException ) @ \\ \hline
\verb@throw( exc : WriteFileException ) @ \\ \hline
\end{tabular}

\vspace*{5ex} \begin{tabular}{| |l ||} \hline
{\tt Iter( T : Any )} \\ \hline \hline
\verb@more() : boolean; @ \\ \hline
\verb@next() : T @ \\ \hline
\verb@reset() @ \\ \hline
\verb@toArray() : array(T)[] @ \\ \hline
\end{tabular}

\vspace*{5ex} \begin{tabular}{| |l ||} \hline
{\tt IterFilter( T : Any )} \\ \hline \hline
\verb@init( piter : Iter(T); pf : Filter(T) ) @ \\ \hline
\verb@more() : boolean @ \\ \hline
\verb@next() : T @ \\ \hline
\verb@toArray() : array(T)[] @ \\ \hline
\verb@reset() @ \\ \hline
\end{tabular}

\vspace*{5ex} \begin{tabular}{| |l ||} \hline
{\tt Filter( T : Any )} \\ \hline \hline
\verb@test( x : T ) : boolean @ \\ \hline
\end{tabular}

\vspace*{5ex} \begin{tabular}{| |l ||} \hline
{\tt Command(T)} \\ \hline \hline
\verb@doIt( x : T ) : T @ \\ \hline
\end{tabular}

\vspace*{5ex} \begin{tabular}{| |l ||} \hline
{\tt Function(T)} \\ \hline \hline
\verb@exec( x : T ) @ \\ \hline
\end{tabular}

\newpage

\vspace*{5ex} \begin{tabular}{| |l ||} \hline
{\tt Container( T : Any )} \\ \hline \hline
\verb@init() @ \\ \hline
\verb@add( elem : T ) @ \\
\verb@  ( exception : CatchOutOfMemoryException )@\\ \hline
\verb@get() : T @ \\
\verb@   ( exception : CatchNotFoundException )@\\ \hline
\verb@remove() : T @ \\
\verb@   ( exception : CatchNotFoundException )@\\ \hline
\verb@removeAll() @ \\ \hline
\verb@getSize() : integer @ \\ \hline
\verb@empty() : boolean @ \\ \hline
\verb@full() : boolean @ \\ \hline
\verb@getIter() : DS.Iter(T) @ \\ \hline
\verb@forEach( f : Function(T) ) @ \\ \hline
\verb@replaceBy( cmd : Command(T) ) @ \\ \hline
\verb@collect( f : Filter(T) ) : Container(T) @ \\ \hline
\verb@remove( f : Filter(T) ) @ \\ \hline
\verb@reset() @ \\ \hline
\verb@next() : T @ \\ \hline
\verb@endIter() @ \\ \hline
\end{tabular}

\vspace*{5ex} \begin{tabular}{| |l ||} \hline
{\tt List( T : Any ) subclassOf Container(T)} \\ \hline \hline
no methods defined in this class\\ \hline

\end{tabular}

\vspace*{5ex} \begin{tabular}{| |l ||} \hline
{\tt DList( T : Any ) subclassOf Container(T) } \\ \hline \hline
no methods defined in this class\\ \hline
\end{tabular}

\vspace*{5ex} \begin{tabular}{| |l ||} \hline
{\tt Stack( T : Any ) subclassOf Container(T) } \\ \hline \hline
no methods defined in this class\\ \hline
\end{tabular}

\vspace*{5ex} \begin{tabular}{| |l ||} \hline

{\tt DStack( T : Any ) subclassOf Container(T)}\\ \hline \hline
no methods defined in this class\\ \hline

\end{tabular}

\vspace*{5ex} \begin{tabular}{| |l ||} \hline

{\tt Queue( T : Any ) subclassOf Container(T)}\\ \hline \hline
no methods defined in this class\\ \hline

\end{tabular}

\newpage
\vspace*{5ex} \begin{tabular}{| |l ||} \hline

{\tt DQueue( T : Any ) subclassOf Container(T)}\\ \hline \hline
no methods defined in this class\\ \hline

\end{tabular}

\vspace*{5ex} \begin{tabular}{| |l ||} \hline

{\tt DoubleQueue( T : Any ) subclassOf Container(T)}\\ \hline \hline

\verb@addFront( elem : T )@\\
\verb@        ( exception : CatchOutOfMemoryException )@\\ \hline
\verb@addBack( elem : T )@\\
\verb@       ( exception : CatchOutOfMemoryException )@\\ \hline
\verb@getFront() : T@\\
\verb@        ( exception : CatchNotFoundException )@\\ \hline
\verb@getBack() : T@\\
\verb@       ( exception : CatchNotFoundException )@\\ \hline
\verb@removeFront() : T@\\
\verb@           ( exception : CatchNotFoundException )@\\ \hline
\verb@removeBack() : T@\\
\verb@          ( exception : CatchNotFoundException )@\\ \hline

\end{tabular}

\vspace*{5ex} \begin{tabular}{| |l ||} \hline

{\tt DDoubleQueue( T : Any ) subclassOf Container(T)}\\ \hline \hline
no methods defined in this class\\ \hline

\end{tabular}

\vspace*{5ex} \begin{tabular}{| |l ||} \hline

{\tt Vector( T : Any )  subclassOf  Container(T)}\\ \hline \hline
\verb@init( p\_size : integer )@\\ \hline
\verb@init()@\\ \hline
\verb@add( elem : T )@\\
\verb@   ( exception : CatchOutOfMemoryException )@\\ \hline
\verb@add( elem : T; i : integer )@\\
\verb@   ( exception : CatchOutOfMemoryException )@\\ \hline
\verb@get() : T@\\
\verb@   ( exception : CatchNotFoundException )@\\ \hline
\verb@get( i : integer ) : T@\\
\verb@   ( exception : CatchNotFoundException )@\\ \hline
\verb@remove() : T@\\
\verb@      ( exception : CatchNotFoundException )@\\ \hline
\verb@remove( i : integer ) : T@\\
\verb@      ( exception : CatchNotFoundException )@\\ \hline
\verb@getNum() : integer@\\ \hline
\verb@setNum( p\_num : integer )@\\ \hline

\end{tabular}

\newpage

\vspace*{5ex} \begin{tabular}{| |l ||} \hline

{\tt HashFunction( T :  Any )}\\ \hline \hline
\verb@setSize( psize : integer )@\\ \hline
\verb@hash( elem : T ) : integer@\\ \hline

\end{tabular}

\vspace*{5ex} \begin{tabular}{| |l ||} \hline

{\tt Compose( T, U : Any )}\\ \hline \hline
\verb@init( pt : T; pu : U )@\\ \hline
\verb@getT() : T @\\ \hline
\verb@getU() : U@\\ \hline

\end{tabular}

\vspace*{5ex} \begin{tabular}{| |l ||} \hline

{\tt Dict( KeyType, ValueType : Any )}\\ \hline \hline
\verb@init()@\\ \hline
\verb@init( pSize : integer )@\\ \hline
\verb@init( pSize : integer; phashFunction : HashFunction(KeyType) )@\\ \hline
\verb@add( key : KeyType; value : ValueType )@\\
\verb@   ( exception : CatchOutOfMemoryException )@\\ \hline
\verb@get( key : KeyType ) : U@\\
\verb@   ( exception : CatchNotFoundException )@\\ \hline
\verb@remove( key : KeyType ) : U@\\
\verb@      ( exception : CatchNotFoundException )@\\ \hline
\verb@getSize() : integer@\\ \hline
\verb@removeAll()@\\ \hline
\verb@getIter() : DS.DictIter(KeyType, ValueType)@\\ \hline

\end{tabular}

\vspace*{5ex} \begin{tabular}{| |l ||} \hline

{\tt DictIter( KeyType, ValueType : Any )}\\ \hline \hline
\verb@init( pdict : Dict(KeyType, ValueType) )@\\ \hline
\verb@more() : boolean@\\ \hline
\verb@next() : Container(KeyType, ValueType)@\\ \hline

\end{tabular}

\vspace*{5ex} \begin{tabular}{| |l ||} \hline

{\tt IntegerSet}\\ \hline \hline
\verb@init()@\\ \hline
\verb@add( n : integer )@\\
\verb@   ( exception : CatchOutOfMemoryException )@\\ \hline
\verb@remove( n : integer )@\\
\verb@      ( exception : CatchNotFoundException )@\\ \hline
\verb@inSet( n : integer ) : boolean@\\ \hline
\verb@empty() : boolean@\\ \hline
\verb@getIter() : DS.Iter(integer)@\\ \hline
\verb@removeAll()@\\ \hline

\end{tabular}

\newpage
\vspace*{5ex} \begin{tabular}{| |l ||} \hline

 {\tt ClassInfo}\\ \hline \hline
\verb@getName() : String@\\ \hline
\verb@toString() : String@\\ \hline
\verb@isSupertypeOf( aSubtype : ClassInfo ) : boolean@\\ \hline
\verb@isSuperclassOf( aSubclass : ClassInfo ) : boolean@\\ \hline
\verb@getSuperclass() : ClassInfo@\\ \hline
\verb@getInstanceVariables() : DS.Iter(ClassInstanceVariableInfo)@\\ \hline
\verb@getMethods() : DS.Iter(ClassMethodInfo)@\\ \hline
\verb@getPublicMethods() : DS.Iter(ClassMethodInfo)@\\ \hline
\verb@getInstanceVariable(name : String) : ClassInstanceVariableInfo@\\ \hline
\verb@getMethod( name : String; paramTypes : array(ClassInfo)[] ) : ClassMethodInfo@\\ \hline
\verb@getMethod\_v( name : String; paramTypes : ... array(AnyClassObject)[]@\\
\verb@              ) : ClassMethodInfo@\\ \hline
\verb@getPublicMethod( name : String ) : DS.Iter(ClassMethodInfo)@\\ \hline
\verb@getThisClassInstanceVariables() : DS.Iter(ClassInstanceVariableInfo)@\\ \hline
\verb@getThisClassMethods() : DS.Iter(ClassMethodInfo)@\\ \hline
\verb@getThisClassPublicMethods() : DS.Iter(ClassMethodInfo)@\\ \hline
\verb@getAssociateClassObject() : AnyClassObject@\\ \hline
\verb@isClassOf( any : Any ) : boolean@\\ \hline
\verb@isAbstract() : boolean@\\ \hline
\verb@getNotes() : DS.Iter(CodeAnnotations)@\\ \hline

\end{tabular}

\vspace*{5ex} \begin{tabular}{| |l ||} \hline

{\tt ValueClassInfo  subclassOf  ClassInfo}\\ \hline \hline
no methods defined in this class\\ \hline

\end{tabular}

\vspace*{5ex} \begin{tabular}{| |l ||} \hline

{\tt RefClassInfo subclassOf ClassInfo}\\ \hline \hline
no methods defined in this class\\ \hline

\end{tabular}

\vspace*{5ex} \begin{tabular}{| |l ||} \hline

{\tt AbstractClassInfo subclassOf RefClassInfo}\\ \hline \hline
no methods defined in this class\\ \hline

\end{tabular}

\vspace*{5ex} \begin{tabular}{| |l ||} \hline

 {\tt  ConcreteClassInfo subclassOf RefClassInfo}\\ \hline \hline
no methods defined in this class\\ \hline

  \end{tabular}

  \vspace*{5ex} \begin{tabular}{| |l ||} \hline

{\tt NormalClassInfo subclassOf ConcreteClassInfo}\\ \hline \hline
no methods defined in this class\\ \hline

\end{tabular}

\newpage

\vspace*{5ex} \begin{tabular}{| |l ||} \hline

{\tt ArrayClassInfo subclassOf ConcreteClassInfo}\\ \hline \hline
\verb@getArrayElementClass() : ClassInfo@\\ \hline
\verb@getNumberOfDimensions() : integer@\\ \hline
\verb@toString() : String@\\ \hline

\end{tabular}

\vspace*{5ex} \begin{tabular}{| |l ||} \hline

{\tt MethodInfo}\\ \hline \hline
\verb@getName() : String@\\ \hline
\verb@toString() : String@\\ \hline
\verb@getBodyInfo() : MethodBodyInfo@\\
\verb@           ( exception : CatchNoReflectiveBodyInfoException )@\\ \hline
\verb@getVisibility() : integer@\\ \hline
\verb@getParameterTypes() : DS.Iter(ClassInfo)@\\ \hline
\verb@getReturnType() : ClassInfo@\\ \hline
\verb@getParameters() : DS.Iter(ParameterInfo)@\\ \hline
\verb@getExceptionClass() : ClassInfo@\\ \hline
\verb@getNotes() : DS.Iter(CodeAnnotations)@\\ \hline

\end{tabular}

\vspace*{5ex} \begin{tabular}{| |l ||} \hline

 {\tt ClassMethodInfo subclassOf  MethodInfo} \\ \hline \hline
\verb@isAbstract() : boolean@\\ \hline
\verb@invoke( obj : Any; args : array(Any)[] ) : Any@\\
\verb@      ( exception : CatchInvokePackedException )@\\ \hline
\verb@invoke\_v( obj : Any; args : ... array(Any)[] ) : Any@\\
\verb@      ( exception : CatchInvokePackedException )@\\ \hline

\end{tabular}

\vspace*{5ex} \begin{tabular}{| |l ||} \hline

{\tt ObjectMethodInfo subclassOf  MethodInfo}\\ \hline \hline
\verb@invoke( args : array(Any)[] ) : Any@\\
\verb@      ( exception : CatchInvokePackedException )@\\ \hline
\verb@invoke\_v( args : ...array(Any)[] ) : Any@\\
\verb@      ( exception : CatchInvokePackedException )@\\ \hline

\end{tabular}

\vspace*{5ex} \begin{tabular}{| |l ||} \hline

{\tt InstanceVariableInfo}\\ \hline \hline
\verb@getName() : String@\\ \hline
\verb@getType() : ClassInfo@\\ \hline
\verb@getNotes() : DS.Iter(CodeAnnotations)@\\ \hline

\end{tabular}

\newpage

\vspace*{5ex} \begin{tabular}{| |l ||} \hline

{\tt ClassInstanceVariableInfo subclassOf InstanceVariableInfo} \\
\hline \hline
\verb@set( obj : Any; v : Any )@\\
\verb@   ( exception : CatchTypeErrorException )@\\ \hline
\verb@get( obj : Any ) : Any@\\
\verb@   ( exception : CatchTypeErrorException )@\\ \hline

\end{tabular}

\vspace*{5ex} \begin{tabular}{| |l ||} \hline

{\tt ObjectInstanceVariableInfo subclassOf  InstanceVariableInfo} \\
\hline \hline
\verb@set( v : Any )@\\
\verb@   ( exception : CatchTypeErrorException )@\\ \hline
\verb@get() : Any@\\ \hline

\end{tabular}

\vspace*{5ex} \begin{tabular}{| |l ||} \hline

{\tt AnyObjectInfo}\\ \hline \hline
\verb@getObject() : Any@\\ \hline
\verb@getInstanceVariables() : DS.Iter(ObjectInstanceVariableInfo)[]@\\ \hline
\verb@getMethods() : DS.Iter(ObjectMethodInfo)@\\ \hline
\verb@getPublicMethods() : DS.Iter(ObjectMethodInfo)@\\ \hline
\verb@getPublicMethod( name : String ) : DS.Iter(ObjectMethodInfo)@\\ \hline
\verb@getMethod( name : String; paramTypes : array(ClassInfo)[] ) : ObjectMethodInfo@\\ \hline
\verb@getMethod( name : String ) : ObjectMethodInfo@\\ \hline
\verb@getInstanceVariable( name : String ) : ObjectInstanceVariableInfo@\\ \hline
\verb@getTypeInfo() : ClassInfo@\\ \hline

\end{tabular}

\vspace*{5ex} \begin{tabular}{| |l ||} \hline

{\tt ObjectInfo subclassOf AnyObjectInfo}\\ \hline \hline
no methods defined in this class\\ \hline

\end{tabular}

\vspace*{5ex} \begin{tabular}{| |l ||} \hline

{\tt ClassObjectInfo subclassOf  AnyObjectInfo}\\ \hline \hline

\verb@new( args : array(Any)[] )@\\
\verb@   ( exception : CatchNewException ) : Any@\\ \hline
\verb@new\_v( args : ... array(Any)[] )@\\
\verb@     ( exception : CatchNewException ) : Any@\\ \hline
\verb@getNotes() : DS.Iter(CodeAnnotations)@\\ \hline

\end{tabular}

\vspace*{5ex} \begin{tabular}{| |l ||} \hline

{\tt MethodBodyInfo}\\ \hline \hline
\verb@getMethodInfo() : MethodInfo@\\ \hline
\verb@getAssertionInfo() : AssertionInfo@\\ \hline
\verb@getLocalVariables() : DS.Iter(VariableInfo)@\\ \hline
\verb@getStatements() : DS.Iter(StatementNo)@\\ \hline

\end{tabular}

\newpage

\vspace*{5ex} \begin{tabular}{| |l ||} \hline

{\tt AssertionInfo}\\ \hline \hline
\verb@getBeforeExpression() : ExprAST@\\ \hline
\verb@getAfterExpression() : ExprAST@\\ \hline
\verb@getStatVarDeclarations() : DS.Iter(StatVarDecNo)@\\ \hline

\end{tabular}

\vspace*{5ex} \begin{tabular}{| |l ||} \hline

{\tt VariableInfo}\\ \hline \hline
\verb@getName() : String@\\ \hline
\verb@getType() : ClassInfo@\\ \hline
\verb@getDeclaringMethod() : MethodInfo@\\ \hline
\verb@getNotes() : DS.Iter(CodeAnnotations)@\\ \hline

\end{tabular}

\vspace*{5ex} \begin{tabular}{| |l ||} \hline

{\tt ParameterInfo subclassOf VariableInfo}\\ \hline \hline
\verb@isVariableNumber() : boolean@\\ \hline

\end{tabular}

\vspace*{5ex} \begin{tabular}{| |l ||} \hline

{\tt MethodCallInfo}\\ \hline \hline
\verb@getMethodInfo() : MethodInfo@\\ \hline
\verb@getLiveLocalVariables() : DS.Iter(LiveLocalVariableInfo)@\\ \hline
\verb@getLiveParameters() : DS.Iter(LiveParameterInfo)@\\ \hline

\end{tabular}

\vspace*{5ex} \begin{tabular}{| |l ||} \hline

{\tt LiveLocalVariableInfo}\\ \hline \hline
\verb@getVariableInfo() : VariableInfo@\\ \hline
\verb@set( v : Any )@\\
\verb@   ( exception : CatchTypeErrorException )@\\ \hline
\verb@get() : Any@\\ \hline

\end{tabular}

\vspace*{5ex} \begin{tabular}{| |l ||} \hline

{\tt LiveParameterInfo  subclassOf  LiveLocalVariableInfo}\\ \hline \hline
no methods defined in this class\\ \hline

\end{tabular}

\end{document}